\documentclass[longauth]{aa}

\usepackage{siunitx}
\usepackage{gensymb}
\usepackage{xcolor}
\usepackage{graphicx}
\usepackage{txfonts,textcomp}
\usepackage[htt]{hyphenat}
\usepackage{amsmath}    
\usepackage{multicol}
\usepackage{multirow}
\usepackage[colorlinks=true,citecolor=blue]{hyperref}
\usepackage{tablefootnote}

\newcommand{\pycheops}{\texttt{PyCHEOPS}\xspace}
\newcommand{\teff}{\ensuremath{T_{\mbox{\scriptsize eff}}}\xspace}
\newcommand{\DRP}{\texttt{DRP}\xspace}
\newcommand{\PIPE}{\texttt{PIPE}\xspace}
\newcommand{\DRT}{\texttt{DRT}\xspace}
\newcommand{\CHEOPS}{\textrm{CHEOPS}\xspace}
\newcommand{\CHEOPSim}{\texttt{CHEOPSim}\xspace}

\begin{document} 

\title{CHEOPS in-flight performance }
\subtitle{A comprehensive look at the first 3.5 years of operations}


\author{
A. Fortier\inst{1,2} $^{\href{https://orcid.org/0000-0001-8450-3374}{\includegraphics[scale=0.01]{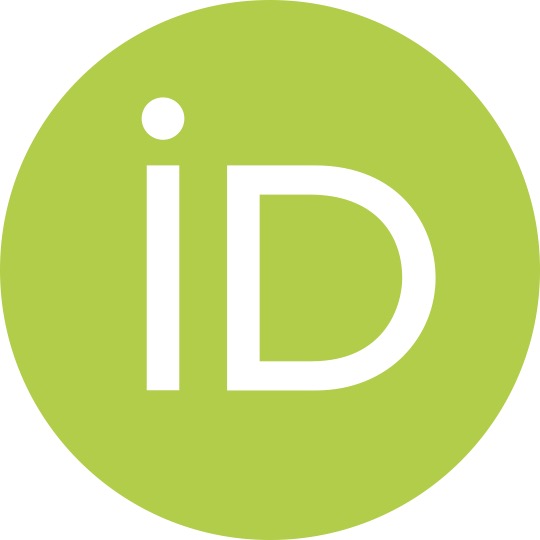}}}$, \thanks{E-mail:andrea.fortier@unibe.ch}\and
A. E. Simon\inst{1,2} $^{\href{https://orcid.org/0000-0001-9773-2600}{\includegraphics[scale=0.01]{figures/orcid.jpg}}}$\and
C. Broeg\inst{1,2} $^{\href{https://orcid.org/0000-0001-5132-2614}{\includegraphics[scale=0.01]{figures/orcid.jpg}}}$\and
G. Olofsson\inst{3} $^{\href{https://orcid.org/0000-0003-3747-7120}{\includegraphics[scale=0.01]{figures/orcid.jpg}}}$\and
A. Deline\inst{4}\and
T. G. Wilson\inst{5} $^{\href{https://orcid.org/0000-0001-8749-1962}{\includegraphics[scale=0.01]{figures/orcid.jpg}}}$\and
P. F. L. Maxted\inst{6} $^{\href{https://orcid.org/0000-0003-3794-1317}{\includegraphics[scale=0.01]{figures/orcid.jpg}}}$\and
A. Brandeker\inst{3} $^{\href{https://orcid.org/0000-0002-7201-7536}{\includegraphics[scale=0.01]{figures/orcid.jpg}}}$\and
A. Collier Cameron\inst{7} $^{\href{https://orcid.org/0000-0002-8863-7828}{\includegraphics[scale=0.01]{figures/orcid.jpg}}}$\and
M. Beck\inst{4} $^{\href{https://orcid.org/0000-0003-3926-0275}{\includegraphics[scale=0.01]{figures/orcid.jpg}}}$\and
A. Bekkelien\inst{4}\and
N. Billot\inst{4} $^{\href{https://orcid.org/0000-0003-3429-3836}{\includegraphics[scale=0.01]{figures/orcid.jpg}}}$\and
A. Bonfanti\inst{8} $^{\href{https://orcid.org/0000-0002-1916-5935}{\includegraphics[scale=0.01]{figures/orcid.jpg}}}$\and
G. Bruno\inst{9} $^{\href{https://orcid.org/0000-0002-3288-0802}{\includegraphics[scale=0.01]{figures/orcid.jpg}}}$\and
J. Cabrera\inst{10} $^{\href{https://orcid.org/0000-0001-6653-5487}{\includegraphics[scale=0.01]{figures/orcid.jpg}}}$\and
L. Delrez\inst{11,12} $^{\href{https://orcid.org/0000-0001-6108-4808}{\includegraphics[scale=0.01]{figures/orcid.jpg}}}$\and
B.-O. Demory\inst{2,1} $^{\href{https://orcid.org/0000-0002-9355-5165}{\includegraphics[scale=0.01]{figures/orcid.jpg}}}$\and
D. Futyan\inst{4}\and
H.-G. Florén\inst{3}\and
M. N. Günther\inst{13} $^{\href{https://orcid.org/0000-0002-3164-9086}{\includegraphics[scale=0.01]{figures/orcid.jpg}}}$\and
A. Heitzmann\inst{4} $^{\href{https://orcid.org/0000-0002-8091-7526}{\includegraphics[scale=0.01]{figures/orcid.jpg}}}$\and
S. Hoyer\inst{14} $^{\href{https://orcid.org/0000-0003-3477-2466}{\includegraphics[scale=0.01]{figures/orcid.jpg}}}$\and
K. G. Isaak\inst{13} $^{\href{https://orcid.org/0000-0001-8585-1717}{\includegraphics[scale=0.01]{figures/orcid.jpg}}}$\and
S. G. Sousa\inst{15} $^{\href{https://orcid.org/0000-0001-9047-2965}{\includegraphics[scale=0.01]{figures/orcid.jpg}}}$\and
M. Stalport\inst{12,11}\and
A. Turin\inst{14}\and
P. Verhoeve\inst{13}\and
B. Akinsanmi\inst{4}\and
Y. Alibert\inst{2,1} $^{\href{https://orcid.org/0000-0002-4644-8818}{\includegraphics[scale=0.01]{figures/orcid.jpg}}}$\and
R. Alonso\inst{16,17} $^{\href{https://orcid.org/0000-0001-8462-8126}{\includegraphics[scale=0.01]{figures/orcid.jpg}}}$\and
D. Bánhidi\inst{18}\and
T. Bárczy\inst{19} $^{\href{https://orcid.org/0000-0002-7822-4413}{\includegraphics[scale=0.01]{figures/orcid.jpg}}}$\and
D. Barrado\inst{20} $^{\href{https://orcid.org/0000-0002-5971-9242}{\includegraphics[scale=0.01]{figures/orcid.jpg}}}$\and
S. C. C. Barros\inst{15,21} $^{\href{https://orcid.org/0000-0003-2434-3625}{\includegraphics[scale=0.01]{figures/orcid.jpg}}}$\and
W. Baumjohann\inst{8} $^{\href{https://orcid.org/0000-0001-6271-0110}{\includegraphics[scale=0.01]{figures/orcid.jpg}}}$\and
T. Baycroft\inst{22}\and
T. Beck\inst{1}\and
W. Benz\inst{1,2} $^{\href{https://orcid.org/0000-0001-7896-6479}{\includegraphics[scale=0.01]{figures/orcid.jpg}}}$\and
B. I. Bíró\inst{18,23} $^{\href{https://orcid.org/0000-0001-9061-2147}{\includegraphics[scale=0.01]{figures/orcid.jpg}}}$\and
A. Bódi\inst{24,25} $^{\href{https://orcid.org/0000-0002-8585-4544}{\includegraphics[scale=0.01]{figures/orcid.jpg}}}$\and
X. Bonfils\inst{26} $^{\href{https://orcid.org/0000-0001-9003-8894}{\includegraphics[scale=0.01]{figures/orcid.jpg}}}$\and
L. Borsato\inst{27} $^{\href{https://orcid.org/0000-0003-0066-9268}{\includegraphics[scale=0.01]{figures/orcid.jpg}}}$\and
S. Charnoz\inst{28} $^{\href{https://orcid.org/0000-0002-7442-491X}{\includegraphics[scale=0.01]{figures/orcid.jpg}}}$\and
B. Cseh\inst{24,29} $^{\href{https://orcid.org/0000-0002-6497-8863}{\includegraphics[scale=0.01]{figures/orcid.jpg}}}$\and
Sz. Csizmadia\inst{10} $^{\href{https://orcid.org/0000-0001-6803-9698}{\includegraphics[scale=0.01]{figures/orcid.jpg}}}$\and
I. Cs{\'a}nyi\inst{18}\and
P. E. Cubillos\inst{30,8}\and
M. B. Davies\inst{31} $^{\href{https://orcid.org/0000-0001-6080-1190}{\includegraphics[scale=0.01]{figures/orcid.jpg}}}$\and
Y. T. Davis\inst{22}\and
M. Deleuil\inst{14} $^{\href{https://orcid.org/0000-0001-6036-0225}{\includegraphics[scale=0.01]{figures/orcid.jpg}}}$\and
O. D. S. Demangeon\inst{15,21} $^{\href{https://orcid.org/0000-0001-7918-0355}{\includegraphics[scale=0.01]{figures/orcid.jpg}}}$\and
A. Derekas\inst{69,74} $^{\href{https://orcid.org/0000-0002-6526-9444}{\includegraphics[scale=0.01]{figures/orcid.jpg}}}$\and
G. Dransfield\inst{22}\and
E. Ducrot\inst{32}\and
D. Ehrenreich\inst{4,33} $^{\href{https://orcid.org/0000-0001-9704-5405}{\includegraphics[scale=0.01]{figures/orcid.jpg}}}$\and
A. Erikson\inst{10}\and
C. Fariña\inst{34,17}\and
L. Fossati\inst{8} $^{\href{https://orcid.org/0000-0003-4426-9530}{\includegraphics[scale=0.01]{figures/orcid.jpg}}}$\and
M. Fridlund\inst{35,36} $^{\href{https://orcid.org/0000-0002-0855-8426}{\includegraphics[scale=0.01]{figures/orcid.jpg}}}$\and
D. Gandolfi\inst{37} $^{\href{https://orcid.org/0000-0001-8627-9628}{\includegraphics[scale=0.01]{figures/orcid.jpg}}}$\and
Z. Garai\inst{70, 69, 73} $^{\href{https://orcid.org/0000-0001-9483-2016}{\includegraphics[scale=0.01]{figures/orcid.jpg}}}$\and
L. Garcia\inst{11}\and
M. Gillon\inst{11} $^{\href{https://orcid.org/0000-0003-1462-7739}{\includegraphics[scale=0.01]{figures/orcid.jpg}}}$\and
Y. Gómez Maqueo Chew\inst{38} $^{\href{https://orcid.org/0000-0002-7486-6726}{\includegraphics[scale=0.01]{figures/orcid.jpg}}}$\and
M.A. Gómez-Muñoz\inst{39}\and
V. Granata\inst{40,27} $^{\href{https://orcid.org/0000-0002-1425-4541}{\includegraphics[scale=0.01]{figures/orcid.jpg}}}$\and
M. Güdel\inst{41}\and
P. Guterman\inst{14,42}\and
T. Heged{\"u}s\inst{18}\and
Ch. Helling\inst{8,43}\and
E. Jehin\inst{12}\and
Cs. Kalup\inst{24,44} $^{\href{https://orcid.org/0000-0002-1663-0707}{\includegraphics[scale=0.01]{figures/orcid.jpg}}}$\and
D. Kilkenny\inst{45}\and
L. L. Kiss\inst{24,46}\and
L. Kriskovics\inst{24} $^{\href{https://orcid.org/0000-0002-1792-546X}{\includegraphics[scale=0.01]{figures/orcid.jpg}}}$\and
K. W. F. Lam\inst{10} $^{\href{https://orcid.org/0000-0002-9910-6088}{\includegraphics[scale=0.01]{figures/orcid.jpg}}}$\and
J. Laskar\inst{47} $^{\href{https://orcid.org/0000-0003-2634-789X}{\includegraphics[scale=0.01]{figures/orcid.jpg}}}$\and
A. Lecavelier des Etangs\inst{48} $^{\href{https://orcid.org/0000-0002-5637-5253}{\includegraphics[scale=0.01]{figures/orcid.jpg}}}$\and
M. Lendl\inst{4} $^{\href{https://orcid.org/0000-0001-9699-1459}{\includegraphics[scale=0.01]{figures/orcid.jpg}}}$\and
A. Lopez Pina\inst{49}\and
A. Luntzer\inst{41}\and
D. Magrin\inst{27} $^{\href{https://orcid.org/0000-0003-0312-313X}{\includegraphics[scale=0.01]{figures/orcid.jpg}}}$\and
N. J. Miller\inst{6,50} $^{\href{https://orcid.org/0000-0001-9550-1198}{\includegraphics[scale=0.01]{figures/orcid.jpg}}}$\and
D. Modrego Contreras\inst{51}\and
C. Mordasini\inst{1,2}\and
M. Munari\inst{9} $^{\href{https://orcid.org/0000-0003-0990-050X}{\includegraphics[scale=0.01]{figures/orcid.jpg}}}$\and
C. A. Murray\inst{52}\and
V. Nascimbeni\inst{27} $^{\href{https://orcid.org/0000-0001-9770-1214}{\includegraphics[scale=0.01]{figures/orcid.jpg}}}$\and
H. Ottacher\inst{8}\and
R. Ottensamer\inst{41}\and
I. Pagano\inst{9} $^{\href{https://orcid.org/0000-0001-9573-4928}{\includegraphics[scale=0.01]{figures/orcid.jpg}}}$\and
A. Pál\inst{24} $^{\href{https://orcid.org/0000-0001-5449-2467}{\includegraphics[scale=0.01]{figures/orcid.jpg}}}$\and
E. Pallé\inst{16,17} $^{\href{https://orcid.org/0000-0003-0987-1593}{\includegraphics[scale=0.01]{figures/orcid.jpg}}}$\and
A. Pasetti\inst{53}\and
P. P. Pedersen\inst{54,55}\and
G. Peter\inst{56} $^{\href{https://orcid.org/0000-0001-6101-2513}{\includegraphics[scale=0.01]{figures/orcid.jpg}}}$\and
R. Petrucci\inst{57,58}\and
G. Piotto\inst{27,59} $^{\href{https://orcid.org/0000-0002-9937-6387}{\includegraphics[scale=0.01]{figures/orcid.jpg}}}$\and
A. Pizarro-Rubio\inst{49}\and
D. Pollacco\inst{5}\and
T. Pribulla\inst{73} $^{\href{https://orcid.org/0000-0003-3599-516X}{\includegraphics[scale=0.01]{figures/orcid.jpg}}}$\and
D. Queloz\inst{55,54} $^{\href{https://orcid.org/0000-0002-3012-0316}{\includegraphics[scale=0.01]{figures/orcid.jpg}}}$\and
R. Ragazzoni\inst{27,59} $^{\href{https://orcid.org/0000-0002-7697-5555}{\includegraphics[scale=0.01]{figures/orcid.jpg}}}$\and
N. Rando\inst{13}\and
H. Rauer\inst{10,60,61} $^{\href{https://orcid.org/0000-0002-6510-1828}{\includegraphics[scale=0.01]{figures/orcid.jpg}}}$\and
I. Ribas\inst{62,63} $^{\href{https://orcid.org/0000-0002-6689-0312}{\includegraphics[scale=0.01]{figures/orcid.jpg}}}$\and
L. Sabin\inst{39}\and
N. C. Santos\inst{15,21} $^{\href{https://orcid.org/0000-0003-4422-2919}{\includegraphics[scale=0.01]{figures/orcid.jpg}}}$\and
G. Scandariato\inst{9} $^{\href{https://orcid.org/0000-0003-2029-0626}{\includegraphics[scale=0.01]{figures/orcid.jpg}}}$\and
N. Schanche\inst{64,65}\and
U. Schroffenegger\inst{2}\and
O. J. Scutt\inst{22}\and
D. Sebastian\inst{22}\and
D. Ségransan\inst{4} $^{\href{https://orcid.org/0000-0003-2355-8034}{\includegraphics[scale=0.01]{figures/orcid.jpg}}}$\and
B. Seli\inst{24,66} $^{\href{https://orcid.org/0000-0002-3658-2175}{\includegraphics[scale=0.01]{figures/orcid.jpg}}}$\and
A. M. S. Smith\inst{10} $^{\href{https://orcid.org/0000-0002-2386-4341}{\includegraphics[scale=0.01]{figures/orcid.jpg}}}$\and
R. Southworth\inst{67}\and
M. R. Standing\inst{22,68}\and
Gy. M. Szabó\inst{69,70} $^{\href{https://orcid.org/0000-0002-0606-7930}{\includegraphics[scale=0.01]{figures/orcid.jpg}}}$\and
R. Szakáts\inst{24} $^{\href{https://orcid.org/0000-0002-1698-605X}{\includegraphics[scale=0.01]{figures/orcid.jpg}}}$\and
N. Thomas\inst{1}\and
M. Timmermans\inst{11}\and
A. H. M. J. Triaud\inst{22}\and
S. Udry\inst{4} $^{\href{https://orcid.org/0000-0001-7576-6236}{\includegraphics[scale=0.01]{figures/orcid.jpg}}}$\and
V. Van Grootel\inst{12} $^{\href{https://orcid.org/0000-0003-2144-4316}{\includegraphics[scale=0.01]{figures/orcid.jpg}}}$\and
J. Venturini\inst{4} $^{\href{https://orcid.org/0000-0001-9527-2903}{\includegraphics[scale=0.01]{figures/orcid.jpg}}}$\and
E. Villaver\inst{16,17}\and
J. Vinkó\inst{24,66,71} $^{\href{https://orcid.org/0000-0001-8764-7832}{\includegraphics[scale=0.01]{figures/orcid.jpg}}}$\and
N. A. Walton\inst{72} $^{\href{https://orcid.org/0000-0003-3983-8778}{\includegraphics[scale=0.01]{figures/orcid.jpg}}}$\and
R. Wells\inst{2}\and
D. Wolter\inst{10}
}
\institute{
\label{inst:1} Weltraumforschung und Planetologie, Physikalisches Institut, University of Bern, Gesellschaftsstrasse 6, 3012 Bern, Switzerland \and
\label{inst:2} Center for Space and Habitability, University of Bern, Gesellschaftsstrasse 6, 3012 Bern, Switzerland \and
\label{inst:3} Department of Astronomy, Stockholm University, AlbaNova University Center, 10691 Stockholm, Sweden \and
\label{inst:4} Observatoire astronomique de l'Université de Genève, Chemin Pegasi 51, 1290 Versoix, Switzerland \and
\label{inst:5} Department of Physics, University of Warwick, Gibbet Hill Road, Coventry CV4 7AL, United Kingdom \and
\label{inst:6} Astrophysics Group, Lennard Jones Building, Keele University, Staffordshire, ST5 5BG, United Kingdom \and
\label{inst:7} Centre for Exoplanet Science, SUPA School of Physics and Astronomy, University of St Andrews, North Haugh, St Andrews KY16 9SS, UK \and
\label{inst:8} Space Research Institute, Austrian Academy of Sciences, Schmiedlstrasse 6, A-8042 Graz, Austria \and
\label{inst:9} INAF, Osservatorio Astrofisico di Catania, Via S. Sofia 78, 95123 Catania, Italy \and
\label{inst:10} Institute of Planetary Research, German Aerospace Center (DLR), Rutherfordstrasse 2, 12489 Berlin, Germany \and
\label{inst:11} Astrobiology Research Unit, Université de Liège, Allée du 6 Août 19C, B-4000 Liège, Belgium \and
\label{inst:12} Space Sciences, Technologies and Astrophysics Research (STAR) Institute, Université de Liège, Allée du 6 Août 19C, 4000 Liège, Belgium \and
\label{inst:13} European Space Agency (ESA), European Space Research and Technology Centre (ESTEC), Keplerlaan 1, 2201 AZ Noordwijk, The Netherlands \and
\label{inst:14} Aix Marseille Univ, CNRS, CNES, LAM, 38 rue Frédéric Joliot-Curie, 13388 Marseille, France \and
\label{inst:15} Instituto de Astrofisica e Ciencias do Espaco, Universidade do Porto, CAUP, Rua das Estrelas, 4150-762 Porto, Portugal \and
\label{inst:16} Instituto de Astrofísica de Canarias, Vía Láctea s/n, 38200 La Laguna, Tenerife, Spain \and
\label{inst:17} Departamento de Astrofísica, Universidad de La Laguna, Astrofísico Francisco Sanchez s/n, 38206 La Laguna, Tenerife, Spain \and
\label{inst:18} Baja Astronomical Observatory of University of Szeged, Szegedi {\'u}t Kt. 766, 6500 Baja, Hungary \and
\label{inst:19} Admatis, 5. Kandó Kálmán Street, 3534 Miskolc, Hungary \and
\label{inst:20} Depto. de Astrofísica, Centro de Astrobiología (CSIC-INTA), ESAC campus, 28692 Villanueva de la Ca\~nada (Madrid), Spain \and
\label{inst:21} Departamento de Fisica e Astronomia, Faculdade de Ciencias, Universidade do Porto, Rua do Campo Alegre, 4169-007 Porto, Portugal \and
\label{inst:22} School of Physics \& Astronomy, University of Birmingham, Edgbaston, Birmingham B15 2TT, United Kingdom \and
\label{inst:23} ELKH-SZTE Stellar Astrophysics Research Group, Szegedi {\'u}t Kt. 766, 6500 Baja, Hungary \and
\label{inst:24} Konkoly Observatory, Research Centre for Astronomy and Earth Sciences, 1121 Budapest, Konkoly Thege Miklós út 15-17, Hungary \and
\label{inst:25} MTA CSFK Lendület Near-Field Cosmology Research Group, 1121, Budapest, Konkoly Thege Miklós út 15-17, Hungary \and
\label{inst:26} Université Grenoble Alpes, CNRS, IPAG, 38000 Grenoble, France \and
\label{inst:27} INAF, Osservatorio Astronomico di Padova, Vicolo dell'Osservatorio 5, 35122 Padova, Italy \and
\label{inst:28} Université de Paris Cité, Institut de physique du globe de Paris, CNRS, 1 Rue Jussieu, F-75005 Paris, France \and
\label{inst:29} MTA CSFK Lendület ``Momentum’’ Milky Way Research Group, Hungary \and
\label{inst:30} INAF, Osservatorio Astrofisico di Torino, Via Osservatorio, 20, I-10025 Pino Torinese To, Italy \and
\label{inst:31} Centre for Mathematical Sciences, Lund University, Box 118, 221 00 Lund, Sweden \and
\label{inst:32} AIM, CEA, CNRS, Université Paris-Saclay, Université de Paris, F-91191 Gif-sur-Yvette, France \and
\label{inst:33} Centre Vie dans l’Univers, Faculté des sciences, Université de Genève, Quai Ernest-Ansermet 30, 1211 Genève 4, Switzerland \and
\label{inst:34} Isaac Newton Group of Telescopes, 38700 La Palma, Spain \and
\label{inst:35} Leiden Observatory, University of Leiden, PO Box 9513, 2300 RA Leiden, The Netherlands \and
\label{inst:36} Department of Space, Earth and Environment, Chalmers University of Technology, Onsala Space Observatory, 439 92 Onsala, Sweden \and
\label{inst:37} Dipartimento di Fisica, Università degli Studi di Torino, via Pietro Giuria 1, I-10125, Torino, Italy \and
\label{inst:38} Universidad Nacional Aut\'onoma de M\'exico, Instituto de Astronom\'ia, AP 70-264, Ciudad de M\'exico,  04510, M\'exico \and
\label{inst:39} Instituto de Astronom\'{i}a, Universidad Nacional Aut\'{o}noma de M\'{e}xico, Apdo. Postal 877, 22860 Ensenada, B.C., Mexico \and
\label{inst:40} Centro di Ateneo di Studi e Attività Spaziali ``Giuseppe Colombo'', Università di Padova, Via Venezia 15, 35131 Padova, Italy \and
\label{inst:41} Department of Astrophysics, University of Vienna, Türkenschanzstrasse 17, 1180 Vienna, Austria \and
\label{inst:42} Division Technique INSU, CS20330, 83507 La Seyne sur Mer cedex, France \and
\label{inst:43} Institute for Theoretical Physics and Computational Physics, Graz University of Technology, Petersgasse 16, 8010 Graz, Austria \and
\label{inst:44} ELTE E\"otv\"os Lor\'and University, Institute of Physics and Astronomy, P\'azm\'any P\'eter s\'et\'any 1/A, Budapest, 1117 Hungary \and
\label{inst:45} University of the Western Cape, Department of Physics, Private Bag X17, Bellville, 7535 CapeTown, South Africa \and
\label{inst:46} ELTE E\"otv\"os Lor\'and University, Institute of Physics, P\'azm\'any P\'eter s\'et\'any 1/A, 1117 Budapest, Hungary \and
\label{inst:47} IMCCE, UMR8028 CNRS, Observatoire de Paris, PSL Univ., Sorbonne Univ., 77 av. Denfert-Rochereau, 75014 Paris, France \and
\label{inst:48} Institut d'astrophysique de Paris, UMR7095 CNRS, Université Pierre \& Marie Curie, 98bis blvd. Arago, 75014 Paris, France \and
\label{inst:49} Airbus Defence and Space, Spain \and
\label{inst:50} Nicolaus Copernicus Astronomical Center, Polish Academy of Sciences, Bartycka 18, 00-716 Warszawa, Poland \and
\label{inst:51} INTA (National Institute of Aerospace Technology), 01 - Campus ``Torrejón de Ardoz’’ \and
\label{inst:52} Department of Astrophysical and Planetary Sciences, University of Colorado Boulder, Boulder, CO 80309, USA \and
\label{inst:53} P\&P Software GmbH, www.pnp-software.com \and
\label{inst:54} Cavendish Laboratory, JJ Thomson Avenue, Cambridge CB3 0HE, UK \and
\label{inst:55} ETH Zurich, Department of Physics, Wolfgang-Pauli-Strasse 2, CH-8093 Zurich, Switzerland \and
\label{inst:56} Institute of Optical Sensor Systems, German Aerospace Center (DLR), Rutherfordstrasse 2, 12489 Berlin, Germany \and
\label{inst:57} Universidad Nacional de Córdoba, Observatorio Astronómico de Córdoba, Laprida 854, X5000BGR Córdoba, Argentina \and
\label{inst:58} Consejo Nacional de Investigaciones Científicas y Técnicas (CONICET), Godoy Cruz 2290, CABA, CPC 1425FQB, Argentina \and
\label{inst:59} Dipartimento di Fisica e Astronomia "Galileo Galilei", Università degli Studi di Padova, Vicolo dell'Osservatorio 3, 35122 Padova, Italy \and
\label{inst:60} Zentrum für Astronomie und Astrophysik, Technische Universität Berlin, Hardenbergstr. 36, D-10623 Berlin, Germany \and
\label{inst:61} Institut fuer Geologische Wissenschaften, Freie Universitaet Berlin, Maltheserstrasse 74-100,12249 Berlin, Germany \and
\label{inst:62} Institut de Ciencies de l'Espai (ICE, CSIC), Campus UAB, Can Magrans s/n, 08193 Bellaterra, Spain \and
\label{inst:63} Institut d’Estudis Espacials de Catalunya (IEEC), Gran Capità 2-4, 08034 Barcelona, Spain \and
\label{inst:64} Department of Astronomy, University of Maryland, College Park, MD  20742, USA \and
\label{inst:65} NASA Goddard Space Flight Center, 8800 Greenbelt Rd, Greenbelt, MD 20771, USA \and
\label{inst:66} ELTE E\"otv\"os Lor\'and University, Institute of Physics and Astronomy, P\'azm\'any P\'eter s\'et\'any 1/A, Budapest, 1117, Hungary \and
\label{inst:67} European Space Operations Centre, Robert Bosch Str 5, 64293 Darmstadt, Germany \and
\label{inst:68} School of Physical Sciences, The Open University, Milton Keynes, MK7 6AA, UK \and
\label{inst:69} ELTE E\"otv\"os Lor\'and University, Gothard Astrophysical Observatory, 9700 Szombathely, Szent Imre h. u. 112, Hungary \and
\label{inst:70} HUN-REN--ELTE Exoplanet Research Group, Szent Imre h. u. 112., Szombathely, H-9700, Hungary \and
\label{inst:71} Department of Experimental Physics, University of Szeged, D\'om t\'er 9, Szeged, 6720, Hungary \and
\label{inst:72} Institute of Astronomy, University of Cambridge, Madingley Road, Cambridge, CB3 0HA, United Kingdom \and
\label{inst:73} Astronomical Institute, Slovak Academy of Sciences, 05960 Tatransk\'a Lomnica, Slovakia \and
\label{inst:74} HUN-REN–SZTE Stellar Astrophysics Research Group
}
 
\date{...}

  \abstract 
   {Since the discovery of the first exoplanet almost three decades ago, the number of known exoplanets has increased dramatically. By beginning of the 2000s it was clear that dedicated facilities to advance our studies in this field were needed. The CHaracterising ExOPlanet Satellite (\CHEOPS) is a space telescope specifically designed to monitor transiting exoplanets orbiting bright stars. In September 2023, \CHEOPS completed its nominal mission duration of 3.5 years and remains in excellent operational conditions. As a testament to this, the mission has been extended until the end of 2026.}
   {Scientific and instrumental data have been collected throughout in-orbit commissioning and nominal operations, enabling a comprehensive analysis of the mission's performance.
   In this article, we present the results of this analysis with a twofold goal. First, we aim to inform the scientific community about the present status of the mission and what can be expected as the instrument ages. Secondly, we intend for this publication to serve as a legacy document for future missions, providing insights and lessons learned from the successful operation of \CHEOPS.}
   {To evaluate the instrument performance in flight, we developed a comprehensive monitoring and characterisation (M\&C) programme. It consists of dedicated observations that allow us to characterise the instrument's response and continuously monitor its behaviour. In addition to the standard collection of nominal science and housekeeping data, these observations provide valuable input for detecting, modelling, and correcting instrument systematics, discovering and addressing anomalies, and comparing the instrument's actual performance with expectations.}
   {The precision of the \CHEOPS measurements has enabled the mission objectives to be met and exceeded. The satellite's performance remains stable and reliable, ensuring accurate data collection throughout its operational life. Careful modelling of the instrumental systematics allows the data quality to be significantly improved during the light curve analysis phase, resulting in more precise scientific measurements. }
   {\CHEOPS is compliant with the driving scientific requirements of the mission. Although visible, the ageing of the instrument has not affected the mission's performance. The satellite's capabilities remain robust, and we are confident that we will continue to acquire high-quality data during the mission extension. }

   \keywords{planets and satellites: detection -- techniques: photometric -- space vehicles: instruments -- instrumentation: photometers
               }

   \maketitle

\section{Introduction}
\label{sec:Intro}

The study of extrasolar planets has emerged as one of the most prominent fields in astrophysics over the last few decades. Merely 30 years ago, while hypothesised, no planets beyond our Solar System had been confirmed. The breakthrough came in 1995 with the discovery of a planet orbiting a solar-like star \citep{MayorQueloz1995}, for which the discoverers were awarded the Nobel Prize in Physics in 2019. Since then, and with more than 5,500 exoplanets known to date\footnote{\url{https://exoplanetarchive.ipac.caltech.edu/}}, remarkable progress has been achieved. Ground-based instruments such as High Accuracy Radial velocity Planetary Search \citep[HARPS;][]{HARPS},  Wide Angle Search for Planets \citep[WASP;][]{WASP}, and Hungarian Automated Telescope Network \citep[HATNet;][]{HAT} and dedicated space missions such as Convection, Rotation and planetary Transits \citep[CoRoT;][]{Corot}, \textit{Kepler} \citep{Kepler}, and Transiting Exoplanet Survey Satellite \citep[TESS;][]{Tess} have been crucial in identifying new worlds orbiting distant stars.

As the number of exoplanets discovered began to rise into the hundreds and scientists were able to draw statistically significant conclusions (e.g.\ \citealt{Mayor2011}, \citealt{Gaudi2021}, \citealt{ZhuDong2021}), the need for additional telescope time to carry out more comprehensive studies of a selected subset of these exoplanets became apparent. Exoplanets with mass and radius measurements became highly valuable, as these quantities allowed a crucial first step towards their characterisation: determining the planet's mean density, a key factor in constraining its fundamental composition.

For this purpose, transiting planets orbiting bright stars (where the radial velocity technique enables precise mass estimation) emerged as ideal targets. However, achieving this goal required high-precision mass and radius measurements to obtain the mean density with relatively low uncertainty. Since the mean density is inversely proportional to the cube of the planet's radius, precise knowledge of the radius was essential to ensure the uncertainty in the mean density was not dominated by the uncertainty in the radius. Therefore, the radius needed to be determined with a precision at least three times higher than that of the mass. A relatively small space telescope was deemed capable of addressing this challenge. Such a telescope could yield the precision required for accurate radius measurements and, therefore, provide insights into the physical properties and composition of exoplanets. On December 18, 2019, the European Space Agency (ESA) launched its first small-class mission, the Characterising ExOPlanet Satellite \citep[\CHEOPS;][]{BenzCHEOPS}, to carry out this task. By design, CHEOPS is a follow-up mission, setting it apart from discovery missions like \textit{Kepler}. Its working strategy involves observing one star at a time, with the targets chosen according to their scientific merit. 

After successfully completing a few months of in-orbit commissioning (IOC) at the beginning of 2020, the satellite transitioned into its scientific operations phase.
Since the start of nominal operations, \CHEOPS has used more than 90\% of its time for scientific observations, with little time for monitoring the instrument and minimal downtime for maintenance activities. As a result of its exceptional precision, the mission has already led to more than 60 publications, making a significant contribution to the field of exoplanetary science. Notably, it detected the asymmetric transit of the hot Jupiter WASP-189 b caused by the gravitational darkening of the star and the inclined orbit of the planet \citep{LendlCHEOPS}, which allowed a determination of the planet's radius with a precision of 1\% \citep{DelineCHEOPS}. Additionally, \CHEOPS revealed two very special planetary systems: the TOI-178 planetary system with six planets in Laplace resonance \citep{LeleuCHEOPS} and the HD 110067 planetary system with six sub-Neptune planets in a chain of first-order mean motion resonance \citep{LuqueCHEOPS}. Another significant achievement includes the transit detection of $\nu^2$ Lupi d, a planet with a 109-day orbital period around a remarkably bright star (\citealt{DelrezCHEOPS}, \citealt{EhrenreichCHEOPS}). Other mission successes include the serendipitous discovery of a further transiting planet in the multi-planet system TOI-1233 \citep{BonfantiCHEOPS}, the detection of tidal deformation in a hot Jupiter \citep{BarrosCHEOPS}, and the detailed analysis of phase curves and albedos (e.g.\ \citealt{MorrisCHEOPS}, \citealt{DelineCHEOPS}, \citealt{BrandekerCHEOPS}, \citealt{JonesCHEOPS}, \citealt{MeierValdesCHEOPS}), providing valuable insights into planetary atmospheres. In addition, intriguing evidence of exocomets was observed, adding to the growing body of knowledge about the dynamic behaviour of planetary systems \citep{KieferCHEOPS}. \CHEOPS also observed the occultation of Quaoar, which contributed to the discovery of its ring, a remarkable finding in our Solar System (see \citeauthor{MorgadoCHEOPS} \citeyear{MorgadoCHEOPS}, \citeyear{MorgadoNature}). The mission has also made a significant contribution to ancillary science, for example in the characterisation of M stars (\citealt{SwayneCHEOPS}, \citealt{MaxtedCHEOPS}, \citealt{SebastianCHEOPS}). In summary, \CHEOPS has achieved numerous scientific milestones, showcasing its ability to make groundbreaking observations and advance our understanding of exoplanets. 

While CHEOPS was initially scheduled to run science operations for 3.5 years, with the mission's end anticipated in September 2023, an extension has been approved by ESA. The extension will allow for the continuation of science operations until the end of 2026. Additionally, there is potential for a second extension, which, if granted, would extend the mission's lifespan until the end of 2029. CHEOPS is the first of ESA's three missions dedicated to the study of exoplanets. Following suit is Planetary Transits and Oscillations of Stars \citep[PLATO;][]{PLATO}, the M3 mission featured in ESA's Cosmic Vision 2015-2025 programme, scheduled for launch in 2026. PLATO's mission objective is to discover and study extrasolar planetary systems, with a particular focus on investigating the characteristics of terrestrial planets in the habitable zone of solar-like stars.
Stepping further into the future, ESA's M4 mission Atmospheric Remote-sensing Infrared Exoplanet Large-survey \citep[ARIEL;][]{ARIEL} is expected to launch in 2029. ARIEL will conduct a survey of the chemical composition of exoplanetary atmospheres. During its extension, CHEOPS will devote up to 15\% of guaranteed time observations (GTOs) to a new programme: synergies with other missions. Both PLATO and ARIEL have made strong proposals to join this collaboration. Pending approval of the second mission extension, CHEOPS and PLATO will operate in parallel. The scheduling flexibility of CHEOPS will facilitate a rapid and seamless follow-up of PLATO's exciting targets.

This paper presents a comprehensive compilation of our findings regarding the capabilities and overall performance of \CHEOPS after nearly four years in orbit. Section~\ref{sec:Overview} provides an overview of the \CHEOPS mission, offering contextual information to set the stage for the subsequent sections. Section~\ref{sec:InstStability} focuses on the stability of the instrument over time, and Sect.~\ref{sec:InstPerfromance} on instrument performance itself. Section~\ref{sec:SciPerformance} presents an analysis of the photometric precision achievable with \CHEOPS and explains instrumental systematics that were encountered and subsequently characterised. In Sect.~\ref{sec:Discussion} we discuss the effects of ageing on the performance of the mission. Lastly, Sect.~\ref{sec:Conclusion} presents our conclusions, summarising the key findings and highlighting the mission's significance in advancing the field of exoplanetary research.

To improve the paper's readability, we have structured the content to present specific topics at varying levels of detail in different sections. This approach allows readers to choose the depth of information based on their specific interests. Additionally, we have included several appendices to provide comprehensive explanations of technical aspects and the methods employed in various calculations (Appendices \ref{pointingPerf} to \ref{sec:NoiseBadPixels}). Appendix \ref{Glossary} includes a comprehensive glossary of all the acronyms used throughout the document. Additionally, to facilitate traceability, the list of \CHEOPS visits analysed in this paper can be found in Table \ref{table:visits}.

\section{Overview of the \CHEOPS mission}
\label{sec:Overview}

\begin{figure}
    \centering
    \includegraphics[scale=0.5]{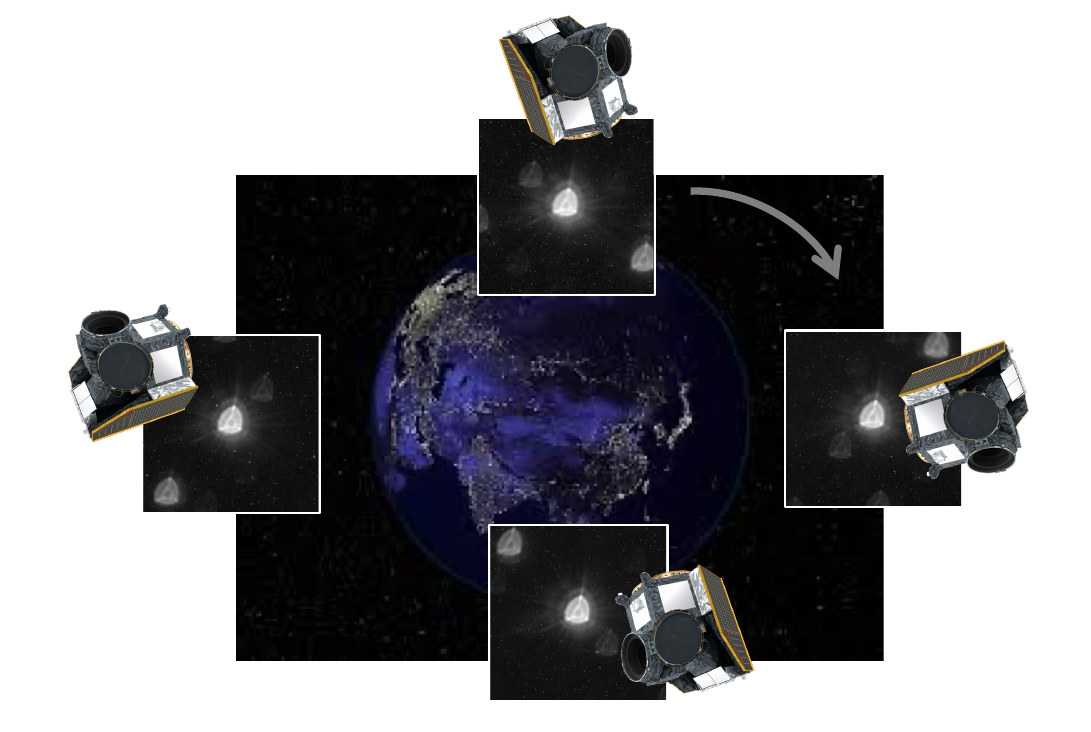}
    \caption{Spacecraft is nadir-locked to ensure thermal stability. As a result, the rotation of the spacecraft around the Earth is reflected in the images, where the stars in the FoV rotate in the indicated direction.}
    \label{fig:Nadir_locked}
\end{figure}

\CHEOPS, led by the University of Bern, is a joint effort between ESA and Switzerland, with significant contributions from ten additional ESA Member States. It is the first small mission in ESA's scientific mission portfolio and was established under stringent financial and temporal constraints. Its cost to ESA, inclusive of launch expenses, was capped at 50 million euros, and the project had to be developed within a four-year time frame. These limitations prompted various tradeoffs in platform (PF) selection, payload design, and ground segment development. While \CHEOPS was tailored to meet budget and time constraints and may not represent the ultimate follow-up mission, it reflects the best achievable outcome within the ESA's framework and the specified conditions. Therefore, the performance of each element of the \CHEOPS mission outlined in this paper should be evaluated in this context.
 
 To provide the necessary context for this paper, the following sections summarise the mission's key aspects. In addition, relevant figures that characterise the mission are summarised in Table \ref{table:CHEOPS_summary}.

\begin{table}
\caption[]{{CHEOPS main characteristics.}}
\label{table:CHEOPS_summary}
\begin{tabular}{ll}
\hline
\noalign{\smallskip}
Type of mission & Exoplanet follow-up\\
Telescope & Reflector, 30 cm aperture\\
Launch Date &  18.12.2019\\
Nominal Mission & 25.03.2020 - 24.09.2023\\
First Mission Extension &  25.09.2023 - 31.12.2026\\
Satellite weight & 275 kg\\
Satellite dimensions & $\sim 1.5\times1.5\times1.5$ m\\
Telemetry rate & 1.2 Gbit/day (S band)\\
Orbit & Sun-synchronous, LTAN 6 am \\
Orbital altitude & 700 km\\
Orbital period & 98.7 min\\
Bandpass & 0.33-1.1 $\mu$m\\
Integrated throughput & 25-50 \%\tablefootnote{It depends on the stellar spectral type; see Fig. \ref{fig:bandpass}.}\\
CCD & E2V AIMO CCD47-20 \\
FoV & $0.32\degr$\\
Plate scale & 1 arcsec/px\\
PSF & Defocused $R_{90} = 16.5$\tablefootnote{90\% of the encircled energy is contained in this radius.}\\
Nominal magnitude range & G = 6 - 12\tablefootnote{Observations outside the nominal magnitude range are possible; see Sect. \ref{sec:MagRange} for details.}\\
Science images & Circular\tablefootnote{Since 05.05.2024 window images are square to favour the study of the CTI effect.}, r = 100 px\\
Image cadence & $\geq$ 1 image/minute\\
Image exposure time & 0.001 - 60 s\tablefootnote{Set by the user.}\\
Sky coverage & 70\% (the ecliptic poles are \\
             &  not observable)\\
\noalign{\smallskip}
\hline
\end{tabular}  
\end{table}

\subsection{Payload}
\label{sec:Instrument}

The \CHEOPS satellite is positioned in a polar, Sun-synchronous, low Earth orbit (LEO) at an altitude of 700 km, with a local time of the ascending node (LTAN) set at 6 am. The spacecraft (SC) orbit has a period of 98.7 minutes and closely follows the day-night terminator, ensuring that the solar panels are nearly always illuminated by the Sun. This orbital configuration was chosen to minimise Earth's stray light contamination, reduce the frequency of Earth occultations and optimise the available energy.

The \CHEOPS SC is equipped with a single instrument, a 30\,cm effective aperture reflecting telescope. The instrument utilises a frame-transfer back-side illuminated charge-coupled device (CCD) E2V AIMO CCD47-20\footnote{Due to the high photometric precision required for CHEOPS, CCDs were chosen as the best option because of their demonstrated photometric stability. In contrast, CMOS detectors had not been proven to have long-term gain stability at the ppm level, and the lack of a fixed reference voltage was a concern. Notably, the CCD used in CHEOPS is very similar to that used in CoRoT.} to record images of bright stars.  To study transiting exoplanets through high-quality light curves, the stability of the instrument is of utmost importance, specifically its thermal, electronic, and pointing stability, as discussed in Sect.~\ref{sec:InstStability}.

Thermal stability is critical for the CCD detector, and efficient cooling is required to maintain it. To achieve the necessary cooling, the payload radiators of the SC must continuously face away from the Earth and the Sun, directing towards cold space. This is accomplished through nadir-locking the SC, which ensures that it maintains its proper orientation. However, this nadir-lock configuration leads to a rotating field of view (FoV) around the line of sight (LoS). The rotation period corresponds to the satellite's orbital period around the Earth.
Fig.~\ref{fig:Nadir_locked} illustrates one complete orbit rotation of the SC and how this rotation is reflected in the images captured by the instrument. Despite this rotation, the instrument is designed to track and observe one star at a time.

\begin{figure}[h]
    \centering
    \includegraphics[width=\columnwidth]{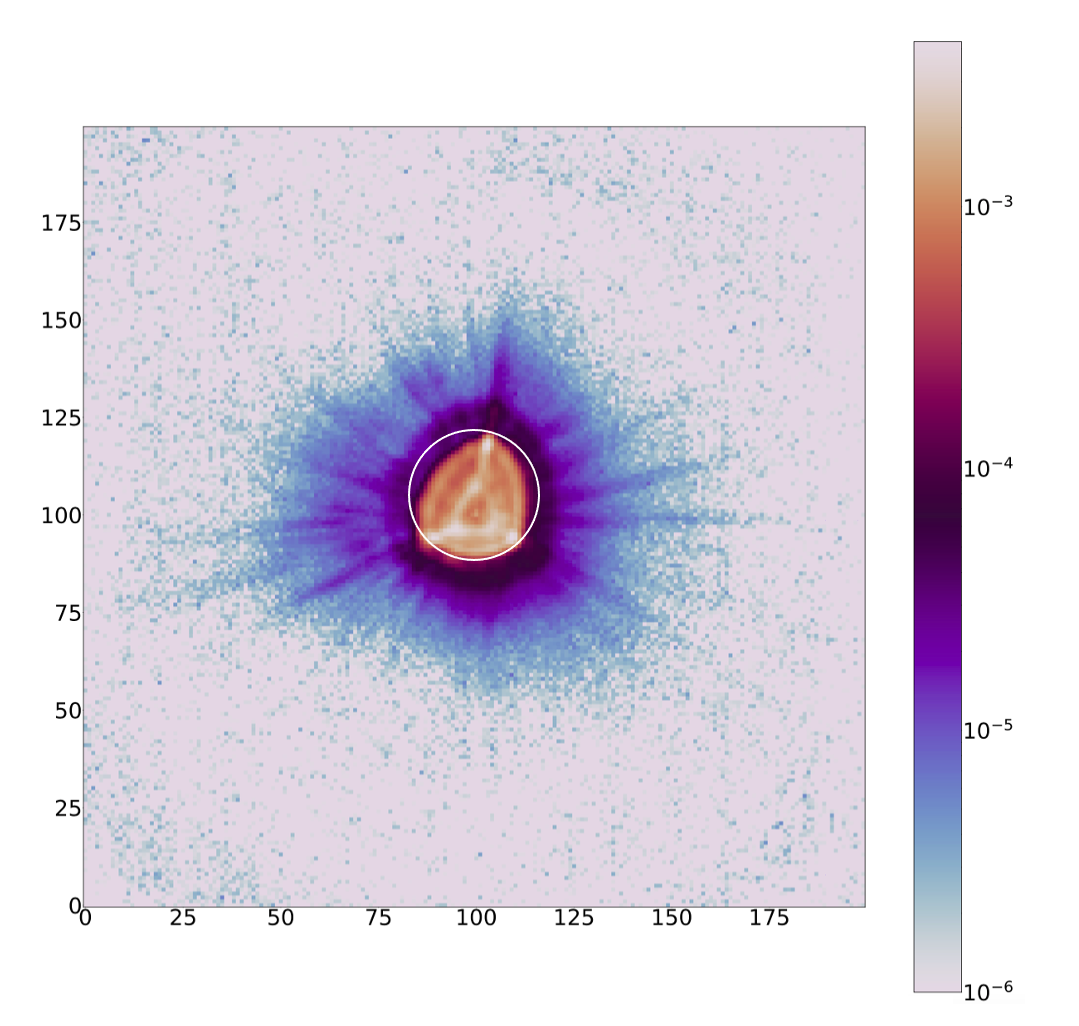}
    \caption{ \CHEOPS PSF in flight. The PSF shown here corresponds to a star in the centre of the CCD. The white circle shows the region inside which 90\% of the energy is concentrated. This corresponds to a radius $R_{90} = 16.5$\,px. The triangular shape of the PSF is related to the mount of the primary mirror. }
    \label{fig:PSF_Spider}
\end{figure}

To achieve precise photometric measurements of bright stars, the point spread function (PSF) was designed to minimise noise caused by the combination of the satellite's pointing jitter and the pixel response non-uniformity of the CCD. Additionally, it was also intended to avoid saturation of the CCD detector when observing bright stars. The PSF is therefore defocused to distribute the energy of the star over multiple pixels (\citealt{Magrin2014}). Approximately 90\% of the encircled energy is contained within a radius of 16.5\,px, as illustrated in Fig.~\ref{fig:PSF_Spider}. The triangular shape of the PSF is a consequence of the primary mirror's mounting at three interface points. The mirror, initially kept at room temperature before launch, cools down to its operational temperature of $-10\degr$C in space. This cooling results in a distortion of the mirror's surface in the direction of the three mounting points. The defocused image of the PSF makes this effect particularly noticeable.

The plate scale is 1 arcsecond to 1 pixel. Consequently, when stars are located close to each other in the FoV, their PSFs partially overlap. This overlapping effect can influence the photometric measurements, especially for closely spaced stars, and must be taken into account during data analysis.

The \CHEOPS bandpass covers the visible to near-infrared range, spanning from 330 nm to 1100 nm. At the blue end, this range is constrained by the transmission of the optical train, whose efficiency is limited for shorter wavelengths. At the red end of the bandpass, the detection limit is set by the quantum efficiency (QE) of the CCD.

\begin{figure*}
    \centering
    \includegraphics[width=\textwidth]{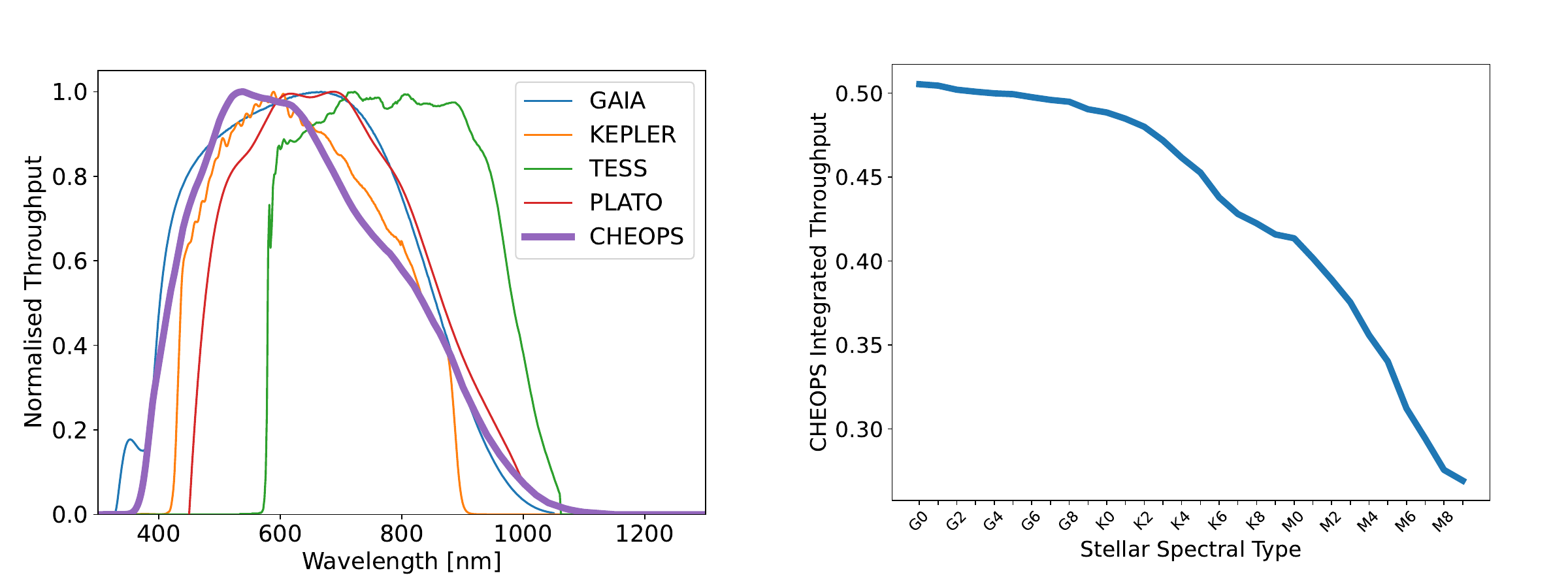}
    \caption{CHEOPS throughput. Left: Normalised \CHEOPS throughput as a function of wavelength. For comparison, normalised \textit{Gaia}, \textit{Kepler}, TESS, and PLATO throughput are also plotted. Note the similarity between \CHEOPS and \textit{Gaia} throughput. Right: \CHEOPS integrated throughput as a function of the star's spectral type using PHOENIX model for the SED (\citealt{Hauschildt1999A}, \citealt{Hauschildt1999}). This represents the fraction of all incident photons in the 330 nm to 1100 nm wavelength range converted into photoelectrons.}
    \label{fig:bandpass}
\end{figure*}

The throughput of the instrument over the entire bandpass is determined by the product of the optical transmission and the CCD's QE. By combining these factors, the instrument's throughput quantifies the fraction of incoming photons per wavelength efficiently converted into photoelectrons, which measures the instrument's sensitivity over the spectral range covered. Fig.~\ref{fig:bandpass} shows, on the left, the throughput as a function of wavelength, highlighting the overall sensitivity of the instrument across the 330 nm to 1100 nm range. \textit{Gaia}, \textit{Kepler}, TESS, and PLATO throughput are shown for comparison. On the right, the throughput is plotted as a function of the star's spectral type, allowing us to understand the instrument's sensitivity to different types of stars.

The CCD detector has an image area of 1024$\times$1024\,px, providing coverage over a FoV that spans $0.32\degr$. The images are transmitted to the ground for further processing to extract light curves for scientific analysis. The actual cadence may vary based on the exposure time set for the observation (i.e. visit). The observer can adjust the image's exposure time according to the target's brightness, with exposure times ranging between 0.001 and 60 seconds (the exposure time has an accuracy better than 2 ppm). However, due to the limited telemetry budget, and even if images are compressed on board, it is not feasible to downlink the full-frame images (1024$\times$1024\,px). Instead, cut-out circular images with a radius of 100\,px, centred on the target star, are sent to the ground. These circular images or `window images' provide enough information for scientific analysis and are more bandwidth-efficient. Furthermore, for exposure times shorter than 23 seconds, a stacking technique is employed on board to stay within the allocated downlink bandwidth. In such cases, multiple exposures are co-added, and only small images, known as `imagettes', containing the individual exposure's PSF in a circle of 30\,px radius, are downlinked to the ground.

For more information about the payload, we refer the reader to \citealt{BenzCHEOPS} and references therein, and to the \CHEOPS Observers Manual\footnote{\url{https://www.cosmos.esa.int/documents/1416855/12025024/CHEOPS-UGE-PSO-MAN-001_d4-Observers_Manual_AO4.pdf/50e2fb40-1553-a778-e519-8a9b0dc7e005?t=1680597342084}}.
\subsection{Ground segment}
\label{sec:GS}

The \CHEOPS Ground Segment comprises two primary centres: the Mission Operation Centre (MOC) in Torrejón de Ardoz, Spain, managed by the Instituto Nacional de Técnica Aeroespacial (INTA), and the Science Operation Centre (SOC) at the University of Geneva, Switzerland. As a small mission, the \CHEOPS ground segment has been developed and managed by the mission consortium, unlike other ESA missions where it is provided by ESA.

The MOC is responsible for mission oversight, SC control, orbit and attitude determination, anomaly management, and end-of-mission disposal procedures. On the other hand, the SOC handles mission planning, data processing, and distribution, and provides essential support to the \CHEOPS science team and guest observers.

The \CHEOPS mission follows a structured weekly schedule, where the SOC generates the activity plan, transmits it to the MOC, and the MOC translates it into commands for the SC. Data downlink occurs during scheduled ground station contacts in Spain, and the collected data undergoes a Quick Look analysis, automatic processing using the data reduction pipeline (\DRP; \citealt{DRPHoyer}), and is made accessible to the scientific community through the Mission Archive\footnote{The processed data are made accessible to the scientific community through the Mission Archive, which can be accessed at the following link: \url{https://cheops.unige.ch/archive_browser/}. Public data can also be accessed through \texttt{DACE}: \url{https://dace.unige.ch/dashboard/} and through the Mission Mirror Archive in Rome: \url{https://cheops-archive.ssdc.asi.it}.}.

During nominal operations, \CHEOPS has observed an average of 18 targets weekly, with observation times varying from one hour to a week (8 hours on average). This collaborative structure between the MOC and SOC ensures the seamless and efficient operation of the \CHEOPS mission, achieving its overarching goals of precise scientific data delivery.

\subsection{Data processing and analysis}
\label{sec:DPandA}

The SOC automatically processes and archives both science and housekeeping (HK) data received from the MOC. This process involves the transformation of telemetry files into \texttt{FITS} data structures through a pre-processing pipeline. Once a specific observational visit is concluded, the \DRP automatically undertakes the processing of the data to generate intermediate and final products.

\paragraph{\bf{Automatic data reduction with \DRP}}
The \DRP systematically follows a series of steps for each visit. It commences with the calibration of the science images, which includes bias subtraction, gain correction, linearisation, dark current subtraction, and flat field correction. Subsequently, the pipeline corrects instrumental and environmental effects on the images, such as smearing, identification and correction of cosmic ray hits, and computation of background contamination. The culmination of the \DRP's process involves conducting aperture photometry on each processed image. This step results in the creation of a time series dataset, also known as a `light curve'. This light curve shows the variations in the star's brightness over time and serves as input for subsequent scientific analysis and interpretation. 
Following the publication of \citealt{DRPHoyer} describing the \DRP, a series of enhancements have been incorporated into the \DRP, spanning across two key releases: versions 13 and 14 (\texttt{v13} and \texttt{v14}, respectively). These updates encompass minor bug fixes and modifications to improve the \DRP's performance and capabilities. We note that after a key release of the \DRP, all previously acquired data (dating back to the beginning of nominal operations) are re-processed and released to the users.

Among the improvements, one salient feature is the use of dark maps, which are regularly acquired through the M\&C programme (discussed in Sect.~\ref{sec:BadPixels}). These dark maps are leveraged for the correction of the dark current component in the data processing. By subtracting the dark map from the images, a significant reduction in the influence of hot pixels on the photometric extraction is achieved, enhancing the accuracy of the data. It is important to note that while the use of dark maps corrects most hot pixels in the images, those not encompassed in the maps are still identified by the \DRP, although they are not subject to correction. These detection outcomes are conveyed in the \DRP bad pixel map, which also contains information about dead and random telegraphic signal (RTS) pixels (see Sects.~\ref{sec:MandC} and \ref{sec:BadPixels}).

The \DRP's cosmic ray detection and correction algorithms have been fine-tuned, and a multi-aperture functionality has been introduced in these later versions. Previously, photometry was provided for four distinct aperture sizes. Now, the photometry is available for a broader range of 26 aperture sizes, spanning radii from 15 to 40\,px. This expanded range of apertures provides greater flexibility and precision in photometric analysis.

\paragraph{\bf{Data analysis with \pycheops}}
In addition to the official \CHEOPS Mission \DRP, the science team has developed an accompanying software toolkit known as \pycheops\footnote{Hosted at \url{https://pypi.org/project/pycheops/}}. This package is specifically designed for the analysis of the light curves generated by the \DRP, focusing on enabling scientific insights and interpretations \citep{Maxtedpycheops}. The current iteration of \pycheops, marked as version \texttt{1.1.0}, introduces several new capabilities and features to enhance its functionality and applicability with respect to \citealt{Maxtedpycheops}. A notable inclusion in this version is the option to incorporate arbitrary basis vectors as part of a linear model fit to the data. This augments the flexibility of \pycheops by allowing users to tailor the analysis to their specific scientific inquiries in addition to the basis vectors available in version \texttt{1.0.0}.  The data from each visit to a target were converted into a \pycheops\ \texttt{Dataset} object that provides methods for visualising, inspecting, and analysing the data, for example to fit a transit model to the data.

\pycheops \texttt{version 1.1.0} also enables the generation of \texttt{Dataset} objects from alternative data sources. This is facilitated by integrating the Python package \texttt{pycdata}, which allows external data to be incorporated into \pycheops workflows\footnote{More information about \texttt{pycdata} can be found at \url{https://github.com/Jayshil/pycdata}}. 

\paragraph{\bf{Manual data reduction with the \DRT or \PIPE}}
The \DRP is designed to process raw data unsupervised automatically, and as such, it needs to be very robust not to fail even for unusual observation configurations and targets. 
To provide the user with a more interactive and flexible option for data reduction, the data reduction tool (\DRT) has been developed. This tool features a graphical user interface that adopts a modular framework, enabling easy replacement and experimentation with new reduction strategies and algorithms. The \DRT's capabilities extend to the processing of window images as well as imagettes. It accommodates standard aperture photometry and mask photometry techniques. Furthermore, the \DRT offers seamless integration with HK parameters, enabling the correlation of these parameters with features identified in the data.

The \DRT is also useful to reduce non-standard datasets where common assumptions used by the \DRP (such as that there is a source centred in the frame) do not apply. Developed in \texttt{Java} and engineered with cross-PF compatibility, the \DRT provides a valuable tool for scientists seeking a more interactive and adaptable approach to data reduction. Access to the \DRT can be arranged upon request\footnote{\href{mailto:floren@astro.su.se}{floren@astro.su.se}}.

PSF Imagette Photometric Extraction (\PIPE) is another processing package specifically developed for \CHEOPS that uses PSF fitting to extract photometry. \PIPE was initially designed to extract photometry from imagettes but has also proven its efficacy in yielding satisfactory outcomes for window images. While the photometric performance derived from the \DRP and \PIPE are comparable for moderately bright stars (\textit{Gaia} magnitude, $G$, in the range 6--10\,mag), \PIPE proves particularly beneficial in several distinct scenarios:

\begin{itemize}
    \item Faint stars ($G > 10$\,mag): \PIPE is advantageous for faint stars where the influence of bad pixels on the photometric precision could, in some cases, be pronounced;
    \item Ultra-bright stars ($G < 6$\,mag): Generally, observations of ultra-bright stars have shorter-cadence imagettes, which \PIPE uses to significantly enhance the data analysis;
    \item Crowded fields: In situations with substantial contamination from chance alignments within the aperture, especially common for fainter targets, \PIPE can help mitigate such effects in the photometric extraction;
    \item Short cadence observations: \PIPE's usage is apt for cases requiring short cadence observations, such as capturing transient events like flares or improving transit timing precision;
    \item Short baseline observations: In scenarios where the baseline is notably short (less than five orbits), de-correlating visits with large roll modulations can lead to spurious correlations. \PIPE's capacity to robustly de-correlate roll modulation is beneficial as \PIPE can produce a light curve where the modulation is very much reduced without the need of de-trending as in the case of light curves obtained with photometric aperture;
    \item PSF variations: The PSF of \CHEOPS is stable but changes measurably with the target location on the detector and the spectral energy distribution (SED) of the source. Temperature variations of the telescope tube also lead to small PSF changes (Sect.~\ref{sec:Ramp}). To cope with these consistent PSF variations, \PIPE uses a library of pre-defined PSFs derived from \CHEOPS science data. The pre-defined PSFs are produced from archival CHEOPS visits. A PSF is defined by simultaneously fitting a 2D-spline to the data from subarrays and imagettes during a single orbit, where background stars have been modelled and subtracted. The rotation of the FoV during an orbit reduces the effects of contamination. The frame-to-frame jitter helps reduce pixel-to-pixel variation. More details can be found in the PIPE manual and an article in preparation (Brandeker et al.).
   
    The best library PSF matches are then decomposed using a principal component analysis (PCA), and the components fit to the data, also taking into account pointing jitter during the exposure.
\end{itemize}

\PIPE is an open-source software package accessible on GitHub\footnote{\url{https://github.com/alphapsa/PIPE}}. This tool has been effectively employed in various publications, such as \citealt{MorrisCHEOPS}, \citealt{SzaboCHEOPS}, and \citealt{BrandekerCHEOPS}. A comprehensive description of \PIPE will be presented in detail in an upcoming publication by Brandeker et al.\ (in preparation). An advantageous feature is the inter-operability between \PIPE and \pycheops, streamlining the data analysis process. \pycheops has been enhanced to incorporate a function \texttt{from\_pipe()}, facilitating the integration of \PIPE's output files. 
\subsection{Calibration, monitoring, and characterisation of the instrument}
\label{cal_m_c}

The \CHEOPS instrument underwent a thorough calibration process in the laboratory prior to launch. The results of this calibration campaign can be found in \citealt{Deline2017} and \citealt{Deline2020}, and are summarised in Sect.~\ref{cal}. In-flight calibration possibilities are, on the other hand, limited, as the instrument does not have features such as a shutter for obtaining dark frames or a lamp for taking flat field images. To compensate for this, we developed a monitoring and characterisation (M\&C) programme  to track the instrument's performance and changes over time. 

The M\&C observations encompass dedicated visits that provide supplementary data for specific instrument calibration objectives. Conceptually akin to standard observations, M\&C observations are treated operationally in a similar manner: they are formulated and outlined within templates resembling observation proposals, scheduled using mission planning software, and subsequently archived in the Mission Archive. Further details pertaining to the programme's objectives and strategies are expounded in Sect.~\ref{sec:MandC}. 
\subsubsection{On-ground calibration assessed in flight}
\label{cal}

Ground calibration of the \CHEOPS payload was an indispensable step to attain the photometric precision set as a goal for the mission. The calibration campaigns were carried out at the Observatory of Geneva and the University of Bern, as comprehensively described in \citealt{DelinePhDT}. The calibration efforts were focused on assessing the performance of the integrated flight payload's nominal and redundant read-out channels of the CCD in diverse system configurations, including variations in CCD temperatures, front-end electronics (FEE) temperatures, and read-out frequencies. Additionally, the characterisation of the PSF entailed the analysis of the PSF's shape at different positions within the FoV, leveraging both monochromatic and broadband illumination sources (\citealt{Chazelas2019}). We note, however, that the PSF shape on the ground differs from the in-flight one due to the zero gravity environment.

The calibration campaign yielded a collection of calibration products, each serving to characterise distinct aspects of the instrument's performance. Whenever possible, the calibration results obtained on the ground were compared with those in flight.

    \paragraph{\bf CCD and electronics} The bias map and read-out noise were recalculated during IOC before the telescope cover was opened. Subsequently, a minor adjustment of the bias reference file computed during the calibration campaign was performed. 
    
    The first hot pixel map was obtained using dark images when the telescope cover was still closed. It already showed a much larger number of hot pixels than those detected on the ground before lunch (2837 vs\ 7 pixels with dark current above 10 $e^-$/s). The hot pixels were randomly distributed, and it was noticed that the generation of new hot pixels happened during crossings of the South Atlantic Anomaly (SAA). Statistics from CoRoT \citep{Pinheiro2008} indicate that CHEOPS, at the beginning of IOC, was above the expected number of hot pixels by about 70\%. There is no explanation for this significant difference. 
    
    The system gain, non-linearity correction, and full-well capacity, initially determined in the laboratory, were confirmed to retain accuracy in the flight environment. The flat field and PSF shape were characterised very accurately in the laboratory \citep{Deline2020}, and their in-flight characterisation is part of the M\&C programme, which is further discussed in Sect.~\ref{sec:MandC}.
    
    \paragraph{\bf Throughput}  The throughput measurement in the laboratory proved challenging due to the high precision requirements and associated cost. Instead, the optical throughput was derived from the combination of measurements of the different optical components and a \texttt{Zemax} model\footnote{\url{https://www.zemax.com/}}, and the manufacturer provided the QE of the detector.  Therefore, the determination of the throughput, though not fully theoretical, relies on models of the optical throughput and the QE. However, during the calibration campaign, an estimation of the relative throughput was possible in the laboratory and was globally in agreement with the expected behaviour. Estimations of the throughput in orbit were also performed and are discussed in  Sect.~\ref{sec:GT}.
    
    \paragraph{\bf Optical distortion} The optical system of \CHEOPS introduces distortion stemming from the lenses within the back-end optics system. This distortion changes the pixel scale across the CCD's FoV, leading to variations in apparent illumination for extended sources. The distortion presents a barrel-like effect, making the sensitivity higher towards the edges of the CCD as compared to the centre. This happens because the pixel scale contracts towards the edges, concentrating more light there. This gain in sensitivity is not observed for stars; it is only observed for extended sources like the uniform illumination used to produce the flat field images in the laboratory. Therefore, these flat fields were corrected for the non-uniformity of the light source and optical distortion to avoid introducing a photometric bias during data reduction. Not accounting for distortion in the flat field correction would affect photometry due to the pointing jitter of the SC; a PSF displacement of one pixel would generate a negligible photometric bias at the centre but up to 150 ppm at the very edge of the detector.
    During IOC, measurements were done to confirm the distortion and the pixel scale in flight by observing a crowded stellar field. The pixel scale perfectly agreed with the one measured during ground calibration (1.002±0.009 arcseconds/pixel in flight vs\ 1.002±0.005 arcseconds/pixel on the ground).  Regarding the optical distortion, it could not be reliably estimated in flight: the obtained absolute distortion signal was consistent with the on-ground calibration although the stars in the centre of the CCD appear to slightly offset away from the optical centre (1 px). However, the position of the optical centre was poorly constrained by the data. 

\subsubsection{The in-flight monitoring and characterisation programme}
\label{sec:MandC}

The M\&C programme was devised already during mission preparation as a response to the need for in-flight M\&C of the instrument's scientific performance. The programme was designed to operate solely based on sky images, without the option of taking completely dark images or images illuminated by a calibrated light source. The \CHEOPS mission's M\&C programme contains the following activities:

    \begin{figure*}
\centering
    \includegraphics[width=\textwidth]{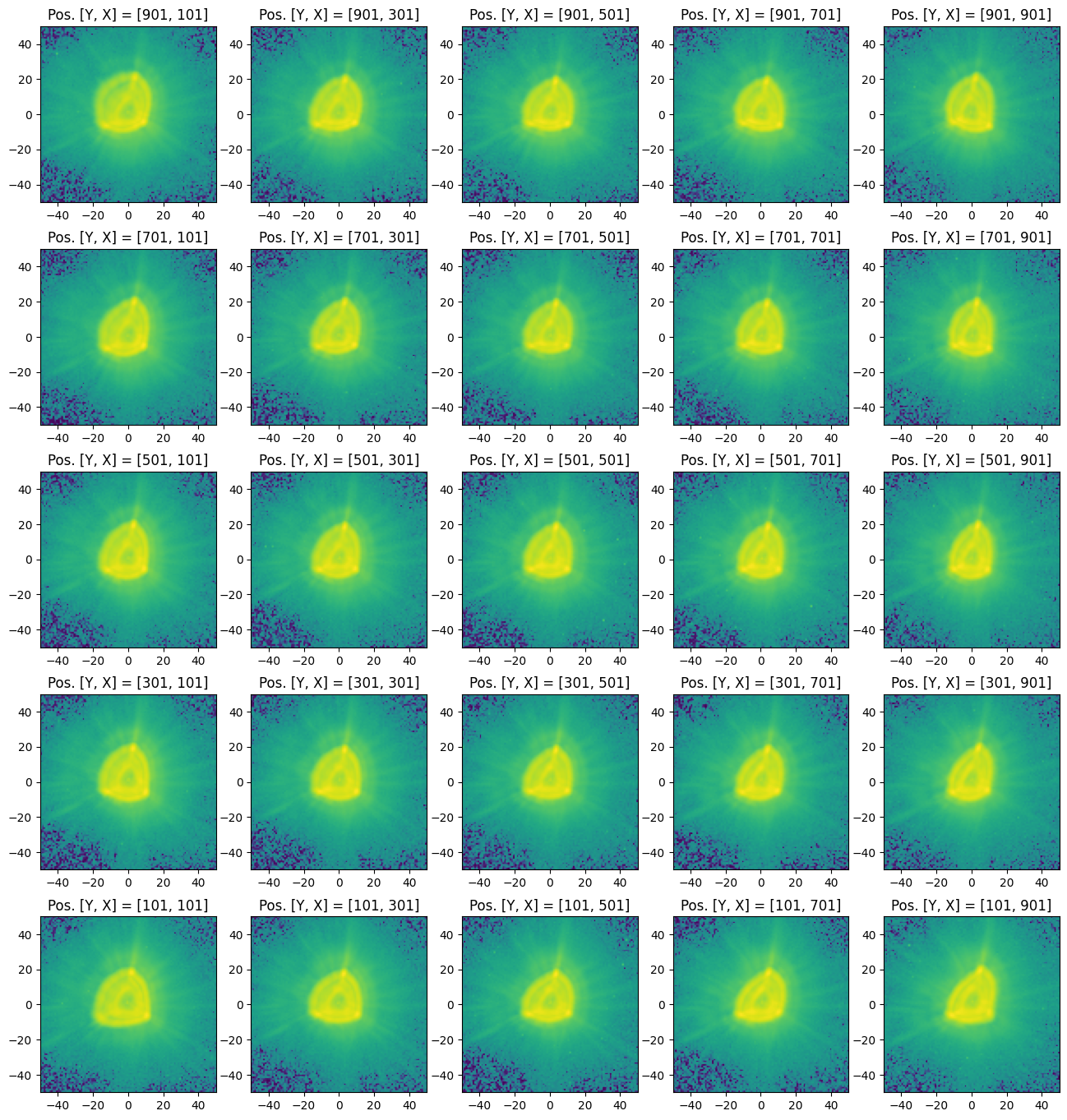}
    \caption{CHEOPS PSF shape as a function of the target location in the CCD. Each PSF is at the centre of a $200\times200$ window, where the coordinates of the bottom-left corner of the window are given in the figure for each PSF. Each image was taken with an exposure time of 1.05\,s. The target used as a reference star is HD\,88742 (visit ID: \texttt{CH\_PR330096\_TG0007}). }
    \label{fig:PSFs}
\end{figure*}

     \paragraph{\bf Bad pixels M\&C} This activity is at the heart of the M\&C programme. Its primary objective is to monitor the CCD detector's thermal electrons or `dark current'. Since \CHEOPS does not have a shutter to obtain true `dark' images, this activity relies on capturing images of relatively empty regions of the sky to assess the behaviour of individual pixels. Essential for the regular diagnostic of the CCD, this activity, performed weekly, detects and characterises hot pixels, which are pixels with elevated dark current, often generated by cosmic ray hits. Hot pixels contribute noise to the photometric data, so monitoring their behaviour is crucial to maintaining the quality of the observations. Also within the scope of this activity is the detection of hot pixels that exhibit varying dark current levels over time, a phenomenon known as RTSs. This behaviour is attributed to the presence of `charge traps' within the pixels that retain and release charges after a certain time. Another type of bad pixels is the dark pixels, which have lower sensitivity to light and produce lower signal output than neighbour pixels.  A detailed description of the bad pixels M\&C activity and its methodologies can be found in Sect.~\ref{sec:BadPixels}. 

    \paragraph{\bf Stray light M\&C} Stray light refers to unwanted light that enters the instrument's optical system and affects the quality of observations.  This M\&C activity aims to monitor the instrument's ability to reject stray light. Two methods were envisioned for this: for small off-axis angles, the light from the full moon is used as a source to measure the transmission of the system; for larger angles, the illuminated Earth limb is observed. During IOC, full Moon observations were carried out with varying angles between the optical axis of the instrument and the Moon, up to 25\degr\ off-axis. With pointing directions differing on average 3.5\degr, these observations lasted one orbit each. The analysis showed that the measured point source transmission function (PST) was only 2--3 times larger than predicted, while its shape was very similar to the predicted one. This outcome was considered satisfactory, indicating that the instrument's stray light rejection, at least for small angles, was close to expectations. Earth limb visits involve observing the illuminated Earth limb at larger off-axis angles. However, this method encountered challenges due to unexpected bright atmospheric glow at low altitudes. The airglow emission was orders of magnitude larger than the anticipated stray light contamination from the illuminated Earth, making these measurements unusable. This led to the cancellation of the attempts to measure the PST for large off-axis angles.
     
    \paragraph{\bf PSF M\&C} This activity was designed to monitor and characterise the PSF's shape in different CCD regions. The objective of this M\&C activity is to understand how the PSF shape varies across the detector's FoV and to ensure that any evident fluctuation in time in the PSF shape is detected. The CCD detector is divided into a grid of 25 windows of 200$\times$200\,px. For each window, a bright star is placed at the centre, and a set of 10 images is taken. The displayed PSFs in Fig.~\ref{fig:PSFs} are the median PSFs obtained for each location. This activity is done at least once a year. During the nominal mission, no significant variations were detected in any location of the CCD.
     
    \paragraph{\bf Timing precision M\&C}  The accuracy and precision of the \CHEOPS time stamps were determined in flight and found to be better than 1 second and better than 0.3\,s, respectively. More details about these measurements can be found in Sect.~\ref{sec:Timing}.
    
    \paragraph{\bf Flat Field M\&C} The flat field (i.e.\ the spatial variation in the sensitivity of the detector pixels) was measured in the laboratory during the calibration campaign. Although it was not expected to change over time, one of the objectives of the M\&C activities was to calculate the flat field in flight. Early in the project, the Kuhn algorithm (\citealt{Kuhn1991}) was identified as the most practical method for measuring the flat field without a calibration lamp on board. This method relies on taking several images of a bright and crowded region of the sky, each of which is shifted with respect to the other. In the case of CHEOPS, the idea was to exploit the small, fast shifts between images caused by the expected jitter of the satellite's pointing without the payload in the loop (PITL; see Sect. \ref{sec:PointStability}). Despite the effort put into developing this technique, it became clear towards the end of the project that the satellite's pointing, even without the PITL, would be remarkable and, therefore, too stable to produce the required shifts of 10--15\,px between images.
     Nevertheless, the flat field M\&C was performed during IOC with the globular cluster NGC3201 to verify the image acquisition procedure. The procedure was successful, but, as expected, the precision of the estimated flat field was more than one order of magnitude worse than the requirements. We calculated the flat field in a region of 130$\times$130\,px using 150 images. The pixel response non-uniformity (PRNU) was 7\% (standard deviation of the signal of all the pixels in the calculated flat field). For reference, the worst measurement of this kind in the laboratory, for the whole CCD, corresponded to the wavelength 1100 nm and was 5\%; the best measurement was 0.5\% for 600 nm. The precision or uncertainty of the PRNU was 2.35\% (this is the average of the standard deviation of the signal measured in each pixel, considering the 150 images, for all the pixels in the 130$\times$130\,px image). This figure was far from the 0.1\% precision requirement and the 0.06\% achieved in the laboratory. Based on these results and many simulations performed prior to launch, it was concluded that this method would not provide the desired precision, and no further observing time was allocated for this purpose. The flat field M\&C was dropped from the M\&C programme.

Combining the information gathered from these M\&C activities, the instrument's performance can be tracked over time, and any potential deviations or anomalies can be identified. This approach enables the \CHEOPS mission to ensure the integrity of its photometric data despite the constraints on in-flight calibration options. 

 Nominal science visits of recurrently observed targets are also used to monitor photometric precision. The root mean square (RMS) values of the light curve residuals (i.e.\ observed transit minus transit model) provide information about the time evolution of the photometric precision. However, it is important to observe some reference targets periodically for the purpose of photometric monitoring only. The M\&C programme includes two activities dedicated to this issue, which are briefly described below.

     \paragraph{\bf Flux Sensitivity M\&C} There are several possible causes for flux variations to occur during the mission, resulting in an overall decrease in the sensitivity; for example, the degradation of optical surface coatings due to the high radiation environment. This evolution is usually slow, so it only affects the absolute photometry and the magnitude range of the targets that can be observed. On the other hand, if the sensitivity decreases with time because \CHEOPS collects fewer photons, the noise in the relative measurements will increase, affecting the relative photometric accuracy. Analyses of the sensitivity evolution during the nominal mission are described in Sects.~\ref{sec:GT} and~\ref{sec:MCperformance}.
    
    \paragraph{\bf Reference Transit M\&C} The main purpose of \CHEOPS is to observe relative flux variations such as those occurring during planetary transits. Many science programmes require that the same target is observed at different epochs. It is therefore important that the photometric precision is such that the transit depths are not affected by the ageing of the CCD. In particular, non-corrected hot pixels within the aperture photometry can sensitively change the base flux level of faint stars, leading to a relatively shallow transit. We used regularly observed GTO transiting planets and eclipsing binaries to monitor changes in transit depths due to instrumental effects. 

\subsection{Health and mission status}
\label{sec:healthandmissionstatus}

\begin{figure}
    \centering
    \includegraphics[width=\columnwidth]{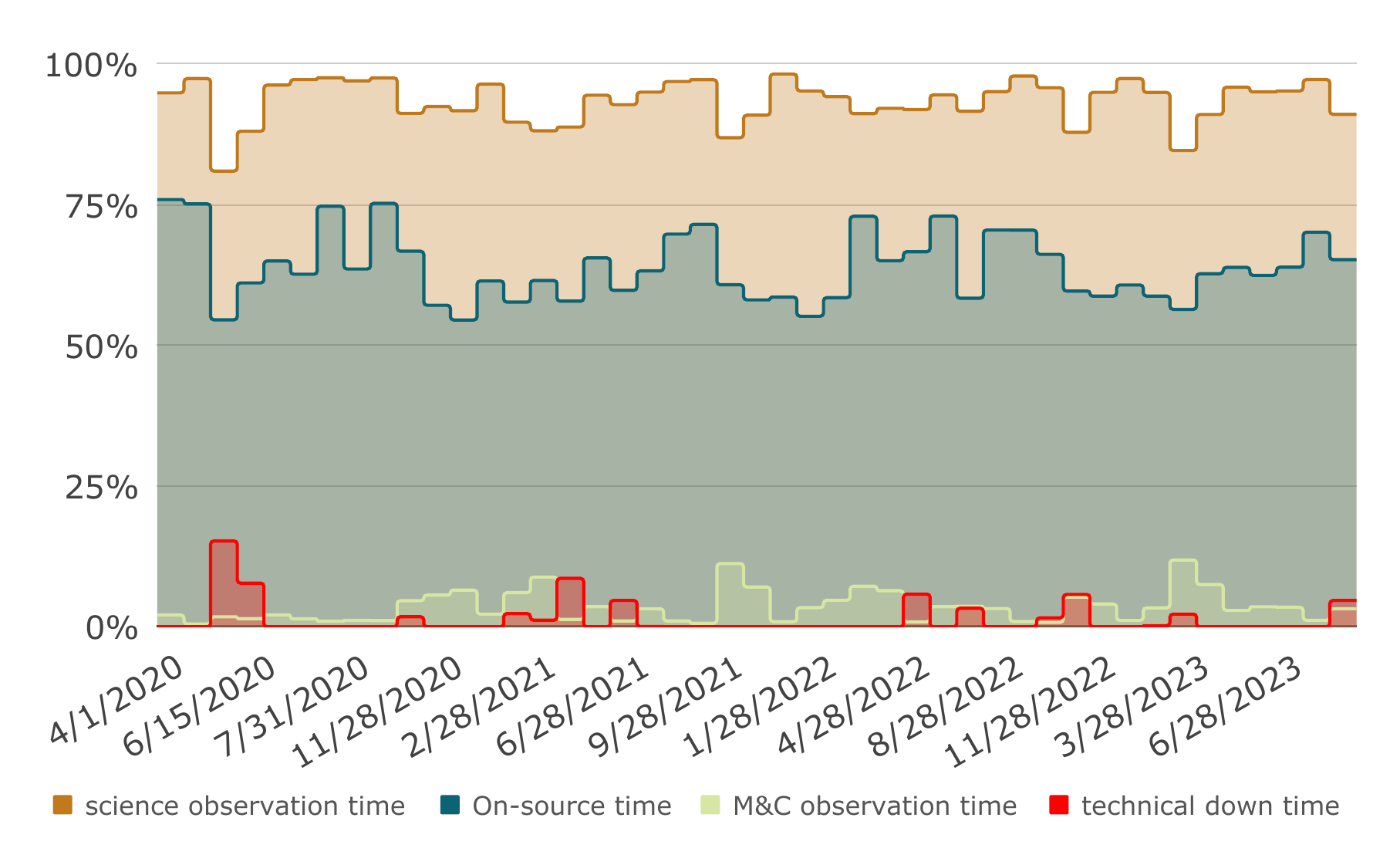}
    \caption{Excellent CHEOPS mission operational performances  for a LEO mission: 93\% science time, 67\% on-source, 3.5\% calibration, less than 1.5\% technical downtime, and 0.65\% slew time.}
    \label{fig:operational_performance}
\end{figure}

After nearly 4 years in orbit, the overall condition of the \CHEOPS mission's space segment remains excellent. All units are operating within normal parameters, and redundancy is fully functional.
The MOC continuously monitors the health of the SC, and the data indicates that all PF subsystems are still performing with no signs of deterioration. The redundant units of the PF are inspected annually and continue to operate nominally. Only one minor software update has been performed in flight to adjust the plate scale to match the observed value.

The instrument is also in excellent health. To maintain the instrument's optimal performance, the sensor electronics module (SEM) software was updated once since launch, and minor patches have been applied to both the SEM and the instrument flight software. Ongoing monitoring of HK data reveals no significant deterioration except for two specific observations: (1) the detector is gradually accumulating defects, including hot pixels and less obvious issues like charge transfer inefficiency (CTI; see subsequent sections); (2) there is a slight decrease in the instrument's sensitivity, as discussed in Sect.~\ref{sec:GT}.
In addition,  a slow drift in bias voltages powering the CCD has been observed. Still, this phenomenon was also present during the ground calibration phase and is not considered a sign of deterioration. In fact, this drift has changed over time to a slower upward trend and is expected to have no impact on the instrument's scientific performance. Detailed information about this phenomenon can be found in Sect.~\ref{sec:ElecStability}.

On another note, CHEOPS has been managing its propellant effectively. Propellant is the only consumable used in the mission, and the available $\Delta v$ (change in velocity) is currently measured at 244\,m\,s$^{-1}$. A very small fraction of the available propellant has been expended up to this point in the mission, equivalent to $\Delta v=6.5$\,m\,s$^{-1}$. This means the mission has an excellent propellant margin, even accounting for the (non-mandatory) maximal disposal variant (reentry of the SC in 5 months, equivalent to $\Delta v = 192$\,m\,s$^{-1}$). This provides a significant level of flexibility and reassurance for the mission's continued operations and potential extensions.


CHEOPS has maintained a high level of operational performance with very few anomalies and down times. The fact that there have been only three instrument anomalies caused by an uncorrectable memory error in the instrument electronics and two operational anomalies leading to instrument shutdown due to erroneous slew computations\footnote{These were due to inadequate time allocation between consecutive activity plans (APs). The mission planning system was therefore updated to give enough time for the anti-Sun pointing mode to converge between activity plans before starting the first slew of the next activity plan. Until implementation was completed, the SOC maintained a margin between APs. The second occurrence could also be fixed by better aligning the slew computation models of the mission planning system and the onboard software.}, all of which were relatively short-lived, speaks for the robustness and reliability of the SC and its subsystems. The operational performance graph in Fig.~\ref{fig:operational_performance} showcases the mission's remarkable track record. 

\section{Instrument stability}
\label{sec:InstStability}
During IOC and nominal operations, the stability of \CHEOPS was continuously monitored. We can confidently conclude that the mission has met all stability requirements in flight. In the following sections, we report on the main stability requirements and their actual values as measured in flight.   
\subsection{Stability requirements}
\label{sec:requirements}
During the design phase, the following key stability requirements were established:
\begin{itemize}
\item{Detector temperature stability:} The detector shall have a temperature stability of better than 10\,mK (RMS) over the duration of a visit (48h was set as the nominal duration of a visit). Rationale: derived from QE and CCD on-chip amplifier sensitivity. CCD calibration confirmed a gain sensitivity of -0.9 parts-per-million per milli-kelvin (ppm/mK). To keep the gain stable at better than 10\,ppm, a stability of 9\,mK is required. 
\item{Electronics stability - gain:} Analogue-to-digital converter -- gain stability: The gain stability shall be better than 10\,ppm (goal: 5\,ppm) over a period of 30\,min -- 48\,h. Rationale: Error allocation in noise budget.
\item{Electronics stability - temperature:} The thermal stability of analogue electronics shall be better than 50\,mK (RMS). Rationale: Derived from gain sensitivity requirement.
\item{Telescope tube temperature:} The Optical Telescope Assembly inner tube shall be maintained at 263\,K with $\pm$ 5\,K over all the nominal orbits defined in the mission manual. Rationale: to keep the combination of the telescope breathing and flat field-induced noise below 5 ppm.
\item{Pointing stability:} The \CHEOPS system shall be capable of pointing with a stability (in yaw and pitch) of at least 8\arcsec\ (RMS, goal: 4\arcsec) over a 48-hour observation. This is equivalent to Actual Pointing Error (APE $\epsilon$) < 4\arcsec\ (68\% confidence, temporal interpretation), where $\epsilon$ is the angle between the direction of the target (LoS$_{\mathrm{target}}$) and the actual direction of the LoS (LoS$_{\mathrm{actual}}$). Rationale: to keep the combination of the pointing jitter and flat field noise below 10\,ppm. We note that the jitter in roll was not constrained. Although worse than the yaw and pitch, it is irrelevant to the photometric performance as the PSF of the star itself does not rotate. Only the background rotation will be `jittery', but the jitter is small compared to the roll rate.
\end{itemize}

The thermal requirements can be directly verified through measurement. It is important to note that these requirements pertain to relative stability rather than absolute accuracy. Nevertheless, it is worth highlighting that the FEE temperature measurements, especially the voltage measurements, have undergone extensive calibration against state-of-the-art reference meters. As such, they can also be deemed reliable for this specific purpose in an absolute sense. The CCD is kept in continuous readout, also when idle, to be in a steady state and avoid thermal perturbations. Only after a change of readout mode (i.e. between two visits), a perturbation of 5-7 mK affects the CCD temperature for 20 to 40 seconds.

Validation of the gain requirements can only be achieved indirectly through gain sensitivity measurements conducted during instrument calibration. For this, it is important to maintain the bias voltages (voltages that power the CCD) very stable. The bias voltages are the substrate voltage (VSS), the output gate voltage (VOG), the output transistor drain voltage (VOD), and the reset transistor drain voltage (VRD).
The gain sensitivities are approximately 35\,ppm/mV for VOD and VSS and less than 10\,ppm/mV for VRD and VOG. The allowable variation depends on the specified gain stability of the CCD. The left (L) and right (R) amplifier gain sensitivities are:

\begin{description}
    \item[VOD (ppm/mV):] 32 (L), 31 (R)
    \item[VRD (ppm/mV):] 6 (L), -5 (R)
    \item[VOG (ppm/mV):] -5 (L), -12 (R)
    \item[VSS (ppm/mV):] -29 (L), -29 (R)
\end{description}

Allowing 5 ppm variation (goal, requirement: 10\,ppm), the allowed voltage change is calculated by considering the larger of the two left and right values:
\begin{description}  
\item[VOD:] 0.16 mV 
\item[VRD:] 0.83 mV 
\item[VOG:] 0.41 mV 
\item[VSS:] 0.16 mV 
\end{description}

\subsection{Stability in orbit}
Voltage and temperature excursions during a visit predominantly occur over the orbital period. Inter-visit changes can be larger for large changes in SC attitude. Significant parameters such as CCD temperature and bias voltages are continuously monitored through trend reports, which the SOC automatically generates. 
The values presented in the subsequent paragraphs are extracted from the long-term trend report, encompassing the time span from April 16, 2020, to September 25, 2023, covering the entire nominal operations phase.

\subsubsection{Electronics stability: Temperatures}
\label{sec:ElecStability}

Stable thermal balance is achieved by a proportional integral derivative (PID)-controlled heating of the FEE and CCD to compensate for its varying cooling via radiators, which are unexposed to the Sun and with minimal viewing angle of the Earth. Both the CCD and focal plane assembly temperatures have displayed exceptional stability, surpassing initial expectations. 

\begin{figure*}
    \centering
    \includegraphics[width=\textwidth]{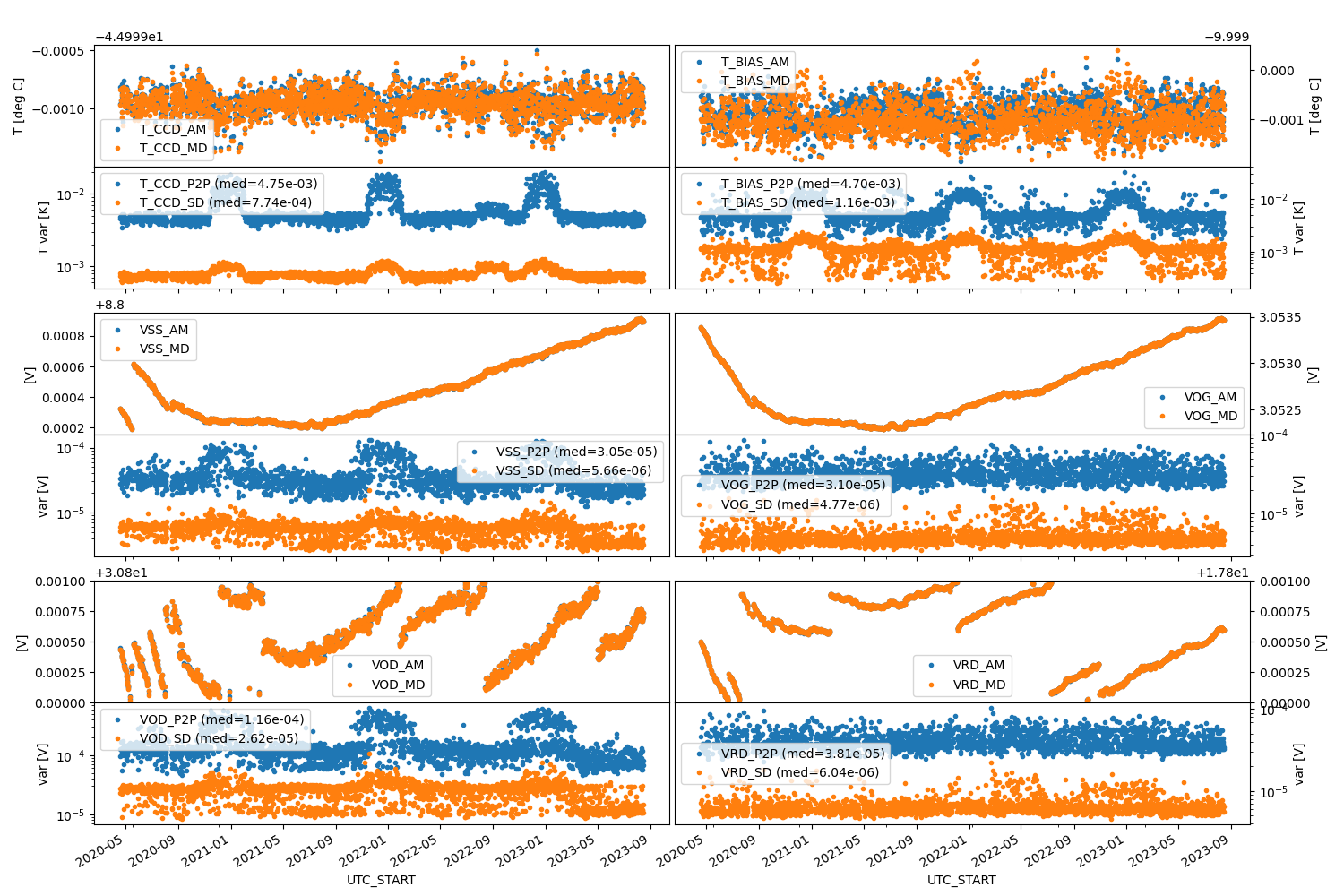}
    \caption{Stability of the CCD and the readout electronics since the start of nominal operations. Data are aggregated to one point per visit. Top: Arithmetic mean (AM) and median (MD). Bottom: Peak-to-peak variation (P2P) and standard deviation (SD). When the SC enters Earth's shadow in winter (eclipse season), the negative effect on thermal stability is clearly visible.}
    \label{fig:cis_stability}
\end{figure*}

The CCD is stabilised at -45\textdegree C with an RMS precision better than 2\,mK. In worst-case scenarios during eclipse seasons, peak-to-peak temperature variations can reach up to 20\,mK when entering and exiting the Earth's shadow. The CCD is controlled to the 1 mK level and connected to the radiator via a high-conductivity capacitor. Despite this capacitor, significant changes in the radiator's heat flux on orbital timescales (like eclipses) propagate to the CCD and FEE, leading to reduced temperature stability as seen in the top panels of Fig.~\ref{fig:cis_stability}.
At these low temperatures, the average dark current remains insignificant, measuring less than 0.08\,electrons/pixel/second at the beginning of the mission, although this value is now estimated to be approximately one order of magnitude higher.
 
The thermal control for the analogue electronics is based on the sensor closest to the bias components, as these are responsible for generating CCD bias voltages. Referred to as the FEE bias temperature, this parameter exhibits a median peak-to-peak amplitude of 4.72\,mK (with a standard deviation of 1.17\,mK), well below the specified requirement. While this value can vary from one pointing to another, due to exposure to varying heat fluxes on the radiator, the recorded numbers are more than one order of magnitude beneath the requirements, even for peak-to-peak variations.

\subsubsection{Electronics stability: Bias voltages} 

Maintaining stable bias voltages is crucial for ensuring CCD gain stability. These values are influenced by various factors, including temperature stability, and are thus anticipated to exhibit exceptional performance. The actual measurements substantiate this. The VOD shows a standard deviation of the voltage of 26.2\,$\mu$V with a peak-to-peak value of 116\,$\mu$V. This is below the threshold (requirement) of 160\,$\mu$V, which translates to a 5\,ppm gain change. The VRD shows a standard deviation of the voltage of 6.04\,$\mu$V with a peak-to-peak value of 38.1\,$\mu$V. This is well below the 830\,$\mu$V threshold for VRD. The VOG shows a standard deviation of the voltage of 4.77\,$\mu$V with a peak-to-peak value  of 31.0\,$\mu$V. This is well below the 160\,$\mu$V threshold for VOG. Lastly, the VSS shows a standard deviation of the voltage of 5.66 $\mu$V with a peak-to-peak value  of 31.5\,$\mu$V. This is well below the 410\,$\mu$V threshold for VSS.

When observing longer timescales, a bias drift becomes noticeable. The sensor electronics system currently fine-tunes the bias voltages for each observation, but only with a threshold accuracy of 1 mV. No re-tuning takes place during a science observation. Looking at the example of VOD and VRD over approximately 30 days during the early mission phase, the following trends are observed: over a span of 35 days, VOD drops from 30.8007\,V to 30.8001\,V. This indicates a bias voltage drift in VOD of 17\,$\mu$V per day. For VRD, over 29 days, the values change from 17.80095\,$\mu$V to 17.8007\,$\mu$V, implying a drift of 9\,$\mu$V per day. These drift rates align with values projected during calibration, falling within the range of 10 to 30\,$\mu$V per day for datasets spanning tens of hours. Specifically, focusing on the VOD value suggests a drift exceeding 160\,$\mu$V in 10 days, which is the threshold at which a measurable gain change (5\,ppm) is anticipated.

The comprehensive evolution of temperatures and bias voltages over the nominal mission is depicted in Fig.~\ref{fig:cis_stability}. After the first year in orbit, the bias voltage drift ceased and gradually reversed to exhibit increasing voltages with a smaller slope. As of the third year in orbit, all bias voltages exhibit a minor increase of a few $\mu$V per day.
\subsubsection{Telescope temperature stability}
\label{sec:TempStability}

\begin{figure*}
    \centering
    \includegraphics[width=\textwidth]{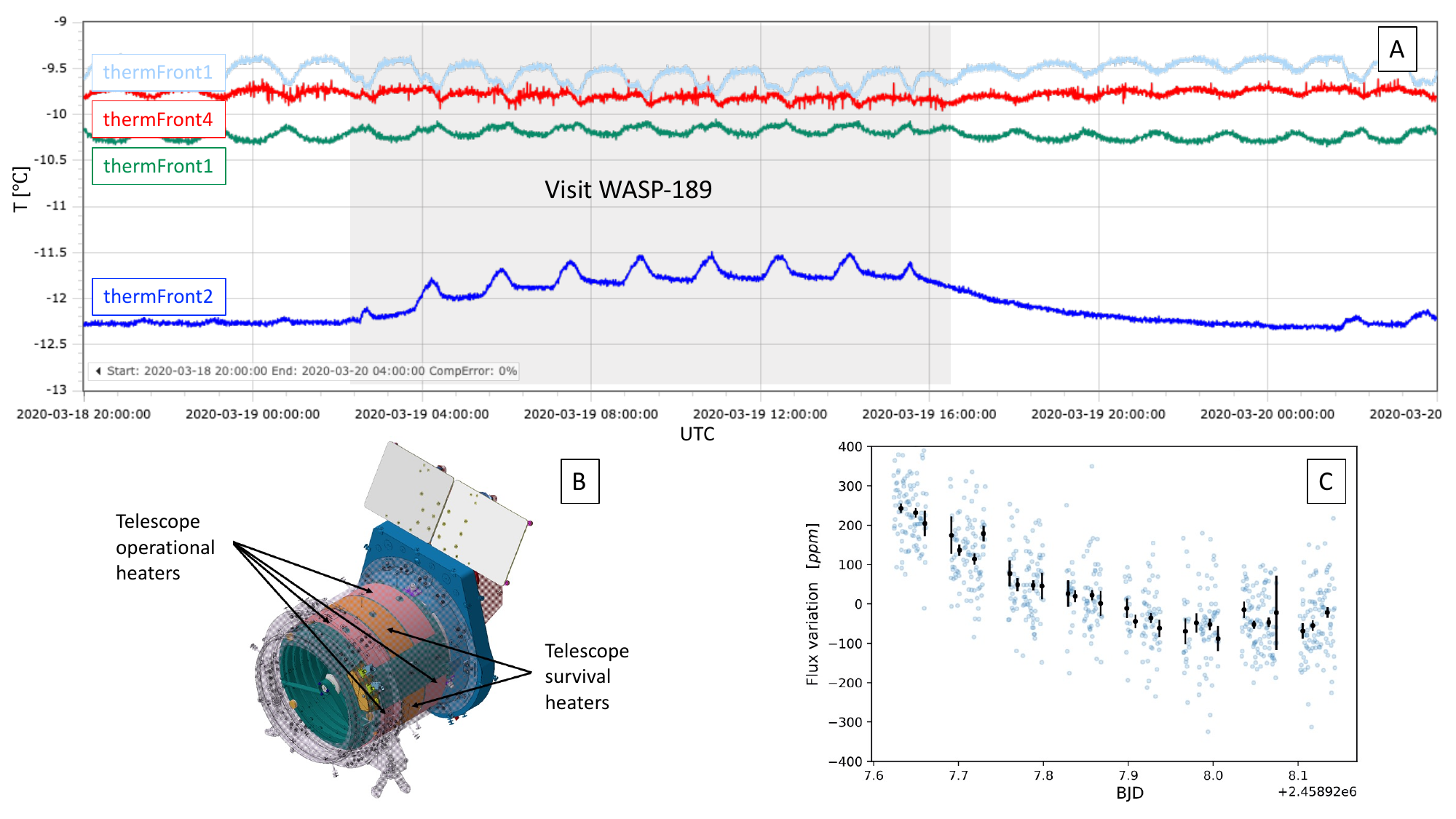}
    \caption{Flux ramp. Panel A: Temperatures measured by the four front sensors in the telescope tube.  A different colour represents each quadrant: green corresponds to the top quadrant, light blue to the bottom quadrant, red to the left quadrant, and blue to the right quadrant. The shaded area represents the duration of the first visit of WASP-189 (visit ID: \texttt{CH\_PR100041\_TG000201}). The temperature variation in the right quadrant is notably distinct from the other three quadrants, exhibiting a strong dependence on the pointing direction. While orbital oscillations are clearly discernible as spikes, the alteration in pointing direction gives rise to the more significant fluctuations in temperature. Panel B: Heater rings (operational, front and rear, and survival) in the telescope tube. A gap or discontinuity in the heater rings can be seen on the right side when one is standing in front of the telescope tube. Panel C: Deviations in flux from the average value during the visit of WASP-189. At the beginning, there is a downward flux ramp, a consequence of the telescope temperature variation (as discussed in Sect.~\ref{sec:Ramp}).}
    \label{fig:ramp1}
\end{figure*}

The thermal control of the telescope tube is achieved through a PID controller, which manages two separate zones: the front and rear of the telescope. Each of these zones is equipped with four temperature sensors positioned in different quadrants of the tube (top, left, bottom, and right). Additionally, there is a heater ring encircling the tube. The controller employs a majority voting strategy for temperature regulation: it disregards the maximum and minimum sensor readings and calculates the average from the remaining two sensors. This approach ensures stability by preventing erroneous readings from adversely affecting the thermal control system. As shown in panel A of Fig.~\ref{fig:ramp1}, the temperatures recorded by the four front sensors over a 32\,h period illustrate that the temperature in the bottom quadrant (\textit{thermFront\_2}) is consistently lower by 1.5--2\degr\ compared to the other quadrants and very sensitive to changes in the pointing direction.
Notably, the heater rings for both the front and rear zones feature a gap on the right side of the telescope tube. This feature is illustrated in Fig.~\ref{fig:ramp1}, panel B.

Throughout nominal operations, the temperature of the telescope tube is controlled by the heater rings to maintain stability. However, the presence of the `gap' on the right quadrant introduces a limitation: this specific telescope area remains at a lower temperature due to the inability to be actively heated. The external heat load on the telescope shifts with changes in pointing direction as the orientation relative to the Sun changes in accordance with the LoS.

The temperature fluctuations observed in the right quadrant after a slew have an effect on the materials of the telescope tube, leading to expansion or contraction. Consequently, the focus and alignment of the telescope are impacted during thermal stabilisation, which in turn directly influences the extracted light curve, as seen in panel C and elaborated in Sect.~\ref{sec:Ramp}.

\begin{figure*}
    \centering
    \includegraphics[width=\textwidth]{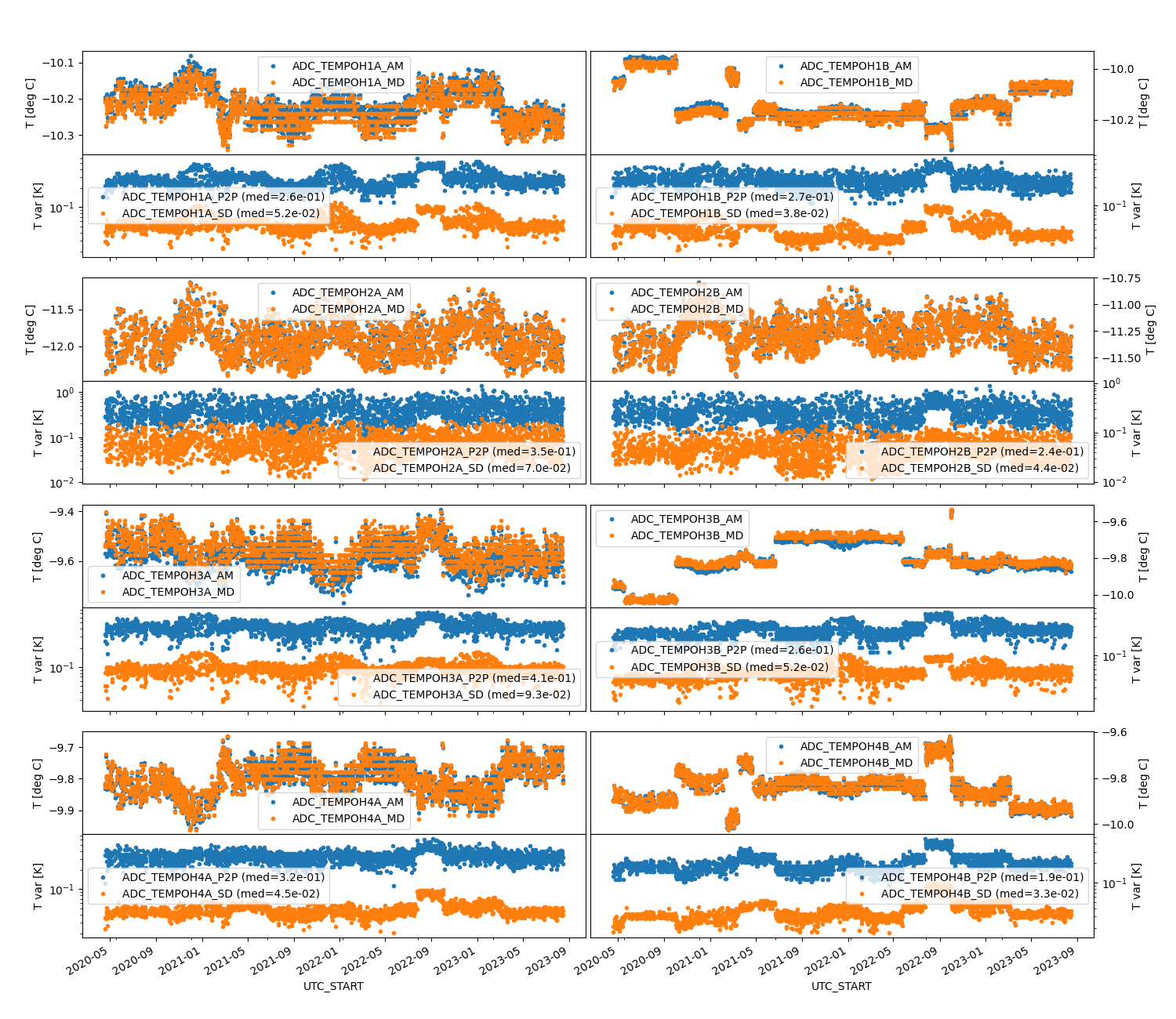}
    \caption{Stability of the telescope tube since the start of nominal operations. Data are aggregated to one point per visit. The values of the four front and rear quadrants are shown. Top: Arithmetic mean (AM) and median (MD). Bottom: Peak-to-peak variation (P2P) and standard deviation (SD). Since thermal stability requirements on the telescope tube were not stringent (5 K), an accuracy of less than 1\,K was deemed satisfactory, and the thermistors have not undergone extensive calibration. Jumps in the curves can be seen when power cycling the instrument as voltage levels change.}
    \label{fig:TEL_stability}
\end{figure*}

The heaters efficiently counteract temperature fluctuations resulting from changes in pointing, with the exception of the gap region. When the relative position of the telescope tube to the Sun undergoes substantial alteration between two pointings, a higher heater power is required to uphold stability, resulting in lower temperatures in the bottom quadrants. 
This behaviour is further highlighted in Fig.~\ref{fig:TEL_stability}, which presents the temperature data from all 8 telescope temperature sensors throughout the mission. Sensors 2A and 2B consistently read cooler temperatures and exhibit more significant variations in both different pointings (top panels) and within a single visit (standard deviation and peak-to-peak variations, lower panel).

The incomplete circular design of the heater rings has a rationale: the heater wiring is arranged to avoid direct contact with the heaters themselves. This design decision seemed unproblematic at the time, as the requirement for telescope thermal stability was established at 5\,K. The actual measured stability has consistently surpassed this requirement, allowing the instrument to remain comfortably within specifications. However, practical experience has demonstrated that the initial requirement for thermal stability of the telescope tube falls short of the demands of ultra-high-precision photometry. Remarkably, the effects on the light curve can be removed with detrending techniques (see Sect.~\ref{sec:RampCorr}).

An additional source of thermal instability occurs when light from the Earth enters the telescope tube, which is common during a substantial portion of visits, particularly during Earth occultations. The interior of the tube lacks significant thermal insulation and possesses minimal thermal inertia, resulting in direct heating by the Earth. This situation can create heat gradients between quadrants that thermal control cannot effectively balance, perturbing the tube's equilibrium temperature. The temperature variations in this scenario are notably smaller than those arising from changes in pointing direction. In panel A of Fig.~\ref{fig:ramp1}, these variations are evident as periodic spikes corresponding to the orbit's frequency.

\paragraph{\bf{Flux ramp}}
\label{sec:Ramp}

The thermal instabilities in the telescope tube induce changes in the optical system, resulting in what is known as `PSF breathing'. This phenomenon causes the PSF to either contract or expand based on whether the telescope's temperature decreases or increases. This PSF breathing, in turn, affects the extracted photometry, leading to variations in the measured flux within the aperture.

The characteristic signature of this effect can be observed as a gradual increase or decrease in the star's light curve, depicted in panel C of Fig.~\ref{fig:ramp1}. It is important to note that the flux ramp is not present in all light curves; rather, it appears only in cases where the amplitude of temperature variation significantly influences the PSF shape. As the telescope temperature takes time to stabilise in the new pointing direction, the duration of the flux ramp varies. Based on empirical observations, the ramp's duration spans between 2.5 and 12 hours.

The amplitude of the flux ramp is closely tied to the size of the photometric aperture employed for extracting the light curve. It typically ranges from around 100 to 1000\,ppm, with smaller amplitudes for larger apertures. The occurrence of the ramp is correlated with the temperature readings from the telescope's \textit{thermFront\_2} sensor, which allowed the generation of strategies to mitigate its impact on the light curve. Sect.~\ref{sec:RampCorr} delves into two methods for correcting the ramp, leveraging the measurements from the aforementioned sensor.
\subsubsection{Pointing stability}
\label{sec:PointStability}

The PF offers two fine guidance modes for science observations: No PITL (NOPITL) with autonomous target tracking employing only the SC star trackers, and PITL, where the instrument conveys offsets in the $X$ and $Y$ directions on the detector in arcseconds to the attitude and orbital control system (AOCS) of the PF at a cadence ranging from 1 to 60\,s, contingent upon the exposure time of the ongoing visit. 

Comparatively, PITL tracking is generally superior due to its ability to compensate for thermo-elastic offsets between the star tracker optical heads and the LoS of the instrument. However, there are specific scenarios in which PITL tracking might be less favourable. This includes instances involving contaminating background stars, or even visual binaries, in the centroiding window or strong cosmic rays affecting faint targets, as it can have an adverse impact on pointing stability. Although the use of PITL tracking was originally intended for all targets, after IOC, it was disabled for stars with a magnitude greater than $G = 11$\,mag. This adjustment was made to avoid wrong pointing corrections induced by contamination from background stars and/or cosmic rays.

Appendix~\ref{pointingPerf} offers a detailed evaluation of the pointing performance of the satellite. We can summarise it by saying that the PITL tracking keeps the pointing deviations small, ranging from 0.2 to 0.7\arcsec. The NOPITL, which relies only on the SC star trackers, performs similarly to the PITL mode. The overall pointing stability is better than 1 arcsecond, well below the requirement of 4\arcsec.
\subsubsection{Stability summary}

In summary, all stability requirements have been exceeded except for the issue of flux ramp, which can be observed during the initial 2.5--12\,h of a visit following a significant change in the Sun angle with respect to the previous pointing direction. However, this ramp can be effectively mitigated through calibration efforts.

While the telescope's thermal stability meets its requirement, it is evident that the thermal stability requirement itself was not stringent enough. Notably, no specification was set for the temperature homogeneity across the four quadrants of the telescope. As a result, the presence of a heater blanket gap and the application of a majority voting scheme were considered acceptable. For future missions, it is advisable to include such a requirement, especially if utilising defocused optics, as it has been demonstrated to be considerably more sensitive to changes in focal distance when compared to a focused system.
\begin{figure}
    \centering
    \includegraphics[scale=0.3]{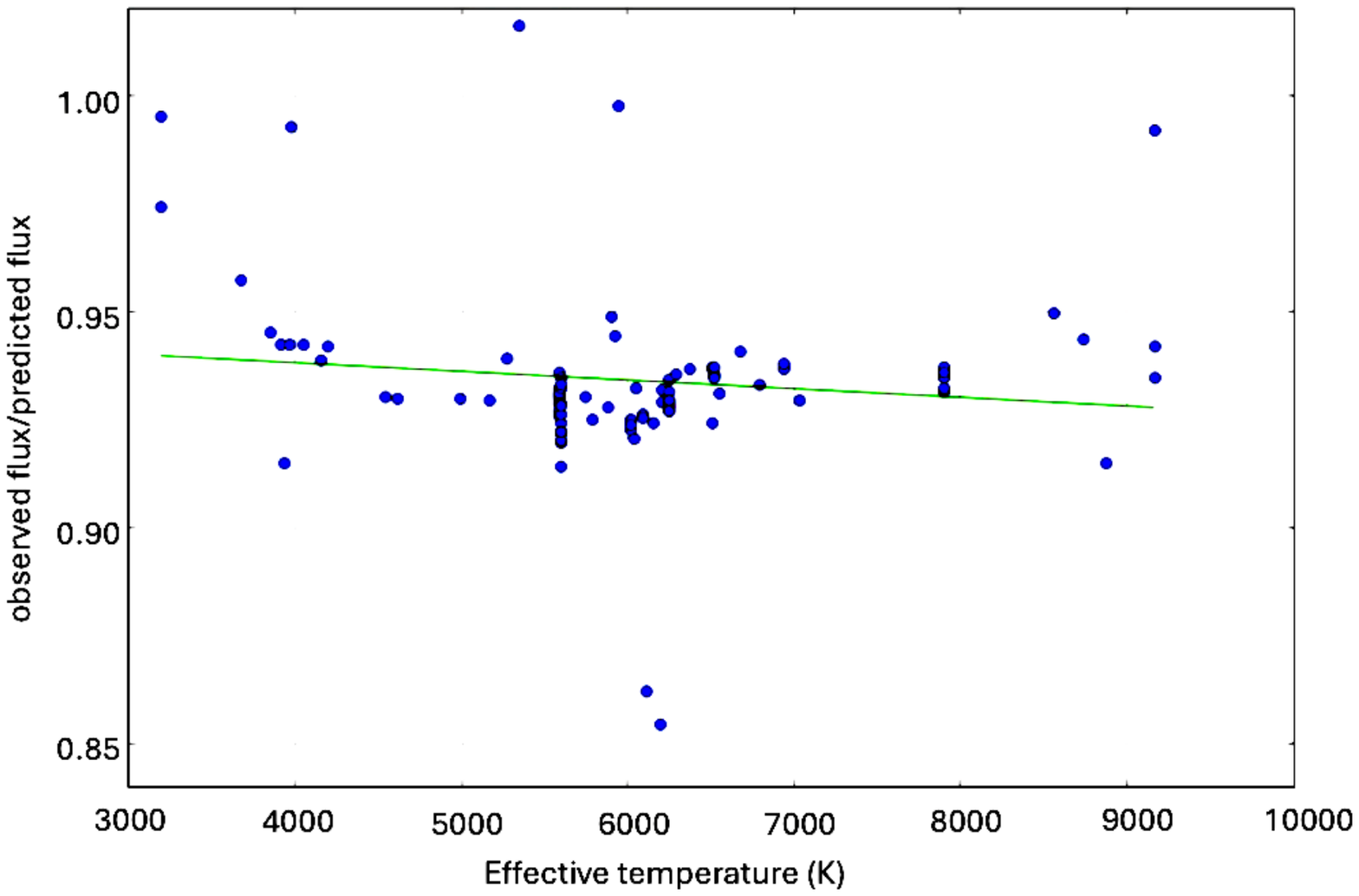}
    \caption{ Ratio between the observed and predicted flux as a function of the stellar effective temperature. The blue dots in the plot represent over 100 measurements conducted during the IOC. Notably, the observed flux is roughly 7\% lower than what was predicted by the instrument model. While the exact cause of this discrepancy remains uncertain, it is apparent that the model encompasses several sources of uncertainty that could potentially contribute to this difference. Some of these sources include uncertainties in the CALSPEC flux scale, variations in the measured optical transmission, and discrepancies in the QE, among others.}
    \label{fig:GT}
\end{figure}
\section{Instrument performance}
\label{sec:InstPerfromance}
\begin{figure*}
    \centering
    \includegraphics[width=\columnwidth]{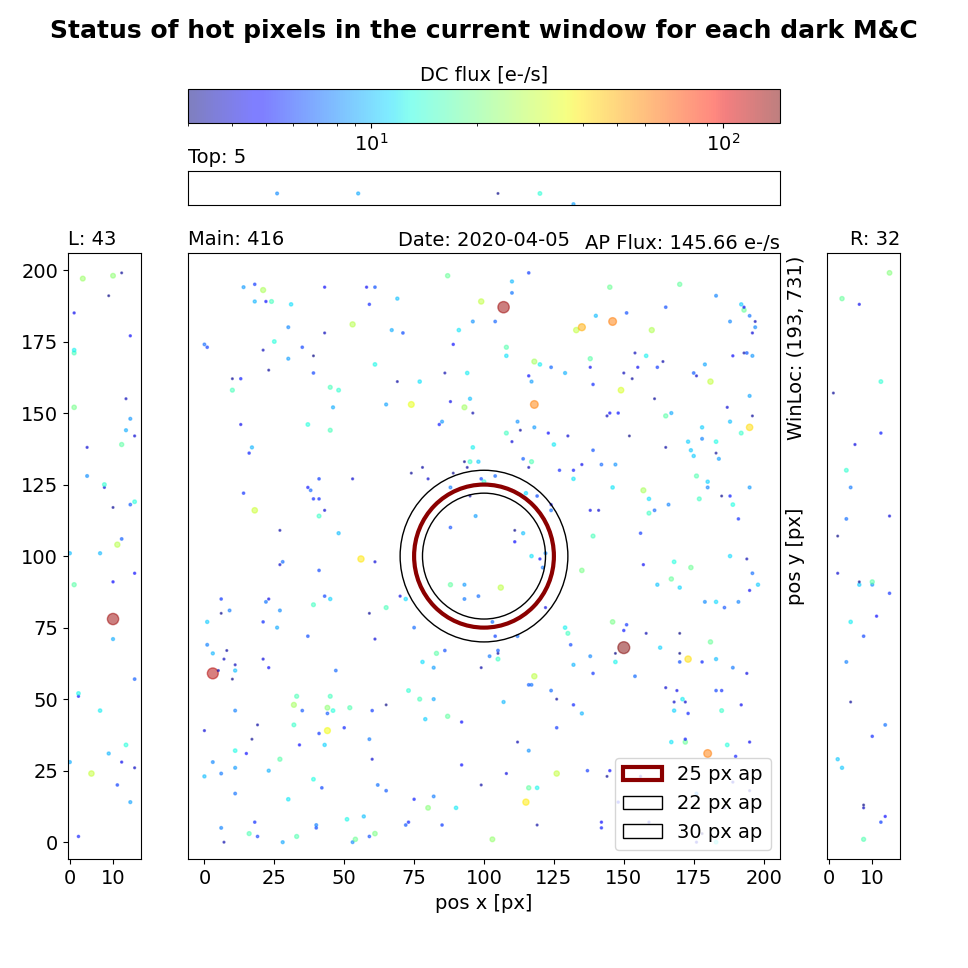}
    \includegraphics[width=\columnwidth]{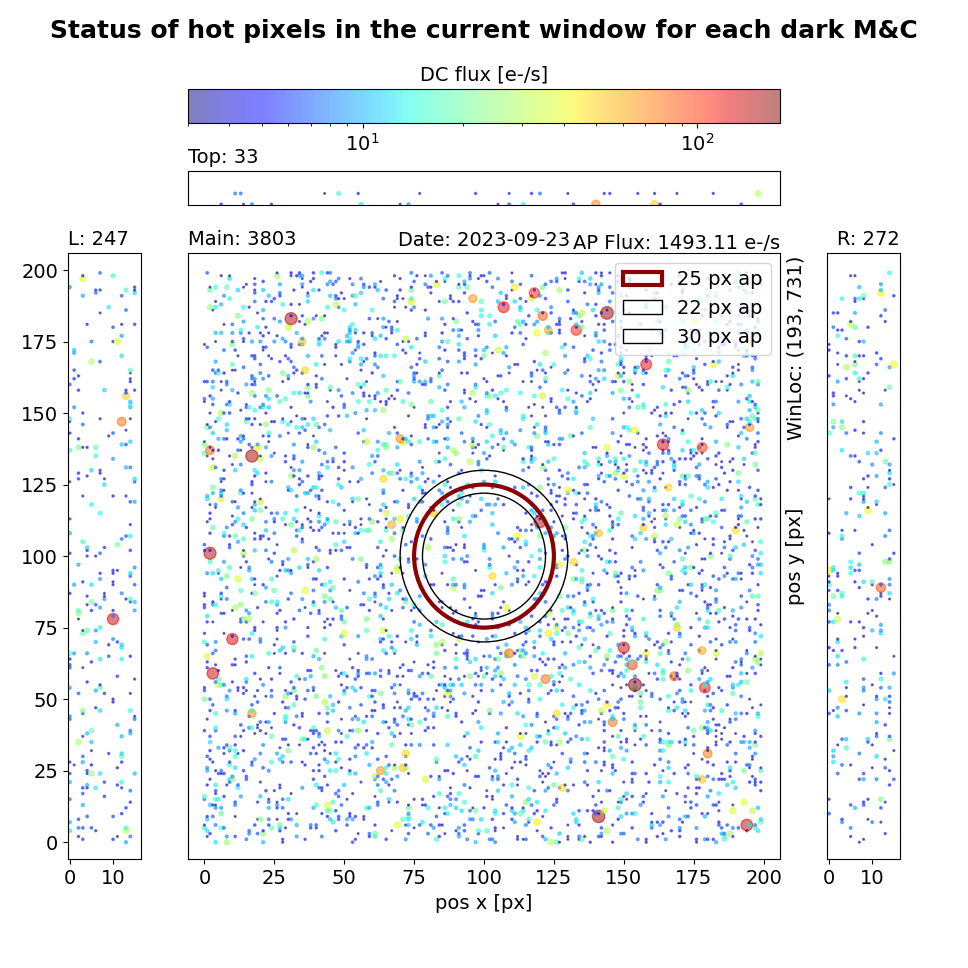}
    \caption{Hot pixels in the window frame and respective margins at the beginning and end of the nominal mission. The points' size and colour emphasise the dark current level of the hot pixels: the hotter the pixel, the bigger and redder the point is. The red and black circles represent three apertures of sizes $r=$ 22.5, 25, and 30\,px.}
    \label{fig:hp_win}
\end{figure*} 
\subsection{Throughput}
\label{sec:GT}
The throughput represents the relationship between the incoming photons at the entrance of the telescope tube and the resulting measured photoelectrons by the CCD. Mathematically, the throughput is the product of the optical throughput and the QE of the camera. As discussed in Sect.~\ref{cal}, conducting direct measurements of the throughput during the ground calibration campaign posed challenges for a small class mission like \CHEOPS. Instead, the approach relied on quantifying the transmission characteristics of the optical components, as well as the QE, along with computational modelling.

From the very first observations after the first light, it became evident that the measured flux was notably lower than anticipated, exhibiting a deficit of around 25-30\%. As the IOC phase progressed, the primary factors contributing to this discrepancy were investigated and are outlined as follows:

    \paragraph{\bf Telescope aperture difference} The initially estimated total flux was higher than expected due to a miscommunication between teams. The calculation of incident flux used a telescope aperture of 32\,cm instead of the effective aperture of 30\,cm\footnote{Although the diameter of the telescope aperture is 32\,cm, the `effective aperture' (the equivalent collecting area of the telescope) is smaller due to the presence of blocking elements in the optical path, like the secondary mirror mount. Sometimes, this is included in the optical throughput.}. This difference in collecting area accounted for approximately 14\% of the observed `missing flux'.
    \paragraph{\bf QE Calibration} The absolute QE of the \CHEOPS detector could not be reliably determined by the manufacturer due to technical challenges during measurements. Only relative QE measurements were available at the consortium level, and these measurements had technical issues at shorter wavelengths, making direct comparison with expected values difficult. During IOC, ESA conducted careful QE measurements in the laboratory on a spare flight CCD, which was a twin of the actual flight CCD. These measurements indicated that the QE was lower than initially assumed for the throughput calculations before flight. The systematic difference between the assumed and actual QE accounted for a 4\% difference. This translates into a reduction of 3.8\% of the throughput and contributes to the observed `flux deficit'.
    \paragraph{\bf Use of \textit{Gaia} magnitudes} \textit{Gaia} magnitudes were determined to be better suited for \CHEOPS than V-magnitudes. The similarity in shape between the \CHEOPS and \textit{Gaia} passbands, as seen in Fig.~\ref{fig:bandpass}, allows for a more accurate determination of the relationship between \textit{Gaia} magnitudes and \CHEOPS magnitudes, as well as their respective photoelectron fluxes. This accuracy surpasses that achieved by using traditional $V$-band magnitudes, especially when considering cool stars (\teff$<4000$\,K). For such stars, only a marginal fraction of the flux within the \CHEOPS passband aligns with the $V$ band, leading to substantial uncertainties. Furthermore, the disparity between $V$-band magnitude and \CHEOPS magnitude displays strong dependence on \teff for cool stars, intensifying sensitivity to uncertainties in \teff, which are generally sizeable. Consequently, this exacerbates the uncertainty associated with projected electron flux predictions. Consequently, an operational adjustment was made, adopting \textit{Gaia} magnitude as the default reference for star brightness. For detailed information about the relationship between \textit{Gaia} magnitude and \CHEOPS magnitude, refer to Appendix~\ref{cheops_mag}.
    \paragraph{\bf Correction to \textit{Gaia} magnitudes} The correction to \textit{Gaia} magnitudes, as derived by \citealt{Casagrande} based on CALSPEC flux measurements, was implemented ($G^{\mathrm{corr}} = 0.0505 + 0.9966G$).
    \paragraph{\bf Improved SEDs} Black body approximations for stellar flux were replaced by more accurate SEDs (\citealt{Husser2013}), leading to improved estimations of expected flux.
    
The similarity in shape between the \CHEOPS and \textit{Gaia} passbands, as seen in Fig.~\ref{fig:bandpass}, allows for a more accurate determination of the relationship between \textit{Gaia} magnitudes and \CHEOPS magnitudes, as well as their respective photoelectron fluxes. This accuracy surpasses that achieved by using traditional $V$-band magnitudes, especially when considering cool stars (\teff$<4000$\,K). For such stars, only a marginal fraction of the flux within the \CHEOPS passband aligns with the $V$ band, leading to substantial uncertainties. Furthermore, the disparity between $V$-band magnitude and \CHEOPS magnitude displays strong dependence on \teff for cool stars, intensifying sensitivity to uncertainties in \teff, which are generally sizeable. Consequently, this exacerbates the uncertainty associated with projected electron flux predictions.

Collectively, these factors accounted for part of the observed flux deficit, which could be reduced to around 7\% as can be seen in Fig.~\ref{fig:GT}. The source of the discrepancy between the newly calculated expected and measured flux is not entirely clear, but it is not a cause for concern either. Notably, approximately one-third (2.2\%) of the observed discrepancy can be attributed to inaccuracies in \textit{Gaia} magnitudes. A correction for these magnitudes has been derived by \citealt{Casagrande} (refer to equation 3 of their paper). However, for simplicity, this correction was not separately applied in this context. It was observed that applying the Casagrande \& VandenBerg correction and then performing the fit resulted in a correction similar to the one derived by performing the fit without initially applying the correction. Overall, achieving accuracy within 7\% between the expected and measured flux is considered acceptable as it does not affect the overall relative photometry performance.

\subsection{Sensitivity loss }
\label{sec:SensitivityLoss}

Monitoring the instrument's throughput over time is essential for understanding and quantifying any sensitivity loss. To achieve this, the flux of all observed targets at multiple epochs has been analysed, and trend lines have been fitted to each target. For stars observed throughout the mission lifetime, the data are further split into observation seasons, and each season is fitted individually. However, the data should be handled with care. Variable stars need to be rejected to ensure the accuracy of the analysis. Therefore, only trends with errors smaller than 3 ppm/day are considered. Also, stars fainter than magnitude $G = 9$ show a positive trend in flux if the hot pixels correction is not applied. This is because the dark current of the hot pixels increases the flux in the aperture, while the background estimate is not yet affected by them. Therefore, stars brighter than magnitude $G = 9$ are exclusively considered for this analysis.

\begin{figure}
    \centering
    \includegraphics[width=\columnwidth]{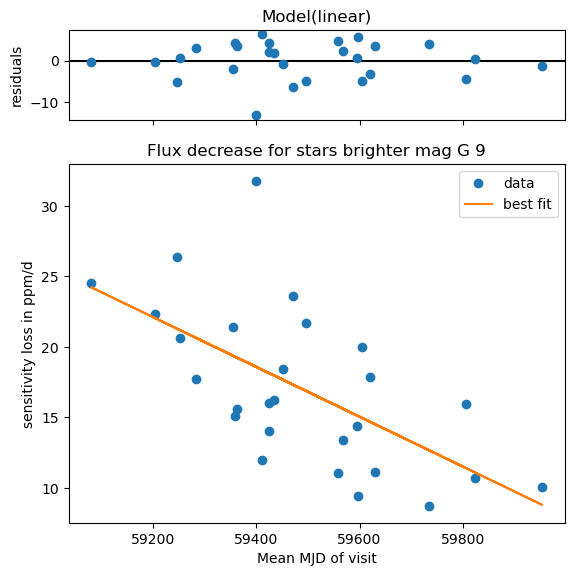}
    \caption{Decrease in flux for all constant stars observed during the first 3 years of operations. Only stars brighter than $G = 9$ are taken into account to avoid the effect of hot pixel dark current increase. We can see a clear trend towards a reduced degradation in transmission from 25 to 10\,ppm/day.}
    \label{fig:flux_over_time}
\end{figure}

As shown in Fig.~\ref{fig:flux_over_time}, results indicate an initial trend of $-25$\,ppm/day in flux, which has now decreased to approximately $-10$\,ppm/day. Three main components likely influence the observed trend:
\begin{itemize}
    \item Outgassing and condensation on the mirrors/CCD: The presence of outgassing and contamination is expected to gradually diminish over time, resulting in a slowdown of the sensitivity loss.
    \item Radiation damage to the optics and mirror coatings: The radiation damage is expected to continue with some saturation effects. 
    \item CTI effect: CTI leads to a trail after bright pixels. There are two kinds of trails: a short one that appears during readout and should not affect aperture photometry absolute levels and a long one from the fast shift to the readout section. The long trails are ~170 px long and remove part of the flux from the aperture, leading to a perceived sensitivity loss. This effect will increase linearly over time but can be corrected using CTI correction. 
\end{itemize}
The total sensitivity loss is consistent with the predicted worst-case transmission reduction from radiation damage. However, this is considered an absolute worst-case scenario that should typically overestimate the actual degradation by an order of magnitude. A more plausible scenario is that, at the beginning of the mission, the sensitivity loss was likely dominated by outgassing, while the continued linear decrease is a combination of radiation damage to the optics and the CTI effect, with the latter probably being dominant.

\subsection{Timing precision}
\label{sec:Timing}

The science requirements for the mission stipulate that the precision of the time stamp (UTC) attached to the photometric measurements' exposure time should be better than 1 second. This level of timestamp accuracy is deemed suitable for determining the central time of transits, which can last from one to several hours, facilitating the analysis of transit timing variations. To meet this criterion, it is necessary to ensure the timing accuracy of various instrument components in relation to the satellite's onboard time (OBT; which distributes the time reference for the satellite) while also synchronising the PF's OBT with UTC. The timing error budget encompasses all contributors to timing uncertainty, which, when combined, should remain below 1 second to fulfil the requirement. During ground testing, the overall precision of the time tag was assessed to be under 650\,ms, comfortably meeting the desired accuracy. 

In the mission's operational phase, \CHEOPS' timing precision was validated against an astrophysical clock represented by the short-period eclipsing binary HW~Vir.
This binary system exhibits eclipses resulting from the subdwarf-B star being occulted by a low-mass companion, with an orbital period of roughly 2 hours and 48 minutes. Over the span of years, long-term investigations into the eclipse times of HW Vir have unveiled pseudo-periodic variations around a linear ephemeris, with fluctuations of the order of a few minutes. These variations could potentially arise from one or more substellar companions in the binary system \citep{2003Obs...123...31K,2021MNRAS.506.2122B, 2022MNRAS.514.5725P}. In a study by \citealt{2018MNRAS.481.2721B}, the light curve of HW Vir collected by the \textit{Kepler} K2 mission was analysed, revealing that the eclipse times exhibited periodic behaviour to a precision better than 0.5 seconds during the 2-month observation period. Characterised by deep eclipses (approximately 50\% depth) with narrow duration (around 20 minutes) and with a moderately bright magnitude ($G=10.6$\,mag), the star allows for the measurement of mid-eclipse times with a precision of about 1 second using ground-based 1-metre class telescopes. Observations of HW Vir were acquired with several ground-based telescopes and \CHEOPS. The time of mid-eclipse was then compared, indicating that the mean time offset measured by \CHEOPS is better than 1\,s (the detailed calculation is available in Appendix~\ref{HWVirtiming}).

It is noted that the maximum deviation of the OBT to UTC correction during the nominal mission so far has been estimated at 188\,ms in an exceptional case, and a typical value is 50\,ms. Synchronisation of instrument time to OBT is expected to be better than 1 ms. However, there is no independent confirmation of these numbers. 
\subsection{Bad pixels}
\label{sec:BadPixels}

Monitoring and characterising the degradation of the CCD is essential to ensure and understand the quality and reliability of the data obtained from the instrument. This process involves tracking the evolution of bad pixels, which includes hot pixels and RTS pixels, in addition to analysing changes in their dark current and flux distribution. The objective is to comprehend the impact of bad pixels on measurement quality, quantify their contribution to noise, and implement appropriate measures to reduce their effects.

The bad pixels M\&C programme involves several steps, and a \texttt{Python} code is utilised to automate the generation of overview reports as new dark M\&C data are acquired. Additional reports are generated manually using \texttt{Jupyter} notebooks based on the latest data available in the archive.  The weekly summary report provides visualisations of new hot pixels near the PSF and summarises the impact of these hot pixels on photometric performance. Other reports include more comprehensive analyses, focusing on various aspects of hot and RTS pixels. While this summary does not delve into the details of each analysis, it provides an overview of the current status of hot pixels and their influence on photometric performance.

Monitoring hot pixels can be divided into three main steps. The first involves visualising the locations and characteristics of hot pixels and identifying any changes over time. Then comes noise estimation, i.e., determining the expected noise introduced by hot pixels in the photometric aperture. This assessment is crucial for avoiding areas on the CCD where these pixels could significantly affect photometric performance. The last step delves into the long-term behaviour of hot pixels, tracking any trends or patterns in their intensity or behaviour.

It is worth highlighting that no dead (i.e.\, unresponsive to light) or permanently saturated pixels have been found to date. Pixels less responsive to light (with lower QE, for example) are tricky to characterise. While a detailed analysis is not included here, some pixels (and even a column) have become `darker' in the course of the mission (as discussed in Sect.~\ref{sec:VisBadPixels}). 

\subsubsection{Visualisation of hot, RTS, and dark pixels}
\label{sec:VisBadPixels}

 The identification of the hot pixels is done using `dark' images obtained during dark M\&Cs and following the procedure described in Appendix~\ref{sec:CalBadPixels}. A pixel is classified as `hot' if the measured flux exceeds 3 electrons per second when it is expected to be close to zero because no source is illuminating it. The `self-luminosity' of a pixel is attributed to the excess in dark current. During the nominal mission, the first dark M\&C was conducted on April 5, 2020, with more than 11\,000 hot pixels detected in the full frame and approximately 400 in the window image. The last dark M\&C taken at the end of the nominal mission reveals that around 110\,000 hot pixels exist in the full frame and 3800 in the active 200$\times$200\,px window location (Fig.~\ref{fig:hp_win}).

An interesting observation is that the total count of pixels identified at least once as hot pixels exceeds 170\,000. Comparing this number with the count of hot pixels detected in the last dark M\&C reveals that approximately one-third of the detected hot pixels (about 60\,000) are inactive. These inactive pixels can be either RTS pixels in a normal state with a long-time constant or hot pixels that were active only briefly and then returned to a normal state.

\begin{figure}
    \centering
    \includegraphics[width=\columnwidth]{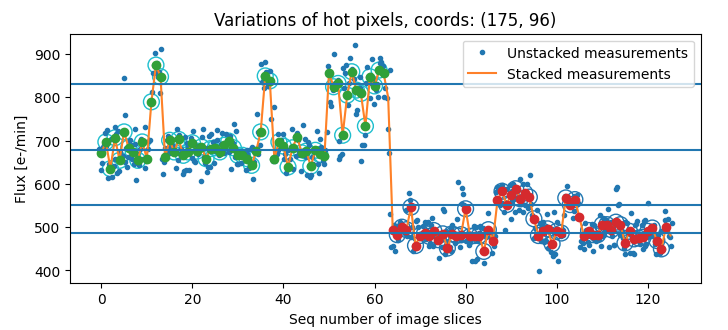}
    \caption{Example of multi-level variation of the dark current of a hot pixel in the latest RTS observation. Most RTS pixels exhibit oscillations between three or four dark current levels (see also Fig.~\ref{fig:rts_dc_levels}).}
    \label{fig:RTS_jumps}
\end{figure}

 A hot pixel is classified as an RTS pixel if variations in its dark current happen in a timescale smaller than that of an average science visit duration (8 h).  The detection and characterisation of RTS pixels is done through dedicated dark M\&C visits, scheduled 4 times a year with a duration of 10 orbits each. Assuming that 40\% of the time of the visit cannot be used for RTS characterisation (due to SAA crossings, Earth occultations, etc.), the analysis occurs over an effective time of 8\,h. Due to the high number of images involved in these visits, only the active window is analysed. Identifying RTS pixels involves searching for potential jumps in the light curves of hot pixels. The presence of a single jump is sufficient to classify a hot pixel as an RTS pixel. The light curve of an RTS pixel is exemplified in Fig.~\ref{fig:RTS_jumps}.

\begin{figure}
    \centering
    \includegraphics[scale=0.25]{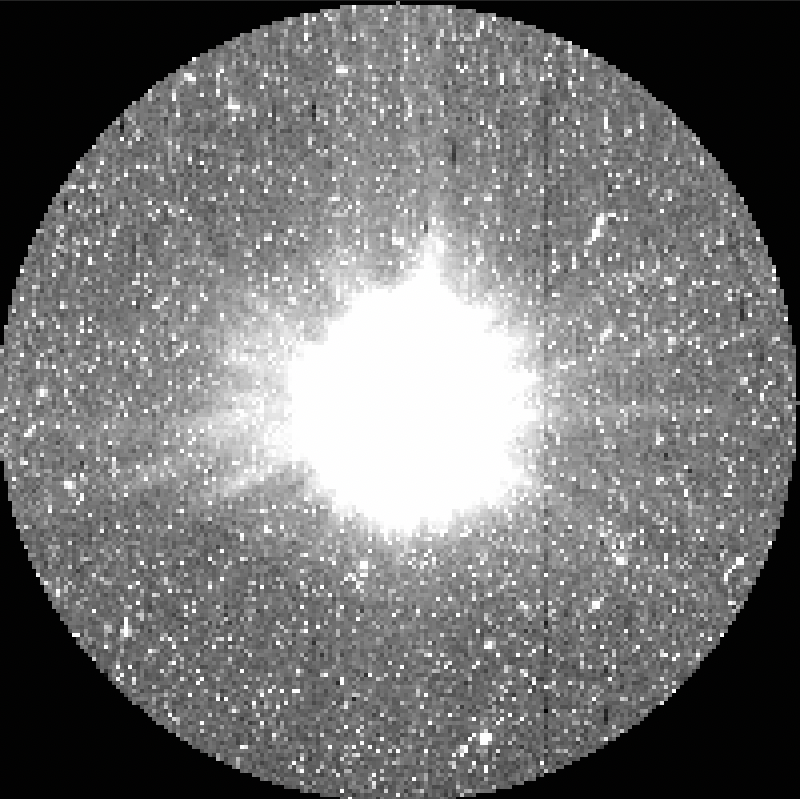}
    \caption{Presence of a `darker column' in the images of GJ~908 to the right of the star (visit ID: \texttt{CH\_PR100018\_TG040003}). Its absolute coordinate (with respect to the left edge of the full frame) is x = 327.}
    \label{fig:dark_column}
\end{figure}

In September 2022, a new CCD artefact was found: a dark column appeared in the CCD (Fig.~\ref{fig:dark_column}), which was detected first in an image of an observation of GJ~908. Further investigations were carried out, which confirmed its presence in all the visits after the first appearance and revealed more unusual behaviour. The column is not dead; the pixels still respond to light, but the flux of these pixels is lower than those of the surrounding ones. This lower response is dependent on the incident flux; for high pixel values (flux $> 10\,000$\,ADU), the loss of flux is relatively small, $\sim$ 0.14\%, while low flux cases (flux $< 5$\,ADU), it can reach more than 50\%, which mean that the higher the incident flux, the larger the absolute flux loss, but the relationship is not linear. The other outcome of the analysis is that many short columns with a length of 10--20\,px and the same feature can be found in many places in the CCD. Even though there were numerous discussions involving various parties, including those from the CCD manufacturer, the actual reason for this phenomenon remains unknown. 

\subsubsection{Noise estimation and best window location}
\label{sec:win_loc}

\begin{figure*}[]
    \centering
    \includegraphics[scale = 0.60]{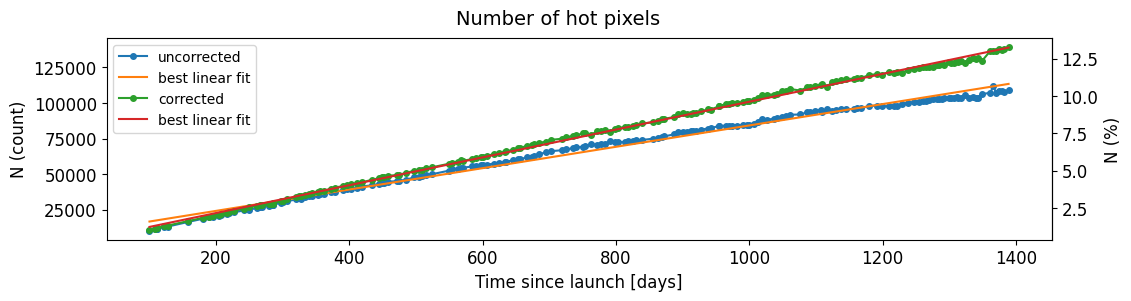}
    \caption{Generation rate of hot pixels. Considering the whole mission timeline, it is around 98\,px/day, while at the beginning of the mission, the rate was approximately 100\,px/day. Blue dots represent the raw data with background correction and green dots correspond to corrected data accounting for CTI  and statistical effects.}
    \label{fig:hp_gen_rate}
\end{figure*}

Due to their potential to significantly impact photometric performance, addressing hot pixels is crucial to ensure accurate measurements. Hot pixels can lead to complications such as the dilution of star flux and the introduction of noise. To mitigate these effects, a best practice is to locate a region on the CCD devoid of hot pixels and use it to place the target. However, the increasing number of hot pixels made finding a single uncontaminated spot impossible. Consequently, a method was devised to estimate the noise contribution of hot pixels and determine an optimal window location that minimises their impact on photometric precision.

This problem was tackled in two stages: (i) evaluating the error contribution from shot noise caused by hot pixels and (ii) quantifying the noise stemming from PSF jitter. In the first scenario, hot pixels fall within the aperture, augmenting the contaminating flux and overall shot noise. In the second, as the telescope experiences jitter, the PSF moves, causing the aperture to jitter (following the centroid) and resulting in flux variations due to hot pixels entering and exiting the aperture. These individual noise values are combined in quadrature to yield the total noise contribution at each possible target location. The result of this can be imagined as a noise map (details on the noise calculation can be found in Appendix~\ref{sec:NoiseBadPixels}).

The dark current correction is a feature that became available with the release of \DRP \texttt{v14}\footnote{All visits, since the beginning of the mission have been reprocessed with \DRP \texttt{v14} meaning that they are all corrected for hot pixels. }. It is then possible to compare the noise due to hot pixels with and without the correction. For our reference star ($G = 12$\,mag), we can assume 120\,ppm/3h photometric noise without hot pixels in the CCD (see, for example, Fig.~\ref{fig:PhotomPerf}, mid-panel) and an anticipated maximum noise of 45\,ppm/3h (600\,ppm/min) due to hot pixels (since we performed window location adjustment if the noise exceeded this limit). \text{Combining them in quadrature} yields 128\,ppm/3h, consistent with the most recent results from the simulations described in Appendix~\ref{sec:NoiseBadPixels}. The dark current correction is now essential to maintain this low noise level caused by hot pixels. The number of good locations without dark correction where the expected maximum noise is below 600\,ppm/min has decreased significantly. However, dark correction can reduce the expected maximum noise due to hot pixels to as low as  $\sim 200$\,ppm/min independently of the window location, releasing the pressure of finding a sweet spot in the CCD to minimise the noise due to hot pixels. 

\begin{figure*}
    \centering
    \includegraphics[scale = 0.45]{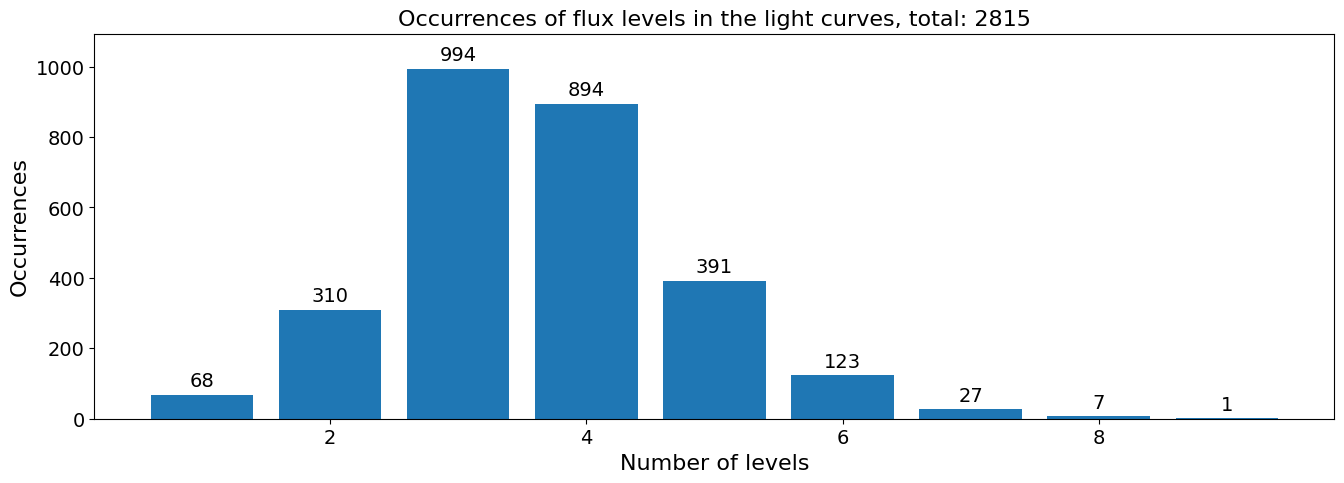}
    \caption{Distribution of long-term variable hot pixels based on the number of their dark current levels. A significant proportion of these pixels exhibit oscillations between three or four distinct dark current levels. This finding underscores the complexity of the dark current behaviour in these pixels, with multiple levels contributing to the variations observed over time.}
    \label{fig:rts_dc_levels}
\end{figure*}

\subsubsection{Dark current, hot, and RTS pixels: Long-term evolution}

\begin{figure*}
    \centering
    \includegraphics[scale = 0.45]{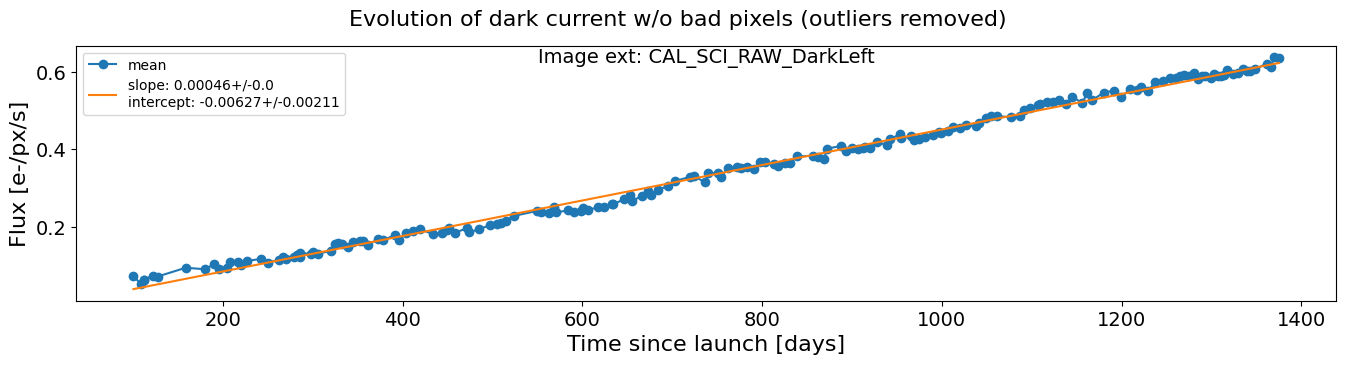}
    \caption{Evolution of the bulk dark current in the left margin of the CCD. The dark current is increasing by 0.5 milli-electrons per day.}
    \label{fig:dc_evo_left}
\end{figure*}

\paragraph{\bf{Number and generation rate of hot pixels}}
The number of hot pixels in the CCD is continually increasing. At the onset of the nominal mission, over 10\,000 hot pixels were already counted. At the end of the nominal mission, this number has escalated one order of magnitude, as demonstrated in Fig.~\ref{fig:hp_gen_rate}. Their growth follows a relatively linear pattern over the long term, but on a monthly basis, it displays large fluctuations. Space weather factors into the count and energy of cosmic rays that impact the detector. Satellite passages through the SAA, occurring approximately 10 times daily, also significantly contribute to pixel excitation. It is crucial to note that hot pixels are not consistently in an excited state; their dark current can decrease, and there is even the possibility for them to revert to a normal state.

The average rate of hot pixel generation throughout the entire mission duration is approximately 98 pixels per day, as illustrated in Fig.~\ref{fig:hp_gen_rate}. Notably, during the mission's first year, the rate was slightly higher at around 100 pixels per day. It is noted that to calculate this accurately, one must properly remove the background, account for the CTI effect, which can modify the detection threshold due to the hot pixel at the top of the CCD potentially losing more charges during the frame transfer, and consider the statistical effect, where the number of normal pixels, or free spots for cosmic rays to impact, reduces over time. Without these corrections, the hot pixel generation rate would appear much lower, approximately 75 pixels per day.

\paragraph{\bf Long-term evolution of RTS pixel}
When considering a period of approximately 10 orbits (the duration of the visits used to characterise RTS pixels), 10--12\% of hot pixels exhibit characteristics indicative of RTS behaviour. Among these, a majority manifest two distinct dark current levels that they alternate between. Additionally, our findings reveal that around 40--50\% of these RTS pixels exhibit only one jump (to a higher or lower flux level) between the two distinct dark current levels during this specific time interval.

With a sampling frequency of three months (i.e.\ time between two RTS visits), approximately 80\% of hot pixels in our study exhibit some form of variation in their dark current. The majority of these variations show state changes between three and four distinct dark current levels (as illustrated in Fig.~\ref{fig:rts_dc_levels}). Remarkably, we have not observed any discernible preferred directions in the changes of dark current; jumps in the dark current occur with the same likelihood towards lower or higher flux levels. Instances of pixels possessing only one dark current level indicate that their variations surpass their shot noise, and they cannot be attributed to random fluctuations alone. We note here these hot pixels cannot be purely classified as RTS ones since we do not know the source of the change in their dark current, which could be either RTS activity or cosmic ray impact.

\paragraph{\bf Evolution of the bulk dark current}
We also monitored the progression of the bulk dark current of the CCD, excluding the contribution from hot pixels. As \CHEOPS lacks a shutter, direct measurement of the dark current in the exposed region of the CCD is unfeasible due to contamination from stray and zodiacal light. However, what we can measure is the dark current in the covered side margins that remain shielded from light exposure. In this manner, although we cannot directly determine the dark current in the exposed region, we can infer the direction and magnitude of changes in it. It is important to mention that the increase in the dark current in the covered regions of the CCD may be different to the one in the exposed region, as seen in irradiated CCDs in the laboratory. During the preparation of the mission, ESA analysed the dark current of an irradiated CCD and found that the aluminium-covered areas (storage and margins) show around 3.5 times higher dark signal than the uncovered (image) area (P.\ Verhoeve, private communication). The reason for an increased dark current in the covered section after irradiation remains unclear. Therefore, the characterisation of the dark current in the covered regions can only be considered as indicative of the status in the image section. Fig.~\ref{fig:dc_evo_left} illustrates the increase in dark current observed in the left margin of the CCD. Over the course of the nominal mission, this dark current has increased by a factor of 10. We focused on the left margin because the right one exhibits significantly larger oscillations in the bias (see \citealt{Deline2020}), rendering precise measurements evasive.

\subsubsection{Bad pixels: Mitigation strategies}

At the current stage, hot pixels are not expected to impact observations significantly. However, light curves of faint stars may see an increase in noise, which could be 10\% for $G = 12$ stars by 2029. To address this issue, the latest release of the \DRP, \DRP~\texttt{v14}, corrects hot pixels based on the weekly update of the hot pixel map. This approach substantially mitigates the impact of hot pixels on the data quality.

Moreover, an additional mitigation measure is to anneal the CCD detector by heating it up to $+50\degr$C. Simulations and laboratory measurements have indicated that this annealing process can potentially improve the hot pixel count while slightly increasing the CTI (see \citealt{Verhoeve2022}). We note that the effect of the annealing process will be limited compared to devices operating at much colder temperatures (\CHEOPS CCD temperature is $-45\degr$C). Currently, no annealing process is planned and will only be if necessary to maintain optimal science performance.

The bulk dark current increase is expected to be negligible for a considerable period in the future. Adjusting the substrate voltage of the CCD is a viable mitigation strategy to address this issue further. Still, it has to be a trade-off with the saturation behaviour of the pixels.

Monitoring the RTS pixels is crucial, as they can introduce complexities in the data analysis, especially in aperture photometry. However, it is encouraging to note that PSF fitting photometry (such as the \PIPE described in Sect.~\ref{sec:Overview}) is relatively insensitive to RTS and hot pixels.

Additionally, other defects on the detector, such as pixels or groups of pixels with reduced sensitivity in the column direction, are being observed. The sensitivity loss in these regions depends on the amount of illumination, and further investigation is ongoing to understand the underlying cause and how it progresses.

\subsection{Charge transfer inefficiency}

The radiation-induced decrease in charge transfer efficiency (or CTI) is another critical ageing effect. Radiation favours the appearance of charge traps in the chip, which hold some of the charges in a pixel to release them after a certain time. This leads to charge trails in the column direction and, to a lesser extent, in the row direction (see \citealt{Verhoeve2022}). For example, CTI trails can be seen in \textit{Hubble} Space Telescope (see \citealt{AndersonBedin2010}) and CoRoT images, in laboratory images of irradiated CCDs, and in CHEOPS images from mid-2021 onwards. While CTI does not significantly affect aperture photometry in defocused extended images, it can introduce a signal in cases where the background illumination is very inhomogeneous, such as in images taken immediately before or after an Earth occultation (see Sect. \ref{sec:Systematics}). As these images are generally excluded from the analysis, the implementation of the CTI correction has been a lower priority. Nevertheless, the team is actively working on quantifying the CTI effect in terms of trap density and trap timescales to develop correction methods based on \citealt{Masseyetal2010} and \citealt{Massey2010}. We used images taken in the laboratory with irradiated CCDs under controlled illumination conditions as well as CHEOPS science images of targets repeatedly observed throughout the mission lifetime. The CTI correction algorithm that will eventually be implemented in the DRP will be subject of a future publication.

\section{Science performance}
\label{sec:SciPerformance}
\subsection{\CHEOPS ETC versus measured performance}
\label{sec:ETC}

\begin{figure*}
    \centering
    \includegraphics[width=\textwidth]{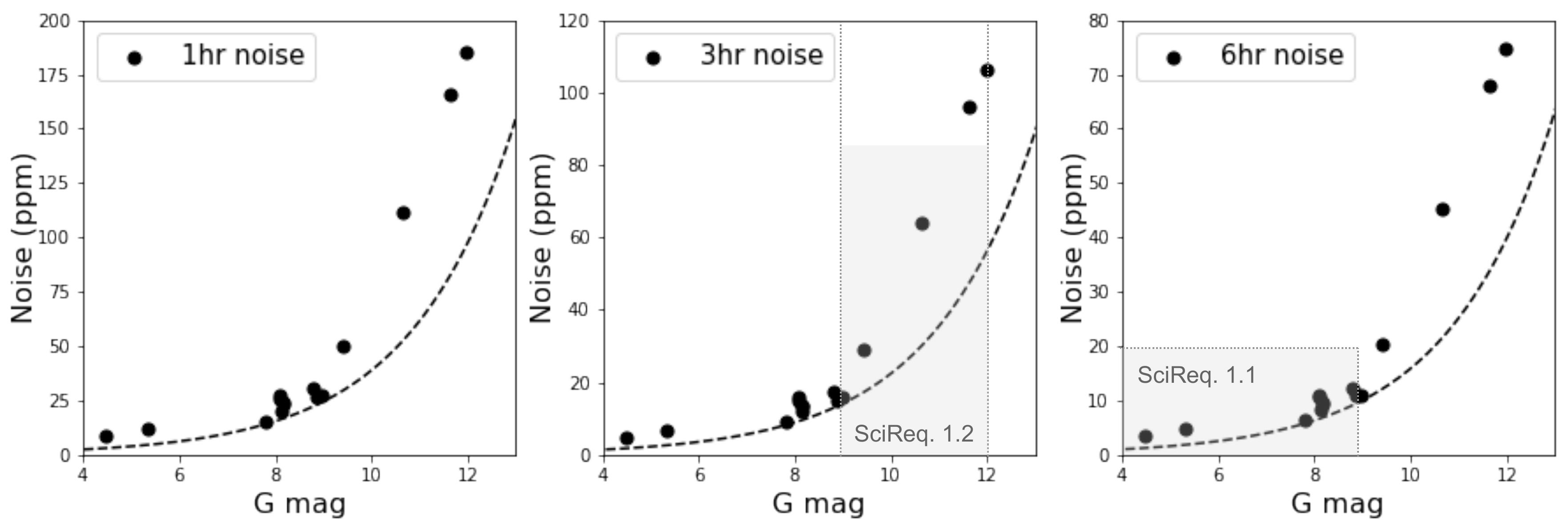}
    \caption{Measured photometric precision as a function of the stellar \textit{Gaia} magnitude and integration time. The dashed curve represents the photon noise. Results obtained with \pycheops (\texttt{v1.0.0}) for light curves extracted using \DRP~\texttt{v13} with the default photometric aperture of $r=25$\,px. The noise is estimated using the minimum error method in all the cases (following \citealt{Maxtedpycheops}, this is a conservative estimation). The grey region highlights the expected performance for bright stars (SciReq.~1.1) and faint stars (SciReq.~1.2).}
    \label{fig:PhotomPerf}
\end{figure*}

The \CHEOPS exposure time calculator (ETC\footnote{\url{https://cheops.unige.ch/pht2/exposure-time-calculator/}}) serves as a tool for predicting the flux uncertainty in a typical \CHEOPS light curve. Grounded in the instrument's noise budget and simulations using \CHEOPSim \citep{Futyan2020}, the ETC also incorporates the stellar noise contribution arising from granulation. Within this tool, primary instrumental noise components encompass:
\begin{itemize}
    \item White noise components: They comprise target photon noise, background photon noise stemming from zodiacal and stray light, photon noise due to dark current (including hot pixels), read-out noise, quantisation noise, and cosmic rays noise.
    \item Systematic noise components: This category entails contributions from the flat-field uncertainty linked to pointing jitter, the extended PSF, and Earth's stray light. Additionally, it accounts for variations in parameters like dark current, gain, QE, and analogue electronics influenced by temperature and voltage fluctuations. Worth noting is that the total systematic noise (excluding Earth's stray light) remains below 5 ppm within a single orbit.
\end{itemize}
While a valuable and comprehensive tool, the ETC adopts a generalised, semi-analytical approach. This implies that it does not model each observation with its own peculiarities to estimate the noise.  Consequently, it does not account for specific factors such as stellar background contamination (it assumes all targets are isolated in the sky), directional dependence of zodiacal light (maximum contribution considered), or the variable stray light contamination influenced by pointing and observation date (maximum allocated stray light contribution used per image). Effectively, the noise estimation offered by the ETC mirrors the expected precision of the light curve after an `almost perfect' de-correlation during post-processing (refer to Sect.~\ref{sec:Systematics}).

To utilise the ETC, inputs required include the target's \textit{Gaia} magnitude, its effective temperature, the duration of observation, and the exposure time of individual images. Optionally, one can include the stellar spectral type to incorporate stellar noise and provide the pointing direction to factor in the expected gap lengths for each orbit.

In the initial year of the mission, observations were conducted on a selection of `well-behaved' stars -- those exhibiting stability, near-flat light curves, and lacking known transiting planets. These stars spanned the recommended magnitude range of \CHEOPS. Processed with \DRP~\texttt{v13}\footnote{These visits were captured early in the mission before the prevalence of numerous hot pixels. Therefore, using reprocessed visits with \DRP~\texttt{v14} does not alter results.}, their light curves were analysed with \pycheops (\texttt{v1.0.0}). Fig.~\ref{fig:PhotomPerf} presents the cumulative outcomes of this analysis, illustrating light curve noise estimation based on stellar magnitude for three distinct time intervals (1\,h, 3\,h, 6\,h). Notably, bright stars ($G<9$\,mag) are almost photon noise limited. 

The ETC has proven to be a reliable predictor for expected performance. Nonetheless, due to the limitations outlined earlier, some patterns have been identified. In the case of bright stars, the ETC usually exhibits a slight `pessimistic' bias. This discrepancy likely stems from the noise attributed to cosmic rays within the ETC, which tends to be overestimated. In general, the noise in bright stars' light curves tends to approach the photon noise. Mid-range stars, i.e. stars falling within the middle of the nominal magnitude range ($G\sim 9$\,mag), tend to receive noise estimations from the ETC that closely align with actual measured values. However, for faint stars, the ETC slightly underestimates noise levels for short intervals (less than 3\,h). This may potentially be attributed to the contamination arising from background stars --- a factor not accounted for within the ETC and which cannot be perfectly corrected in post-processing. 

\subsection{Performance in time: Results from the M\&C programme}
\label{sec:MCperformance}

\begin{figure*}
    \centering
    \includegraphics[width=\textwidth]{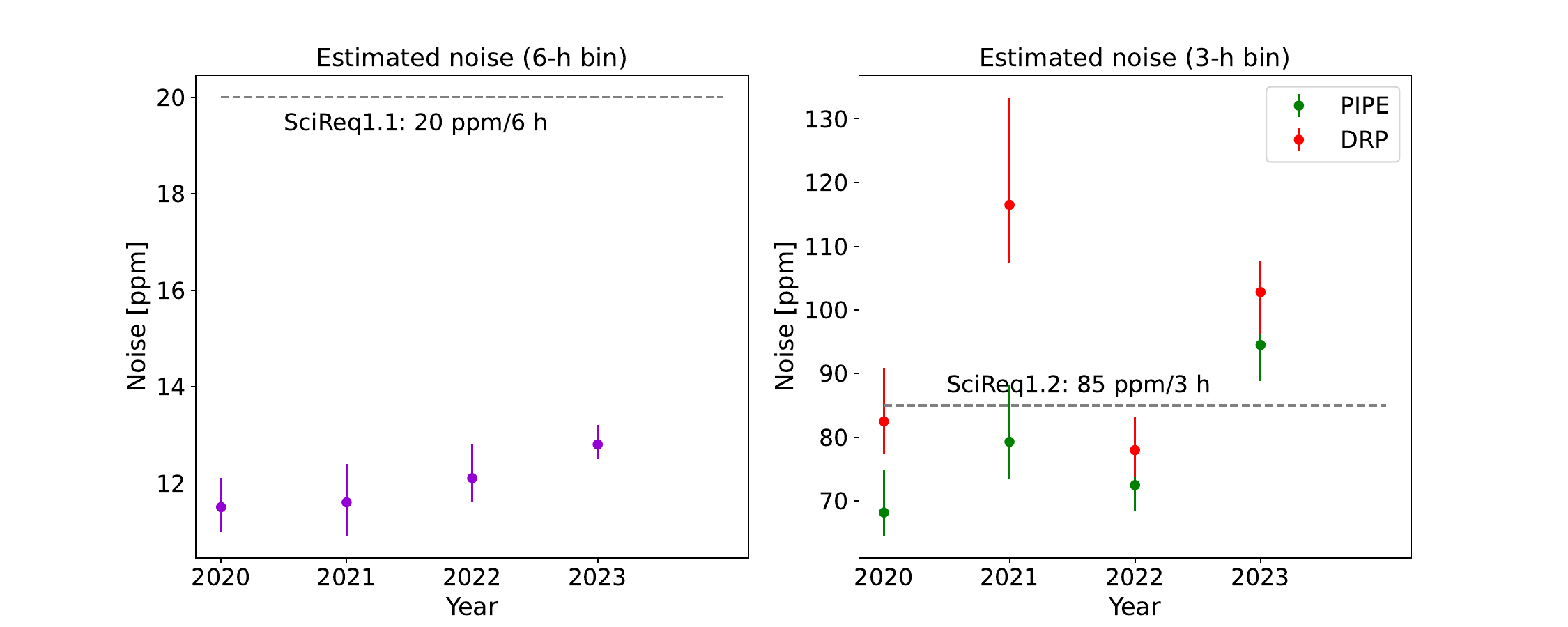}
    \caption{Observation of the stars HD 88111 ($V=9.2$\,mag) and TYC 5502-1037-1 ($V=11.9$\,mag) throughout the duration of the nominal mission, monitored specifically to assess their compliance with SciReq. 1.1 and SciReq. 1.2, respectively. Left: Noise estimation after 6 hours of integration time, derived from the light curves of HD 88111. It is evident that bright targets ($6\leq V\leq 9$ mag) comfortably comply with SciReq 1.1. Right: Noise estimation after 3\,h of integration time, calculated from the light curves of TYC 5502-1037-1 reduced with the \DRP (red) and \PIPE (green). Generally, faint targets with magnitudes around $V \lesssim 12$\,mag either marginally meet or fail to meet the criteria outlined in SciReq 1.2. The precision of light curves for faint targets is more sensitive to contamination effects, including stray light, as well as the duration of interruptions and the presence of bad pixels. These factors contribute to the variability observed in the precision of their light curves, in particular when the light curve is calculated by the \DRP, as \PIPE is less sensitive to these effects. The photometry was obtained with \DRP~\texttt{v14} using the OPTIMAL aperture, and the noise was estimated using the scaled error method (all scaling factors were well below 2; see \citealt{Maxtedpycheops}).}
    \label{fig:Noise1}
\end{figure*}

During IOC, we successfully validated the science requirements of the \CHEOPS mission, as documented in \citealt{BenzCHEOPS}. Among these requirements, two played a pivotal role in ensuring the mission's success:
\begin{itemize}
    \item Science requirement 1.1 (SciReq.~1.1): \CHEOPS shall be able to detect Earth-size planets transiting G5 dwarf stars (stellar radius of $0.9\,R_{\odot}$) with $V$-band magnitudes in the range $6\leq V\leq 9$  mag. Since the depth of such transits is 100\,ppm, this requires achieving a photometric precision of \SI{20} {ppm} (goal: \SI{10} {ppm}) in 6 hours of integration time (at least, signal-to-noise ratio of 5). This time corresponds to the transit duration of a planet with a revolution period of 50 days.
    \item Science requirement 1.2 (SciReq.~1.2): \CHEOPS shall be able to detect Neptune-size planets transiting K-type dwarf stars (stellar radius of $0.7\,R_{\odot}$) with $V$-band magnitudes as faint as $V=12$\,mag (goal: $V=13$\,mag) with a signal-to-noise ratio of 30. Such transits have depths of 2500\,ppm and last for nearly 3\,h, for planets with a revolution period of 13 days. Hence, a photometric precision of 85\,ppm is to be obtained in 3\,h of integration time. 
\end{itemize}

To validate these two requirements, we conducted observations of two well-known stable stars: HD\,88111 ($V=9.2$\,mag, $G=8.97$\,mag, $T_{\text{eff}}=5330$\,K, $R_{\text{star}} = 0.9 R_{\odot}$) for SciReq.~1.1, and TYC 5502-1037-1 ($V=11.9$, $G=11.98$, $T_{\text{eff}}=4750$\,K, $R_{\text{star}}=0.7 R_{\odot}$) for SciReq.~1.2. As detailed in \citealt{BenzCHEOPS}, our observations of both stars were found to be in compliance with the Science Requirements.

Following IOC, we established a yearly monitoring routine of these two stars to assess the instrument's performance over time. We calculated the precision of the light curves obtained from these observations, as shown in Fig.~\ref{fig:Noise1}. Over the years, probably due to the natural ageing of the instrument, the precision of the HD 88111 light curves has experienced a slight decline, going from 11.5\,ppm in 2020 to 12.8\,ppm in 2023 (in a 6\,h bin). Despite this, compliance with SciReq. 1.1 remains comfortably maintained, and we anticipate this trend to continue into the future.

Conversely, the TYC 5502-1037-1 light curves exhibit a more irregular behaviour. When estimating the noise using the \DRP light curves of 2020 and 2022, their precision for the 3-hour bin was better than the requirement of 85 ppm (82.5 ppm and 78 ppm, respectively). However, in 2021 and 2023, the precision declined to more than 100\,ppm (116.5\,ppm and 102.8\,ppm, respectively). On the other hand, if we use \PIPE extracted light curves, the estimated noise is much smaller (68.2\,ppm, 79.3\,ppm, 72.5\,ppm, and 94.5\,ppm). While this variation could be attributed to various factors, including stellar activity, we ruled out any instrumental systematics (except for bad pixels). It is important to note that faint targets ($11\leq G\leq 12$ mag) are generally more susceptible to variations in background contamination, cosmic ray hits and bad pixels, which can influence the precision of their light curves. \PIPE outperforms the \DRP in reducing the data for faint stars. While \PIPE and the \DRP agree within error bars for years 2022 and 2023, there is an evident discrepancy in years 2020 and 2021. By inspecting the images of the 2021 visit, we see that there is a bright RTS in the PSF and another one close to the limit of the photometric aperture. We also detected a very energetic cosmic ray hit at the beginning of the visit in the middle of the PSF, which could have created unstable bad pixels. It is not possible to identify the bad pixels in the PSF with the \DRP. We know, however, that \PIPE is much more insensitive to bad pixels and the clear improvement in this case when using PIPE suggests that there might be some bad pixels in the PSF responsible for the exaggerated noise. Lastly, looking at the 2020 results, \PIPE also clearly outperforms the \DRP. It is possible that some bad pixels are responsible for the noise in the \DRP photometry, although the noise value is compliant with the requirements and, therefore, did not deserve much attention when it was first computed. At the very beginning of the mission, the CCD was much `cleaner' than today, and PIPE shines, getting the best out of CHEOPS faint targets data.  The `drawback' of PIPE is that it is not yet automated and needs to be run manually on every dataset. However, with the detector ageing and CHEOPS  expected to be in use for some more years, we are working on automatising PIPE. Yet, CHEOPS  being a small mission with a limited budget, this is done on a best-effort basis.

\begin{figure*}
    \centering
    \includegraphics[scale=0.3]{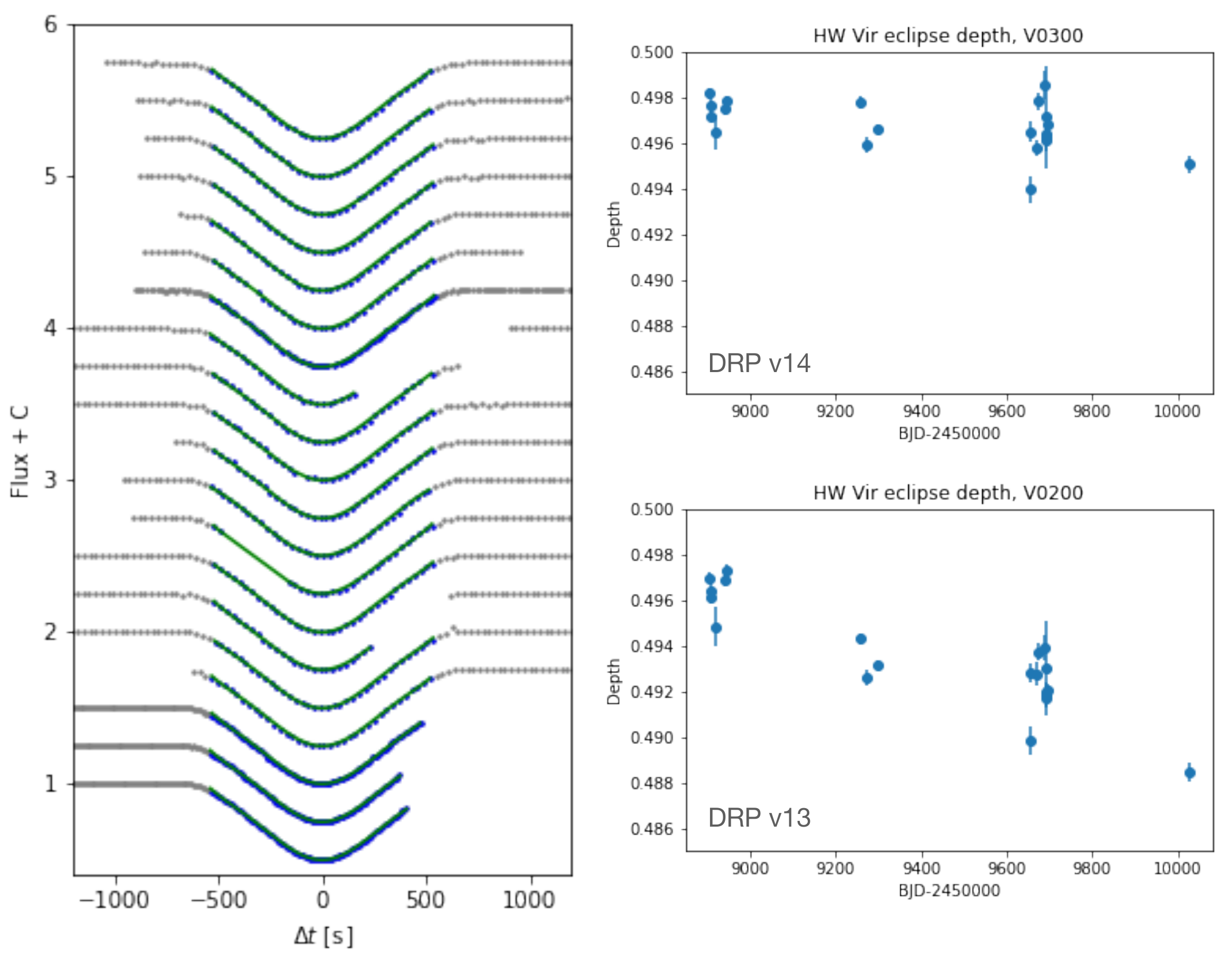}
    \caption{HW~Vir eclipses. Left: Light curves of all eclipses acquired to check the timing precision of \CHEOPS. These light curves can be used to measure the eclipse depth at different epochs. Bottom right: Measured eclipse depth through the years. Data processed with \DRP~\texttt{v13}. Note the decrease in the eclipse depth (2\%) as the instrument ages. The hot pixels increase the flux in the aperture, so the eclipse is slightly diluted. Top right: Same data processed with \DRP~\texttt{v14} where the dark current of the hot pixels is subtracted. The eclipse depth is stable through the years with a dispersion lower than 0.8\%. }
    \label{fig:eclipse_depth}
\end{figure*}

We also conducted annual observations of the primary eclipse of HW~Vir, which serves as a reference to assess the timing precision of \CHEOPS (as discussed in Sect.~\ref{sec:Timing}). Given its relatively faint magnitude ($G \simeq V = 10.6$\,mag), HW~Vir is a suitable candidate to investigate the influence of hot pixels on the photometric aperture. The presence of hot pixels can elevate the measured flux within the aperture, thereby potentially leading to a reduction in the eclipse depth over time if not properly corrected. The left panel of Fig.~\ref{fig:eclipse_depth} displays all the eclipses observed since 2020, arranged chronologically from bottom to top. It is important to note that the light curves are relatively short, spanning just one \CHEOPS orbit, which is not ideal for optimal detrending (so we made no attempt to remove instrumental noise correlated with SC roll angle from these light curves). This short observation duration was primarily intended for monitoring timing precision rather than eclipse depth. In 2023, only one observation was conducted as the focus shifted away from measuring timing precision. In the right panel of the figure, the bottom and top panels exhibit the calculated eclipse depths for each corresponding light curve. In the bottom panel, data reduction was carried out using \DRP~\texttt{v13}, where no hot pixel correction was applied. This resulted in a consistent decrease in eclipse depth over the years. To prove that hot pixels are indeed the source of the eclipse depth variation, the same data were reprocessed using \DRP~\texttt{v14} (i.e.\ hot pixel correction applied).  Fig.~\ref{fig:eclipse_depth}, top-right panel, shows that the eclipse depth remains constant.

\begin{figure*}
    \centering
    \includegraphics[scale=0.55]{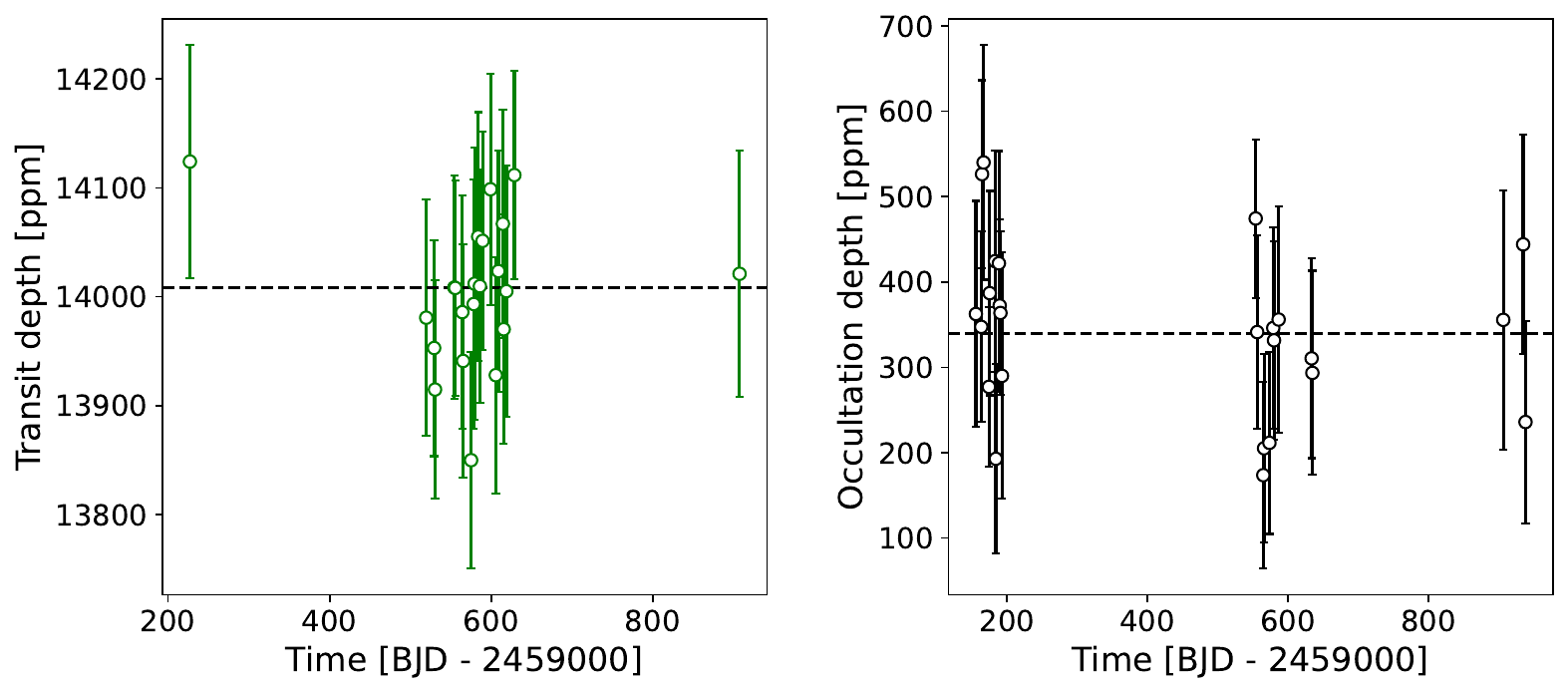}
    \caption{ Left (right): 22 transit depths (26 occultation depths) corresponding to 47 CHEOPS observations of WASP-12b spanning the three first years of the mission. The transit (and occultation) duration is approximately 3 hours. The dashed line indicates the median value of each set of measurements.}
    \label{fig:WASP12depths}
\end{figure*}

\begin{figure*}
    \centering
    \includegraphics[scale=0.5]{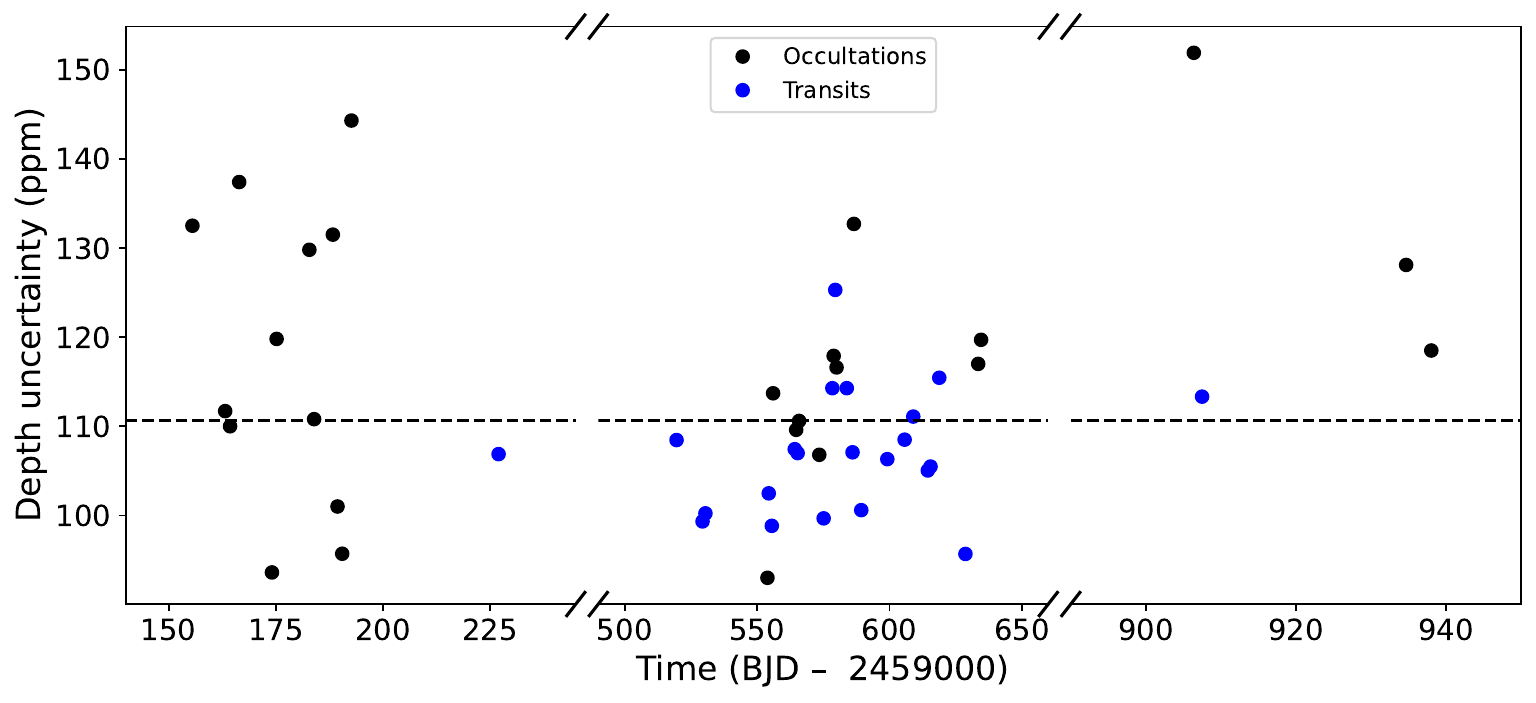}
    \caption{Uncertainties on the transit and occultation depth measurements presented in Fig.~\ref{fig:WASP12depths}. The median value of these measurements is indicated with a dashed line equal to $\sim 110$ ppm. }
    \label{fig:WASP12depths_un}
\end{figure*}

The \CHEOPS GTO programmes have also demonstrated the remarkable precision and stability of the instrument. To highlight a few instances when observing faint stars, consider the measurement of the tidal deformation of WASP-103b around a star with a magnitude of $G=12.2$ (as detailed in \citealt{BarrosCHEOPS}) or the determination of the transit parameters for all the planets within the TOI-178 system ($G=11.2$\, mag as elaborated in \citealt{DelrezCHEOPS2}).
A valuable dataset showcasing the performance of CHEOPS comprises observations of WASP-12b ($G \simeq V=11.6$ mag) obtained over the initial three years of the mission (47 visits in total, spanning from 02/11/2020 to 24/12/2022). Among these, 21 were dedicated to transit observations, 25 to occultations, and one featured a half-phase curve commencing before the occultation and concluding after the transit. This comprehensive campaign sought to identify and quantify tidal deformation and the Love number of this ultra-hot Jupiter, as detailed in the work by \citealt{AkinsanmiCHEOPS}.
Fig.~\ref{fig:WASP12depths} illustrates the computed transit and occultation depths in the left and right plots, respectively. When combining all the light curves, the transit depth is determined to be $14007 \pm 37$ ppm, while the occultation depth is $333 \pm 24$ ppm. In Figure~\ref{fig:WASP12depths_un}, the depth uncertainty for all transit and occultation events is presented, consistently below 155 ppm, with a median value of 110 ppm.

\subsection{Systematics in the light curves}
\label{sec:Systematics}

A few instrumental systematic effects should be considered when analysing \CHEOPS light curves. The ones mentioned in the following sections are the most common and can be easily removed using \pycheops.  
\subsubsection{Correlations with the roll angle}
\label{sec:RollCorr}

For thermal stability reasons, \CHEOPS is nadir locked to its orbit. This means that it rotates around the Earth, keeping constant the relative position to the Earth's surface as much as possible. This is reflected in the images: along one orbit, the background rotates 360\degree\ around the LoS (see Fig.~\ref{fig:Nadir_locked}).  The rotation of the background correlates with the roll angle (as shown, for example, in \citealt{BonfantiCHEOPS}). Other changes in the background also repeat periodically in the orbit, like changes in the reflected light on the Earth's surface that entered off-axis in the telescope, atmospheric airglow and internal reflections in the instrument.  

\paragraph{\bf Stars in the FoV}
Neighbouring stars rotate around the LoS.  Due to the large size and asymmetrical shape of the PSF (the PSF's large size is responsible for the flux contamination inside the aperture, while the PSF asymmetry is responsible for the roll dependence with the stars in the FoV), background stars will contaminate with their flux the photometric aperture. Therefore, the measured flux will depend on the position of the satellite in the orbit; in other words, the photometric extraction will show a correlation with the roll angle. The background contamination is simulated by the \DRP, which provides the background light curve that can be used to correct the target’s light curve. As for other missions that rely on aperture photometry (CoRoT, \textit{Kepler}, and TESS), the decontamination is addressed by calculating the contamination factor of the light curve. This is done by simulating the images corresponding to the visit to determine the fraction of the measured flux that should be attributed to contaminating stars, as described in \citealt{DRPHoyer}. The magnitude and relative position of the known stars expected to be present in the images are obtained using the \textit{Gaia} DR2 catalogue. Their relative contribution to the flux measured in the aperture is then computed. In contrast to the previously mentioned missions, which only estimate one contamination factor per observation sequence, the \DRP computes a time series of contamination factors for each photometric point due to the rotating stellar field and small jitter experienced by CHEOPS.
This contamination estimate can be used to decontaminate the light curve directly or as a detrending vector, which is currently the most common use. In the latter case, the user needs to decontaminate the light curve from the average contamination to correct for the flux dilution effect. \pycheops can be used to improve precision by de-correlating against roll angle. \{\pycheops also has an option to include arbitrary basis functions for detrending the light curves, so users can, in principle, compute their own parameters from the images and include these in the analysis of the light curves.

\begin{figure}
    \centering
    \includegraphics[scale=0.6]{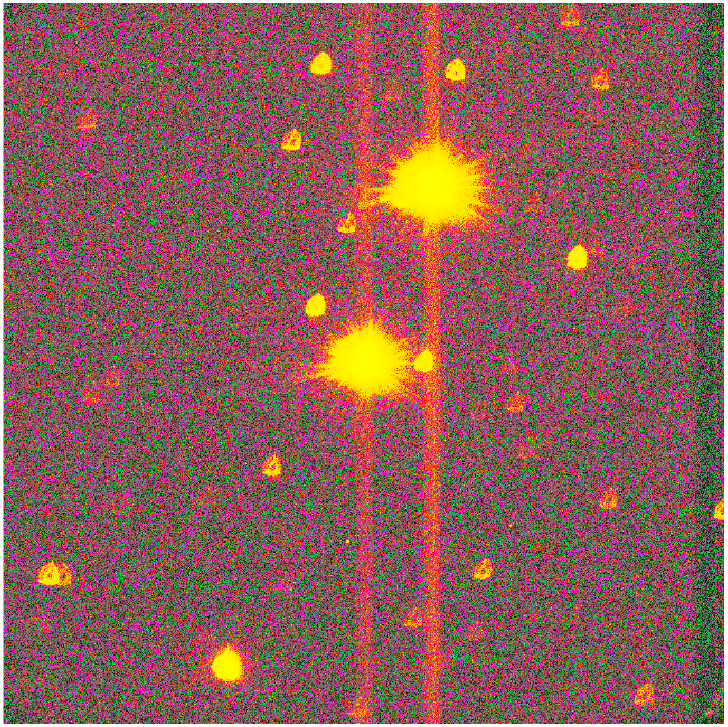}
    \caption{Example of a full-frame image (1024$\times$1024\,px) featuring two bright stars. The central object, 55~Cnc ($G = 5.73$\,mag), is the target of this visit, while the other bright star (positioned above the target) is 53 Cnc ($G = 5.23$\,mag). Both stars exhibit distinct vertical smearing trails (visit ID: \texttt{CH\_PR100041\_TG000601}).}
    \label{fig:smearing_trail55cnc}
\end{figure}

\paragraph{\bf Smearing trails from (bright) stars in the FoV}
CHEOPS has no shutter. After an image is acquired in the exposed section of the CCD, it is fast transferred to the covered section for read-out. During the fast transfer, the pixels of the CCD continue collecting photons. This causes all the stars in the FoV (target star plus neighbour stars) to leave smearing trails in the image, although only for very bright stars this is visually evident (see Fig.~\ref{fig:smearing_trail55cnc}). 

\begin{figure*}
    \centering
    \includegraphics[width=\textwidth]{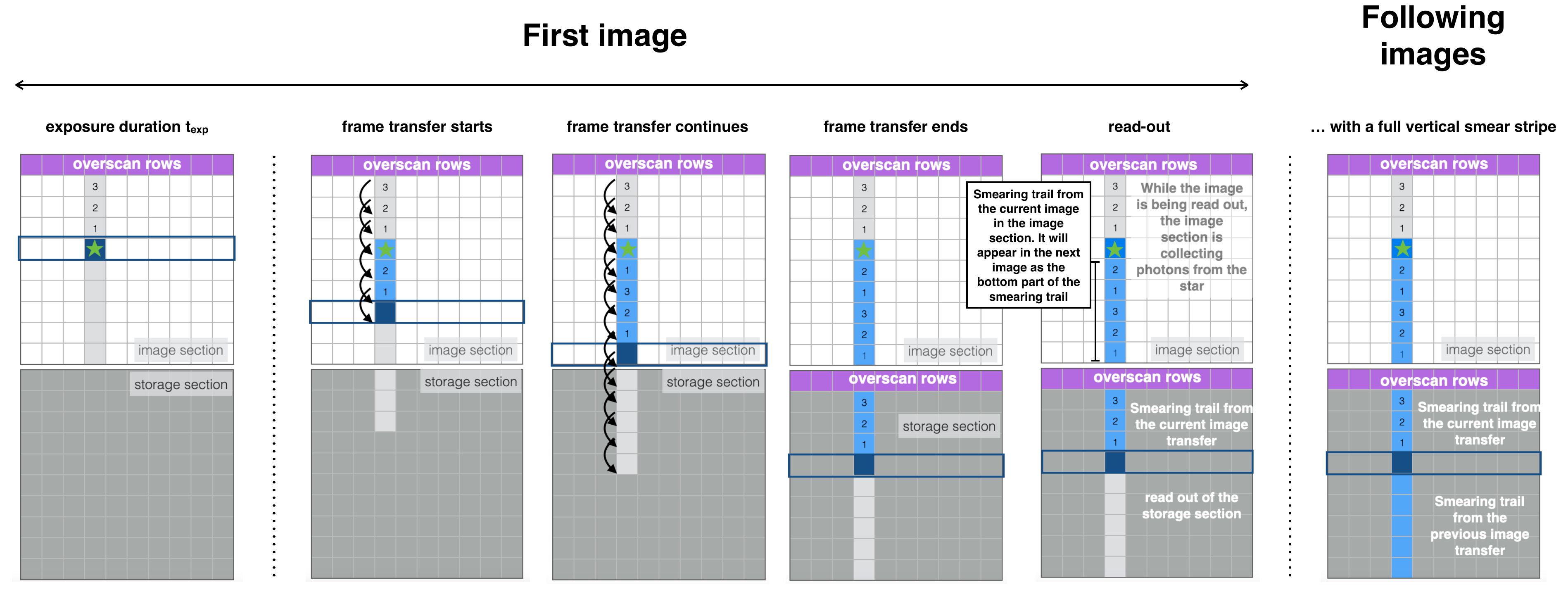}
    \caption{Illustration of the origin and characteristics of the smearing trails. It is assumed, for simplicity, that the CCD was never exposed to light before, that the charge transfer efficiency is perfect, and that only photons coming from the star hit the CCD. }
    \label{fig:smearing.trails}
\end{figure*}

Understanding how these vertical trails are produced is not straightforward, and therefore, it deserves a detailed explanation. 
Fig.~\ref{fig:smearing.trails} offers a schematic representation of the origin of these smearing trails. Consider a simplified observation scenario to enhance clarity: only one star in the FoV, no other light source, and the first image ever acquired by \CHEOPS. During the exposure time, $t_{\mathrm{exp}}$, the CCD collects photons from the star. After the exposure, the CCD's image section is transferred to a covered storage section. This transfer process takes approximately 25\,ms, a time during which pixels within the image section remain exposed to light. Consequently, the pixels in the column below the star undergo transfer without further photon collection. The pixels that collected the photons from the star will transfer the corresponding charge, accumulated during $t_{\mathrm{exp}}$, to the storage section. The pixels in the column above the star, en route to the storage section, encounter starlight for approximately 25\,$\mu$s each, resulting in charge transfer to the storage section. Following the completion of the CCD's full transfer process, the stored image exhibits a vertical charge trail above the star. Meanwhile, in the image section, and owing to the CCD's `rolling' mechanism, the pixels that now lay below the star have also passed through the star's pixels and gathered photons. However, these pixels persist within the image section for the next $t_{\mathrm{exp}}$. 
Subsequent images are formed similarly to the initial one, with the notable difference being the presence of a column of pixels below the star containing charge from the preceding 25\,$\mu$s exposure to the starlight. Consequently, starting from the second image onwards, each image contains a complete vertical illuminated column intersecting with the star.

In reality, the target star is not alone in the CCD. Neighbouring bright stars will leave stronger smearing trails that move horizontally inside and outside the photometric aperture as the background rotates around the target, correlating with the roll angle. The \DRP corrects these smearing trails, and residuals are further corrected by \pycheops using the de-correlation against the roll angle (e.g.\ \citealt{BonfantiCHEOPS}).

\paragraph{\bf Earth's stray light and atmospheric airglow}
The accuracy of photometric extraction can be compromised by the presence of undesirable light sources, such as Earth's stray light, atmospheric airglow, stray light originating from Solar System bodies like the Moon and planets, and internal reflections of off-axis sources. Eliminating this extraneous light is complex due to its often uneven distribution within the background.

\begin{figure*}
    \centering
    \includegraphics[width=\textwidth]{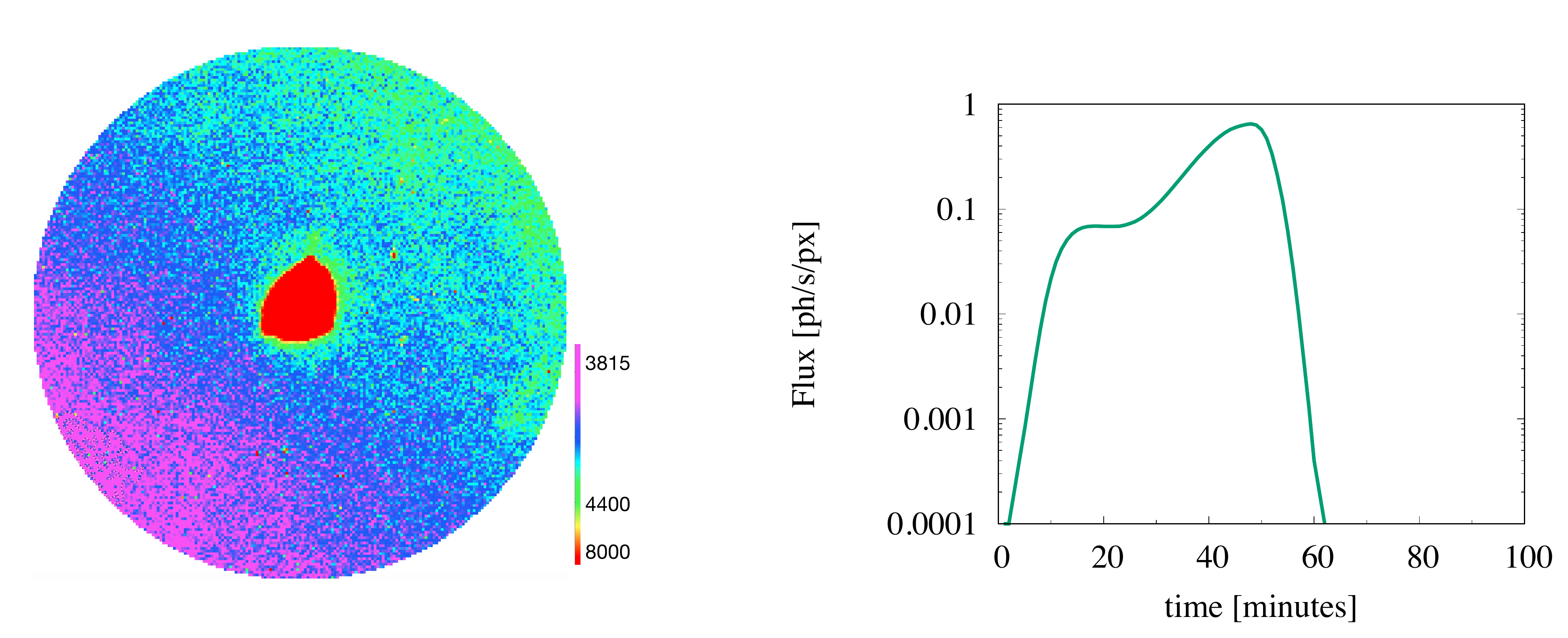}
    \caption{Example of an image with high Earth's stray light contamination (visit ID: \texttt{CH\_PR300059\_TG001201}). Note the inhomogeneous background going from 3815 to 4400 ADUs. In cases like this, the background correction done by the \DRP does not work properly. We recommend ignoring these images during the light curve analysis. The Earth's stray light contamination is a periodic function of the orbit, depending on the time and pointing. A simulation of stray light variation (normalised) during one orbit is given on the right. }
    \label{fig:SL}
\end{figure*}

During a visit, the LoS may approach the illuminated Earth limb as CHEOPS follows its orbit. As a result, the stray light flux is not constant, but varies with the orbital period. It fluctuates, ranging from approximately zero (e.g. when orbiting the night side of the Earth) to its maximum value when grazing the limb, as shown in Figure~\ref{fig:SL}. Consequently, the background of the images can become significantly brighter and exhibit non-uniform illumination. The intensity of the stray light is determined by several factors, including the angle relative to the Sun, the pointing direction, and the nature of the Earth's surface on which the sunlight is reflected (for a detailed explanation, see \citealt{Kuntzer2014}). 

A comparable phenomenon occurs when images are contaminated with atmospheric airglow. The sole distinction between airglow and stray light is that the occurrence of airglow in the atmosphere is unpredictable.

The \DRP performs homogeneous background correction. However, if there are gradients in the image, this correction will not entirely eliminate them. Our recommendation is to exclude images with high background contamination or their corresponding data points from the light curve analysis. 

\paragraph{\bf Internal reflections}
In some cases, for example, when bright objects (stars, planets, the Moon, etc.) within 24\textdegree\ of the LoS reflect in the mounting of the secondary mirror (spiders), images show bright regions that can increase the flux in the aperture. In those cases, the light curve can be corrected. In \citealt{BonfantiCHEOPS}, we see an example of a bright star in the FoV that is reflected in the spiders and produces a `bar' in the images that affect the measured flux\footnote{Note that the disturbing object might be outside the FoV; i.e.\ not even recorded in the image.}. The effect of these internal reflections in the light curve can be removed with \pycheops using the function \texttt{add\_glint}. 
\subsubsection{Correlations with the centroid position}
\label{sec:JitterCorr}

The PSF centroid can jitter around with an amplitude of a couple of pixels when the payload is not in the loop (see Appendix~\ref{pointingPerf}). If the star is fainter than $G=11$\,mag (therefore NOPITL), the centroid can also drift in time. A linear de-correlation against x and y will compensate for this. Such a de-correlation is available in \pycheops.
\subsection{Correlation with the telescope tube temperature}
\label{sec:RampCorr}
As discussed in Sects. \ref{sec:TempStability} and \ref{sec:Ramp}, some light curves present a ramp in the flux at the beginning of the visit (e.g.\ \citealp{MorrisCHEOPS}). The ramp can go upwards or downwards, and it can be more or less pronounced depending on the case. When analysing a light curve, it might not be totally clear if there is a ramp or not, nor if a ramp in the flux is actually linked to the thermal state of the telescope. To check if the root cause of a flux increase/decrease at the beginning of a visit is due to an effect of the telescope tube temperature on the PSF, one should look at the HP parameter \textit{thermFront\_2}. This information can be found in the file containing the window images (\texttt{CH\_PRxxxxxx\_TGxxxxxx\_TU202xx-xx-xxTxx-xx-xx\_SCI\_xxx\_SubArray\_Vxxxx.fits}), in the table \texttt{SCI\_xxx\_ImageMetadata}.
Once it is clear that the ramp is a consequence of the variation in the temperature of the telescope tube, it can be corrected. Two independent ways for correcting the ramp have been proposed and seem to work very well, namely the PCA on the autocorrelation function (ACF) of the PSF (Sect.~\ref{sec:PCA_ACF_PSF}); and the linear de-correlation (Sect.~\ref{sec:Linear decorrelation}).

\subsubsection{Principal component analysis on the autocorrelation function of the PSF}
\label{sec:PCA_ACF_PSF}

When there is a ramp in the light curve linked to thermal instabilities in the telescope tube, it can be measured that the PSF size changes, either increasing or decreasing, depending on the temperature settling. To quantify the relative size, a comparison is made between the PSF size at the beginning of the observation and when it stabilises. A small change in the scale of the PSF can be understood as a slight focus change due to a thermal adaptation of the telescope tube to the new heat load. As a result, the ramp is correlated with the PSF shape.

Inspired by the SCALPELS algorithm, which was developed to differentiate shape-driven radial velocity offsets from genuine stellar Doppler shifts \citep{Cameron2021}, one approach to studying the ramp is to use the ACF of the PSF (i.e. PSF--SCALPELS;  \citealp{WilsonCHEOPS}). The ACF of an image allows for spatial characterisation and analysis while remaining invariant to image translation (jitter). The ACF of the PSF is sensitive to changes in its shape but not to shifts.

\begin{figure}
    \centering
    \includegraphics[width=\columnwidth]{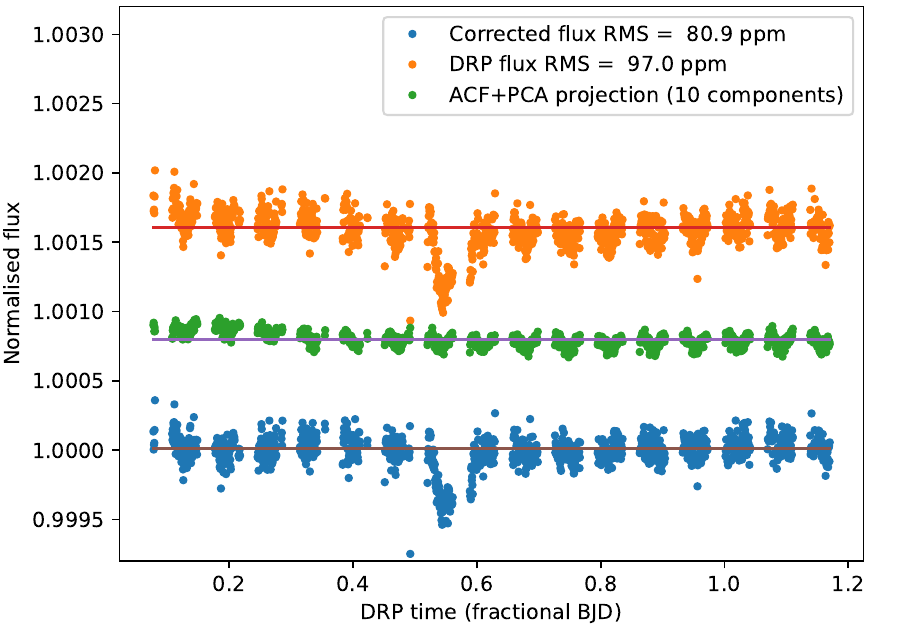}
    \caption{ACF+PCA method used to correct the ramp present in the light curve of 55 Cnc e (visit ID: \texttt{CH\_PR100041\_TG000601}). In orange, the \DRP output light curve (DEFAULT aperture); in green, ACF and PCA projection using the ten largest components; and in blue, the final light curve after correction. Note that the ramp down in the orange light curve is no longer present in the blue one.}
    \label{fig:Ramp_ACF+PCA_55Cnce}
\end{figure}

To analyse the PSF variations, the ACF is calculated for all images in a visit, and then the singular-value decomposition of the ACF provides an orthonormal basis, U, of shape-driven variability patterns (PCA). For each ACF of the PSF, the PCA is applied to obtain the orthonormal basis U. By projecting the \DRP light curve into the U basis, a model of flux variations driven by PSF shape changes is derived. The result of the ACF + PCA is used to correct the light curve, as shown in Fig.~\ref{fig:Ramp_ACF+PCA_55Cnce}. The efficacy of ramp elimination is most pronounced for reduced-rank models, including the 10--20 principal components that exhibit the strongest correlation with the out-of-transit light curve. Further details on the application of the ACF+PCA method are available in \citealp{WilsonCHEOPS}. We note that using an ACF+PCA does not correct for astrophysical signatures. When the method is applied to an astrophysical variable, the ramp is corrected, but the light curve's astrophysical modulation remains. 

\subsubsection{Linear de-correlation}
\label{sec:Linear decorrelation}

\begin{figure}
    \centering
    \includegraphics[width=\columnwidth]{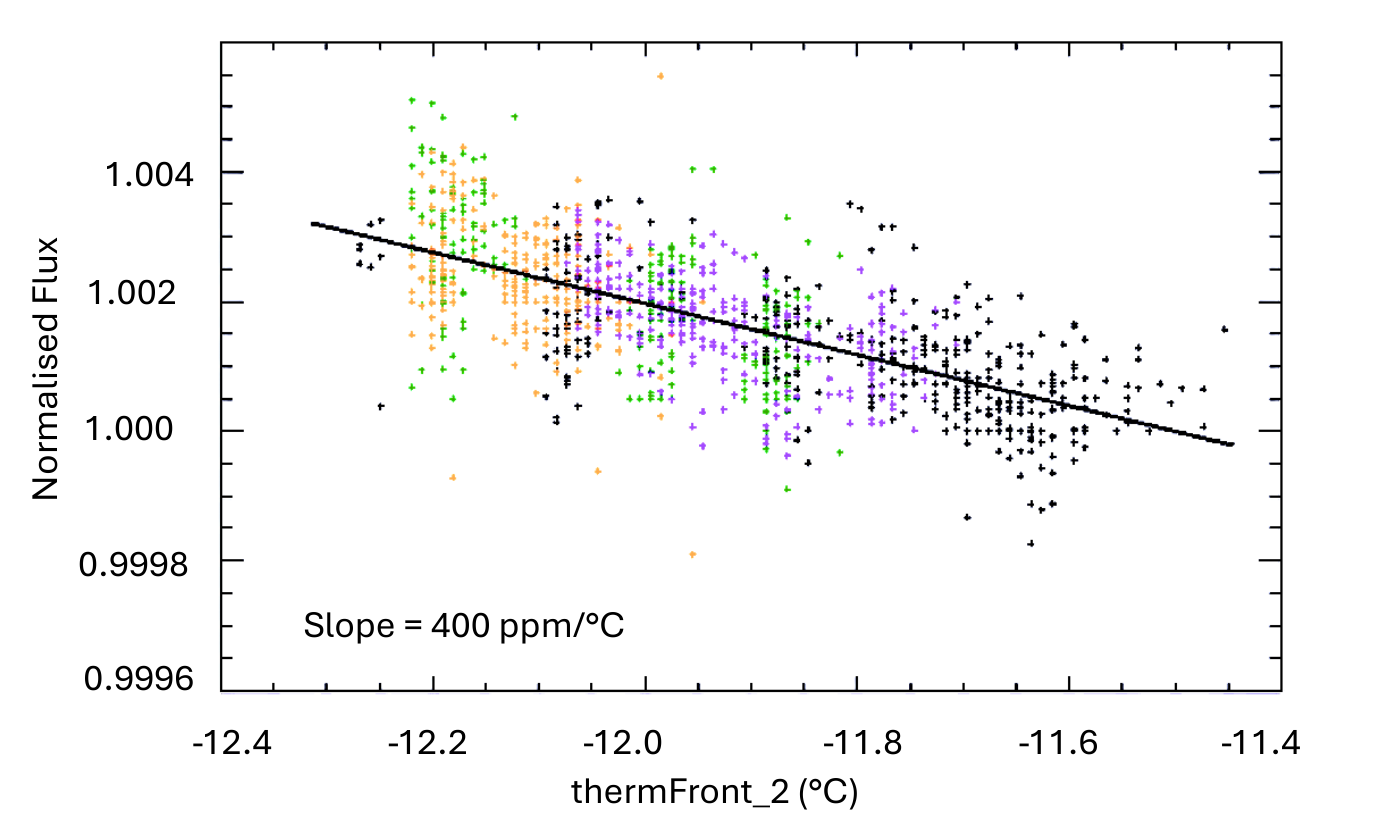}
    \caption{Linear dependence between the telescope tube temperature and the measured flux. Each colour corresponds to a visit with a ramp. These four visits span over the temperature range measured by \textit{thermFront\_2} (x-axis). The y-axis is the normalised flux measured in a photometric aperture of R = 40 px. The flux measured in the aperture decreases linearly as the temperature increases.}
    \label{fig:ramp}
\end{figure}

As mentioned earlier, the ramp is directly linked to temperature variations in the telescope tube after a re-pointing. The most sensitive sensor capturing these temperature changes is represented by the HK parameter \textit{thermFront\_2}. Through the analysis of several visits, it has been established that the measured flux exhibits a linear (anti)correlation with \textit{thermFront\_2} (see Fig. \ref{fig:ramp}). Consequently, it becomes feasible to de-correlate the light curves using the following equation:

$F_{\mathrm{corrected}} = F_{\mathrm{measured}} \times (1+\beta \Delta T)$.

Here, $F_{\mathrm{measured}}$ represents the normalised flux measured inside the photometric aperture, $\beta$ is a constant dependent on the radius of the photometric aperture, and $\Delta T = T (thermFront\_2) + 12\degree C$. To determine the values of $\beta$, interpolation can be employed using the calculated values for various selected apertures, as illustrated in Table \ref{table:beta}. \pycheops provides the option to utilise this de-correlation formula to correct the ramp in the data.

\begin{table}
\caption[]{Linear trend slope ($\beta$) as a function of the aperture radius for correcting the ramp effect.}
\label{table:beta}
\begin{center}
\begin{tabular}{ll}
\hline
\noalign{\smallskip}
\multicolumn{1}{l}{Aperture [px]} &
\multicolumn{1}{l}{$\beta$} [ppm/\degree C]\\
\noalign{\smallskip}
\hline
\noalign{\smallskip}
22.5  & 140    \\
25  &  200   \\
30  &  330   \\
40  &  400   \\
\noalign{\smallskip}
\hline
\end{tabular}  
\end{center}
\end{table}

\subsection{(Potential) correlation with bad pixels}
\label{sec:BPCorr}
\begin{figure*}
    \centering
    \includegraphics[width=\textwidth]{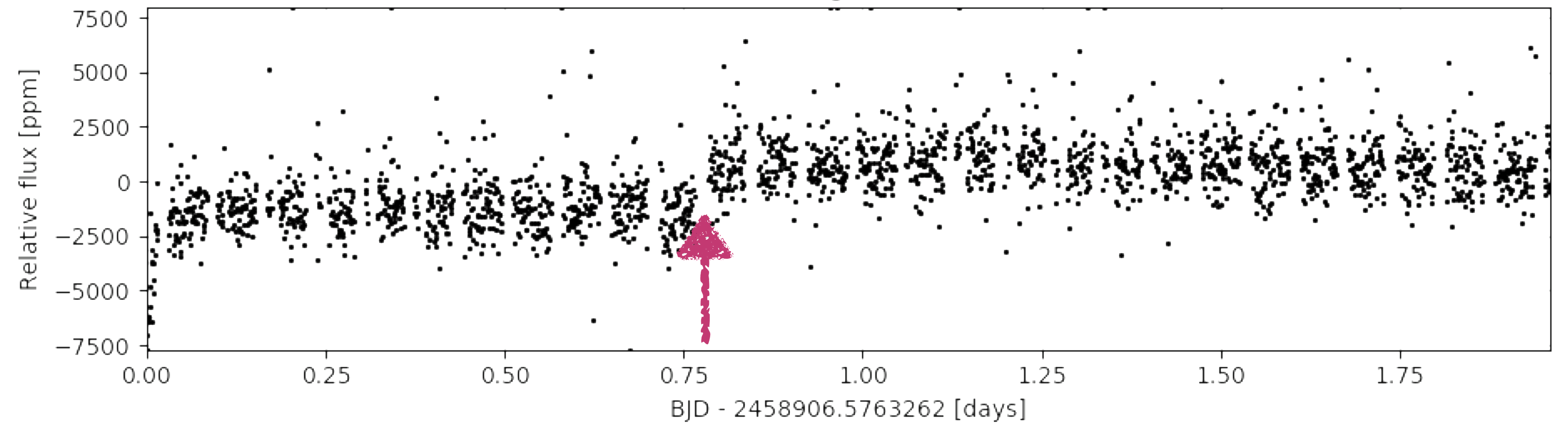}
    \caption{Example of a light curve where a (very) hot pixel appeared inside the aperture in the middle of the visit. We recommend detecting where the hot pixel is located to determine the best approach to correct it (visit ID: \texttt{CH\_PR300005\_TG000201}). }
    \label{fig:HP}
\end{figure*}

Hot pixels can leave noticeable artefacts in the light curve (see, for example, \citealt{LeleuCHEOPS} or \citealt{LacadelliCHEOPS} for the effect of RTS pixels in \CHEOPS light curves). If a hot pixel appears within the photometric aperture during an observation and is sufficiently bright, it can cause a sudden jump in the light curve, as depicted in Fig.~\ref{fig:HP}. Such cases are relatively easy to correct since they generate only an offset in a segment of the light curve (if they are stable hot pixels). To address this, one can identify the hot pixel in the images, estimate its dark current, and subtract it from the light curve if the hot pixel remains stable throughout the observation. However, a hot pixel appearance can be followed by an exponential decay towards a final dark level (whether the end state is `hot' or not), or a hot pixel can become an RTS afterwards.

However, a more challenging scenario arises when a hot pixel appears near the edge of the photometric aperture. In this situation, the hot pixel may intermittently move in and out of the aperture due to pointing jitter. Consequently, the light curve will not display a single jump, as in the previous case, but rather the measured flux will correlate with the pointing jitter. A de-correlation against the ($x$, $y$) coordinates using \pycheops can effectively address the issue. Comparing light curves obtained with different aperture sizes can help diagnose this effect.  

\section{Discussion}
\label{sec:Discussion}
\subsection{Observing outside \CHEOPS' nominal magnitude range}
\label{sec:MagRange}

The design of \CHEOPS was optimised for observing stars within the magnitude range of 6--12\,mag. However, this does not imply that stars brighter or fainter cannot be observed. The key consideration is whether \CHEOPS' under-performing precision justifies the time allocation.

\paragraph{\bf Faint stars}
Observing stars fainter than $G=12$\,mag with \CHEOPS comes with several limitations that need to be considered carefully. These include background contamination from various sources, such as stars in the FoV and Earth's stray light, hot and RTS pixels, and cosmic rays that, while detected and corrected by the \DRP, introduce noise proportionally greater for fainter than for brighter stars.
Despite these limitations, \CHEOPS has performed as predicted by the ETC, even for very faint stars. For `ideal targets', such as isolated and photometrically quiet stars, the noise estimated by the ETC aligns well with the measured noise for stars as faint as $G = 14$\,mag. However, it should be emphasised that ground-based telescopes with a 1-metre aperture can perform photometrically better than \CHEOPS for stars fainter than $G \sim 13$. As a result, unless there are specific reasons (e.g. the need for long, uninterrupted observations), observing such faint targets with \CHEOPS is discouraged, as other facilities may be better suited for studying these stars.

\paragraph{\bf Bright stars}
\CHEOPS demonstrates excellent performance when observing bright stars, and they are much less susceptible to the challenges mentioned earlier. The higher signal-to-noise ratio is primarily due to the increased signal level. However, \CHEOPS' performance becomes sub-optimal when the stars are exceptionally bright (around three to four magnitudes brighter than the nominal bright limit). Observing very bright stars requires exposure times, $t_{\mathrm{exp}}$, significantly below 1 second to prevent saturation. However, the time between two consecutive acquisitions cannot be shorter than 1.1 seconds due to the read-out time. Consequently, for example, for a $G = 2.5$\,mag star, $t_{\mathrm{exp}}$  would be around 0.1 seconds, but the elapsed time between one image and the next will be 1.2 seconds. The observer will therefore receive one stacked image per minute, corresponding to the onboard stacking of 39 individual exposures. Additionally, due to the limited bandwidth, imagettes will be stacked in sets of 3. While there is no inherent issue with this approach, the maximum duty cycle for this observation will be only 9\%. As a result, such short exposure times, needed to avoid saturation, translate into an inefficient utilisation of \CHEOPS' observation time.

Another consideration when observing bright stars is the presence of strong self-smearing trails. Although the \DRP can handle smearing trails within the nominal magnitude range, for very bright stars, these trails become too strong and are not accurately interpreted by the software. Consequently, if the decision is made to observe ultra-bright stars, the consortium will not provide the light curve extraction.

\subsection{Outliers from orbital light pollution}
\label{sec:sat_trails}
\begin{figure*}
    \centering
    \includegraphics[width=\textwidth]{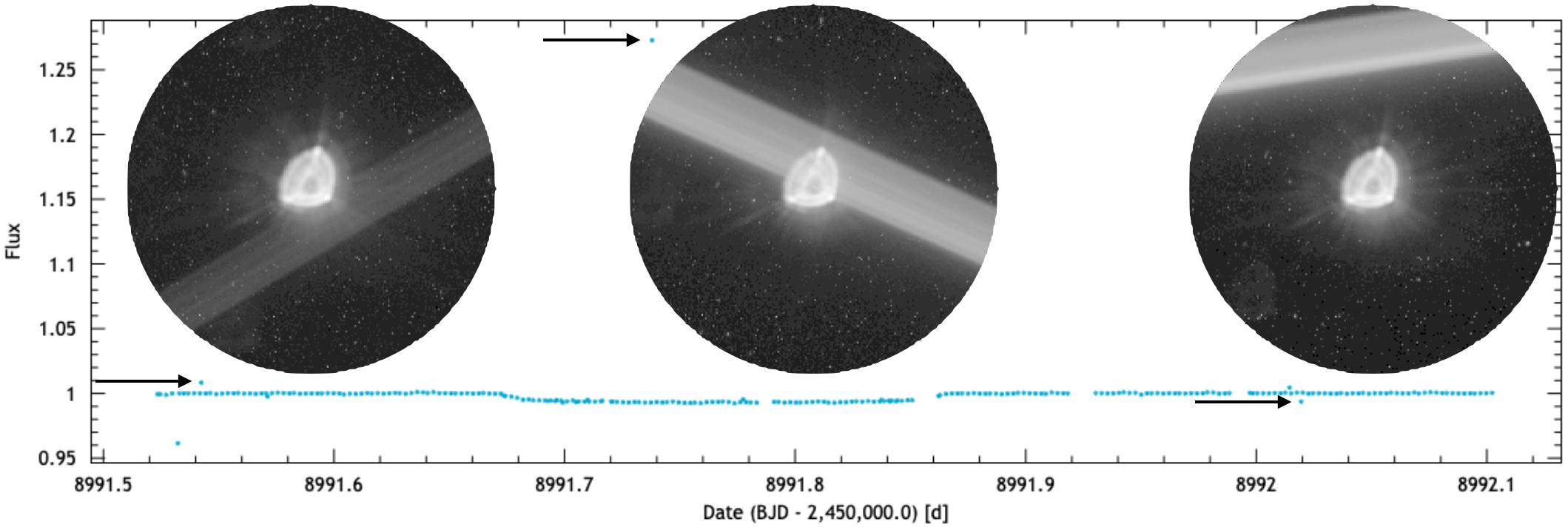}
    \caption{\CHEOPS light curve of WASP-38 (visit ID: \texttt{CH\_PR100015\_TG009101}) showing positive and negative outliers caused by satellites or space debris objects crossing the instrument FoV during an exposure.}
    \label{fig:satTrails}
\end{figure*}

CHEOPS sits on a Sun-synchronous orbit at 700 km altitude. This location is particularly susceptible to contamination from objects on LEO orbits. Indeed, from the \CHEOPS viewpoint, satellites or space debris crossing the instrument FoV are generally illuminated by the Sun, making them relatively bright. These crossing events are short-lived (relative to the typical exposure time of 30 seconds), resulting in a linear trail, or streak, across the entire CCD. Images contaminated by this sort of orbital light pollution translate into outliers in the associated light curves (see Fig.~\ref{fig:satTrails}). When a trail crosses the photometric aperture, the flux is overestimated and leads to a positive outlier. Conversely, the outlier is negative if the trail falls outside the aperture (contaminating the background only).

From the 1.3 million \CHEOPS images acquired as part of the science observing programme, we have identified several thousand such linear trails. This currently represents about 0.3\% of the total number of images collected so far. We have measured a noticeable increase in the number of detected trails over the past couple of years due to the deployment of mega-constellations of communication satellites. A dedicated publication is in preparation to characterise better the population of objects crossing the FoV, thus exploiting some of CHEOPS' unwanted signals and converting those into useful information for the Space Situational Awareness community.

\subsection{Lessons learned}
\label{sec:lessons_lernt}

Throughout nominal operations, we were able to reflect on the successes and mistakes of the mission. We have learned very valuable lessons, some of which are worth sharing with the community as they may be applicable to ongoing or future projects.

    \paragraph{\bf Data reduction and analysis} The \CHEOPS official SW to reduce the raw data, the \DRP, started its design and development very early in the project. A key to its success was \CHEOPSim (\citealt{Futyan2020}). \CHEOPSim provided very realistic simulated data and it was fundamental for the timely development of the different tools for data reduction and analysis. The robust development of the \DRP before launch also allowed the early development of \pycheops. Having the \DRP and \pycheops already working during IOC was fundamental for the data analysis of the IOC activities, and, in parallel, it allowed their deep testing and fine-tuning. Thanks to this, CHEOPS was ready to provide the reduction and analysis tools to the community immediately after IOC. During routine operations, observers highlighted that the {\DRP report} produced for every visit was very informative. Before starting the data analysis, scientists can have a very good idea about the features and quality of an observation. Another key aspect that turned out to be very valuable during the whole nominal mission was the consistency of the data structures: they have been stable throughout the mission, allowing users to develop and/or improve their own pipelines without hiccups or numerous updates. Although the \DRP is flexible and has continued its development as the mission progresses (e.g. the \DRP now outputs light curves using a wide range of apertures as mentioned in Sect. \ref{sec:DPandA}), in general, the stability of the SW was considered more valuable than performing many updates to implement potential new features. In that respect, it has been beneficial to count on a robust \DRP plus other pipelines (\DRT, \PIPE) to perform independent validation, test new algorithms and try new ideas before the release of a new version of the \DRP. This consistency allowed \PIPE, which was first developed to exploit the imagettes, to evolve into an excellent DRT for faint stars. Another important decision when developing the data structures was to include many instrument HK parameters in the products provided to the user. Experience shows that when scientists receive scientific data, some detective work is always needed, so it is good to have a lot of detail available. In this sense, many HK parameters proved to be suitable for detrending light curves, although it is fair to say that it can be challenging for a newcomer to know what to do with all that information. One weak point to mention is the organisation of the data: information is distributed in many files, and quite some guidance is needed to understand this structure. For example, the \DRP output of different levels of image correction may not be required by the regular user and be confusing, but it is essential to produce the {\DRP report}.
    \paragraph{\bf Bad pixels} One of CHEOPS' unexpected features is the high number of hot pixels being generated in the CCD. It is fair, therefore, to wonder if something could have been done to avoid this. Thicker shielding would have provided better protection for the CCD. CHEOPS has a non-isotropic shielding, with 80\% of it ranging from 8 to 80 mm of equivalent aluminium shielding, the thinnest part of which is 4 mm. The total ionising dose (which is essentially the charged particles we observe as cosmic rays in the CCD) that penetrate the thinnest shielding accounts for 6.5\% of the total dose. In the SAA region, the satellite is exposed to higher levels of ionising radiation. Thicker shielding would have been beneficial in limiting the generation of hot pixels, but the question is how much thicker it should have been to have a fundamental impact on the hot pixels generation rate. During the design phase, it was estimated that increasing the minimum shielding to 10 mm would only have reduced the total ionising dose by 7\%. Much thicker shielding would have increased the satellite's weight and, thus, the launch cost. With today's information, we would look more closely at the shielding and aim for a solution that increases the shielding around the CCD. In addition, using a NIMO CCD instead of an AIMO CCD would have been more robust to the effects of cosmic rays but at the cost of a much higher initial dark current.
    \paragraph{\bf On-ground support} Having on-ground hardware and software to reproduce some things we see in flight can be vital. As an example, as the CCD ages, having CCD analogues to the one in orbit on the ground allows further testing in the laboratory. This became very important when we needed to re-measure the absolute QE and currently for the characterisation of the CTI effect.
    \paragraph{\bf A team of dedicated people} Unforeseen things happen. To overcome problems as fast and well as possible, it is crucial to have a team of committed people who got involved in the project before launch and are familiar with different aspects of the mission. In the case of CHEOPS, a clear example is the gap in the telescope tube heater that leads to the ramp in the light curves. It is clear that the requirement that specified the temperature stability of the telescope tube was not stringent enough. However, a quick reaction and understanding of the problem were possible because many key team members knew the instrument very well.

\section{Conclusions}
\label{sec:Conclusion}
CHEOPS has demonstrated exceptional performance throughout its nominal mission, consistently delivering high-quality data. Despite its ageing, the mission's capabilities remained steadfast during its first 3.5 years, providing valuable scientific insights.

To maintain the mission's excellence, it is necessary to continuously assess the satellite's performance. This means conducting regular evaluations and taking steps to minimise any potential impact on data quality and mission objectives. Ageing effects must be taken into account in data analysis and calibration processes. Actively monitoring, understanding, and addressing their impact on the payload through the M\&C programme is essential to ensuring optimal performance and reliable measurements. 
The search for improved correction methods and in-depth analysis of ageing effects, including CTI, will enhance CHEOPS' scientific potential. The CTI poses a potential challenge for the mission extension; while its impact on photometry is currently insignificant, it may become more apparent in the future. The consortium is actively working on characterising and studying techniques for correction.  

The bad pixel M\&C activity plays a crucial role in assessing the condition of the CCD and informing the data reduction process. It maintains the consistent quality of the photometric data over time, even in the absence of traditional dark images acquired through a shutter mechanism.
Regular observations, calibration, filtering, and comprehensive analysis all help us understand the impact of bad pixels on the measurements and enable effective mitigation strategies. This is critical to data reliability, especially considering the large number of hot and RTS pixels.   

In terms of data reduction and data analysis, the constant development of the \DRP demonstrates the commitment of the \CHEOPS consortium to enhancing its efficiency and effectiveness. This ensures that the \DRP continues to be a powerful tool for transforming raw data into valuable scientific datasets, aligning with the overarching goal of exoplanet exploration.
In addition, \PIPE allows for unconventional scenarios, thus broadening the range of science possible with \CHEOPS. With respect to data analysis, \pycheops has become essential to the \CHEOPS community. Its ongoing evolution significantly contributes to the scientific objectives of the mission, providing researchers with a versatile toolkit for extracting physical information from the light curves produced by the \DRP and \PIPE.

By maintaining its compliance with scientific requirements and overcoming the challenges of natural ageing, \CHEOPS has proven its worth as a reliable and invaluable asset to exoplanetary science. The anticipation of more high-quality data during the extended mission further fuels the enthusiasm of the scientific community and promises new discoveries and advances in the field.

\begin{acknowledgements}
CHEOPS is an ESA mission in partnership with Switzerland with important contributions to the payload and the ground segment from Austria, Belgium, France, Germany, Hungary, Italy, Portugal, Spain, Sweden, and the United Kingdom. The \CHEOPS Consortium would like to gratefully acknowledge the support received by all the agencies, offices, universities, and industries involved. Their flexibility and willingness to explore new approaches were essential to the success of this mission. \CHEOPS data analysed in this article will be made available in the \CHEOPS mission archive (\url{https://cheops.unige.ch/archive_browser/}). 

This work has made use of data from the European Space Agency (ESA) mission {\it Gaia} (\url{https://www.cosmos.esa.int/gaia}), processed by
the {\it Gaia} Data Processing and Analysis Consortium (DPAC,\url{https://www.cosmos.esa.int/web/gaia/dpac/consortium}). Funding for the DPAC has been provided by national institutions, in particular the institutions participating in the {\it Gaia} Multilateral Agreement.

AFo, ASi and CBr acknowledge support from the Swiss Space Office through the ESA PRODEX program. 
TWi and ACCa acknowledge support from STFC consolidated grant numbers ST/R000824/1 and ST/V000861/1, and UKSA grant number ST/R003203/1. 
PM acknowledges support from STFC research grant number ST/R000638/1. 
ABr was supported by the SNSA. 
GBr, LBo, VNa, IPa, GPi, RRa, and GSc acknowledge support from \CHEOPS ASI-INAF agreement n. 2019-29-HH.0. 
The Belgian participation to \CHEOPS has been supported by the Belgian Federal Science Policy Office (BELSPO) in the framework of the PRODEX Program, and by the University of Liège through an ARC grant for Concerted Research Actions financed by the Wallonia-Brussels Federation. 
L.D. is an F.R.S.-FNRS Postdoctoral Researcher. 
B.-O. D. acknowledges support from the Swiss State Secretariat for Education, Research and Innovation (SERI) under contract number MB22.00046. 
MNG is the ESA \CHEOPS Project Scientist and Mission Representative, and as such also responsible for the Guest Observers (GO) Programme. MNG does not relay proprietary information between the GO and Guaranteed Time Observation (GTO) Programmes, and does not decide on the definition and target selection of the GTO Programme. 
SH gratefully acknowledges CNES funding through grant 837319. 
KGI is the ESA \CHEOPS Project Scientist and is responsible for the ESA \CHEOPS Guest Observers Programme. She does not participate in, or contribute to, the definition of the Guaranteed Time Programme of the \CHEOPS mission through which observations described in this paper have been taken, nor to any aspect of target selection for the programme. 
S.G.S. acknowledge support from FCT through FCT contract nr. CEECIND/00826/2018 and POPH/FSE (EC). 
YAl acknowledges support from the Swiss National Science Foundation (SNSF) under grant 200020\_192038. 
RAl, DBa, EPa, and IRi acknowledge financial support from the Agencia Estatal de Investigación of the Ministerio de Ciencia e Innovación MCIN/AEI/10.13039/501100011033 and the ERDF “A way of making Europe” through projects PID2019-107061GB-C61, PID2019-107061GB-C66, PID2021-125627OB-C31, and PID2021-125627OB-C32, from the Centre of Excellence ``Severo Ochoa'' award to the Instituto de Astrofísica de Canarias (CEX2019-000920-S), from the Centre of Excellence ``María de Maeztu'' award to the Institut de Ciències de l’Espai (CEX2020-001058-M), and from the Generalitat de Catalunya/CERCA programme. 
S.C.C.B. acknowledges support from FCT through FCT contracts nr. IF/01312/2014/CP1215/CT0004. 
XB, SC, DG, MF and JL acknowledge their role as ESA-appointed \CHEOPS science team members. 
P.E.C. is funded by the Austrian Science Fund (FWF) Erwin Schroedinger Fellowship, program J4595-N. 
This project was supported by the CNES. 
This work was supported by FCT - Fundação para a Ciência e a Tecnologia through national funds and by FEDER through COMPETE2020 - Programa Operacional Competitividade e Internacionalizacão by these grants: UID/FIS/04434/2019, UIDB/04434/2020, UIDP/04434/2020, PTDC/FIS-AST/32113/2017 \& POCI-01-0145-FEDER- 032113, PTDC/FIS-AST/28953/2017 \& POCI-01-0145-FEDER-028953, PTDC/FIS-AST/28987/2017 \& POCI-01-0145-FEDER-028987, O.D.S.D. is supported in the form of work contract (DL 57/2016/CP1364/CT0004) funded by national funds through FCT. 
This project has received funding from the European Research Council (ERC) under the European Union’s Horizon 2020 research and innovation programme (project {\sc Four Aces} grant agreement No 724427). It has also been carried out in the frame of the National Centre for Competence in Research PlanetS supported by the Swiss National Science Foundation (SNSF). DE acknowledges financial support from the Swiss National Science Foundation for project 200021\_200726. 
MF and CMP gratefully acknowledge the support of the Swedish National Space Agency (DNR 65/19, 174/18). 
DG gratefully acknowledges financial support from the CRT Foundation under Grant No. 2018.2323 ``Gaseousor rocky? Unveiling the nature of small worlds''. 
M.G. is an F.R.S.-FNRS Senior Research Associate. 
Some of the observations presented in this paper were carried out at the Observatorio Astronómico Nacional on the Sierra de San Pedro Mártir (OAN-SPM), Baja California, México. YGMC and SCG acknowledges support from UNAM PAPIIT-IG101321 and UNAM PAPIIT-IG101224. 
HW Vir was followed up by the two robotic telescopes in Hungary: RC80 at Konkoly and BRC80 at Baja Observatory. The operation of the RC80 and BRC80 telescopes was supported by the ``Transient Astrophysical Objects’' GINOP 2.3.2-15-2016-00033 project of the National Research, Development and Innovation Office (NKFIH), Hungary, funded by the European Union. 
CHe acknowledges support from the European Union H2020-MSCA-ITN-2019 under Grant Agreement no. 860470 (CHAMELEON). 
Cs. K. was supported by the `SeismoLab' KKP-137523 \'Elvonal and the K-138962 grants of the Hungarian Research, Development and Innovation Office (NKFIH). 
K.W.F.L. was supported by Deutsche Forschungsgemeinschaft grants RA714/14-1 within the DFG Schwerpunkt SPP 1992, Exploring the Diversity of Extrasolar Planets. 
This work was granted access to the HPC resources of MesoPSL financed by the Region Ile de France and the project Equip@Meso (reference ANR-10-EQPX-29-01) of the programme Investissements d'Avenir supervised by the Agence Nationale pour la Recherche. 
ML acknowledges the support of the Swiss National Science Foundation under grant number PCEFP2\_194576. 
This work was also partially supported by a grant from the Simons Foundation (PI Queloz, grant number 327127). 
The material is based upon work supported by NASA under award number 80GSFC21M0002 (NS).
NCSa acknowledges funding by the European Union (ERC, FIERCE, 101052347). Views and opinions expressed are however those of the author(s) only and do not necessarily reflect those of the European Union or the European Research Council. Neither the European Union nor the granting authority can be held responsible for them. 
This work was supported by the Hungarian National Research, Development and Innovation Office (NKFIH) grant K-125015, the PRODEX Experiment Agreement No. 4000137122 between the ELTE University and the European Space Agency (ESA-D/SCI-LE-2021-0025), the City of Szombathely under agreement No. 67.177-21/2016, and by the VEGA grant of the Slovak Academy of Sciences No. 2/0031/22. TP acknowledges support from the Slovak Research and Development Agency -- contract No. APVV-20-0148.
V.V.G. is an F.R.S-FNRS Research Associate. 
NAW acknowledges UKSA grant ST/R004838/1. 
JV acknowledges support from the Swiss National Science Foundation (SNSF) under grant PZ00P2\_208945.
This publication benefits from the support of the French Community of Belgium in the context of the FRIA Doctoral Grant awarded to MT. 
Some of the observations presented in this paper were based on discretionary-time observations made with the William Herschel Telescope operated on the island of La Palma by the Isaac Newton Group of Telescopes in the Spanish Observatorio del Roque de los Muchachos of the Instituto de Astrofísica de Canarias.
Based on observations collected at Copernico telescope (Asiago, Italy) of the INAF - Osservatorio Astronomico di Padova.
This project has received funding from the HUN-REN Hungarian Research Network.
We would like to acknowledge the use of ChatGPT, a language model developed by OpenAI, which was employed to enhance the readability and style of this article.
\end{acknowledgements}

\bibliographystyle{aa}
\bibliography{biblio}
\clearpage
\begin{appendix}

\section{Pointing performance}
\label{pointingPerf}
The onboard centroiding algorithm can be characterised as relatively straightforward: before commencing each visit, the target is first acquired using methods outlined in \citealt{2016SPIE.9913E..2WL} and \citealt{FerstelMaster}. Following a successful acquisition, the science observation ensues, during which centroiding is performed on each captured image according to the procedure elaborated in \citealt{2021SPIE11452E..1FM}:

\begin{enumerate}
    \item Crop a square region of 51$\times$51\,px (configurable) from the centre of the observing window; 
    \item Apply a median filter to the cropped image (can be enabled and configured);
    \item Subtract the smallest pixel value within the cropped square from all pixels;
    \item Calculate the photocentre of the adjusted pixel values.
\end{enumerate}

During IOC, several enhancements were implemented to improve performance. On one hand, the centroid window size was reduced from its original 71$\times$71\,px to 51$\times$51\,px. This change was made based on the reliable performance of absolute pointing and to minimise the influence of background stars. On the other hand, PITL tracking was also disabled during and in the vicinity of Earth occultations. This decision was driven by the fact that the star would move its apparent position in the sky when grazing the Earth's atmosphere due to refraction. The AOCS would try to compensate for this shift, potentially resulting in poor performance and the risk of losing the target after the occultation period.

Furthermore, certain checks are in place to identify and invalidate centroids that might be unreliable. For example, centroids are marked as invalid if no star is present in the FoV or if the image is heavily smeared. In such a case, centroids are sent to the SC but are flagged as invalid, causing the AOCS to ignore the corresponding offset values.

Following the decision to disable centroiding during Earth occultations, the PITL mode has exhibited exceptional performance. Notably, there has been no instance of star loss after a successful acquisition throughout the mission duration. Below, we present statistics pertaining to the tracking performance in both PITL and NOPITL modes. A comparison between onboard centroids and centroids computed by the \DRP underscores the efficacy of the relatively simple onboard centroiding algorithm, comfortably meeting the sub-arcsecond precision requirement. It is important to note that a consistent offset may exist for a specific target due to the shape of the PSF. While this offset is not pertinent to photometric performance, it does need to be considered when determining the optimal target position on the CCD.

\paragraph{\bf{Data filtering}}
Prior to computing statistics on \CHEOPS' tracking performance during the nominal mission, it is necessary to apply data filtering. If over 50\% of the centroids are labelled invalid, the corresponding visit is excluded from the analysis (we note that this criterion does not apply to the NOPITL mode). If the contamination from background stars is excessive, it can potentially degrade the tracking performance. Therefore, visits are classified according to the background contamination.

The first point depends on the in-flight software (IFSW) automatic filtering mechanism. As for the second point, it requires an assessment of the influence of background stars within the FoV. This assessment was achieved by calculating the level of contaminated flux present within the 50$\times$50\,px centroiding window, which was computed using the following procedure:

\begin{itemize}
    \item We utilised the \textit{Gaia} catalogue \citep{2018A&A...616A...1G} to identify stars located within a cone of 50 arcseconds around the target's coordinates. The distances of these stars from the target star, as well as their magnitudes, are readily available in the \textit{StarCatalog}\xspace file \citep{DRPHoyer}.
    \item Next, by taking into account the magnitude differences, we calculated the flux levels for each of these stars relative to the target star. 
    \item Subsequently, using the energy distribution curve of the PSF, we computed the total flux associated with each of the contaminating stars that fall within the confines of the 50$\times$50 centroiding window.
    \item In the final step, we aggregated the fluxes of all the contaminating stars and define the logarithm base 10 of the contamination flux ratio as the contamination factor. For instance, a factor of $-2$ signifies that the contamination level corresponds to 1\% of the flux emitted by the target star.
\end{itemize}

The target stars were categorised into the following bins based on the contamination factor ($\sim$ 3000 visits):
\begin{itemize}
    \item Isolated stars: Contamination $< -2$ (constituting 82\% and 61\% of the visits for PITL and NOPITL data, respectively)
    \item Semi-isolated stars: $-2$ < Contamination $< -1$ (making up 14\% and 29\% of the visits for PITL and NOPITL data, respectively)
    \item Contaminated stars: Contamination $> -1$ (accounting for 4\% and 10\% of the visits for PITL and NOPITL data, respectively)
\end{itemize}

\paragraph{\bf{PITL performances}}
To assess the pointing performance, two main factors were considered: (a) the proximity of the mean position of the PSF to the target's intended location and, (b) the extent to which the PSF of the star moves on the CCD during the course of a visit.

Addressing the first point is crucial for avoiding bad pixels and ensuring that successive visits to the same target consistently position the star on the same region of the detector. While no specific requirement is tied to this aspect, it primarily impacts long-term photometric precision. This is because it is assumed that stellar variability will have a more pronounced effect than $\sim{}50$ ppm flat field variations.
The second question directly influences photometric performance by affecting intra and inter-pixel sensitivity as well as introducing uncertainties in flat field corrections. This is why a requirement regarding pointing stability was established, as detailed in Sect.~\ref{sec:requirements}.

\begin{figure}
    \centering
    \includegraphics[width=\columnwidth]{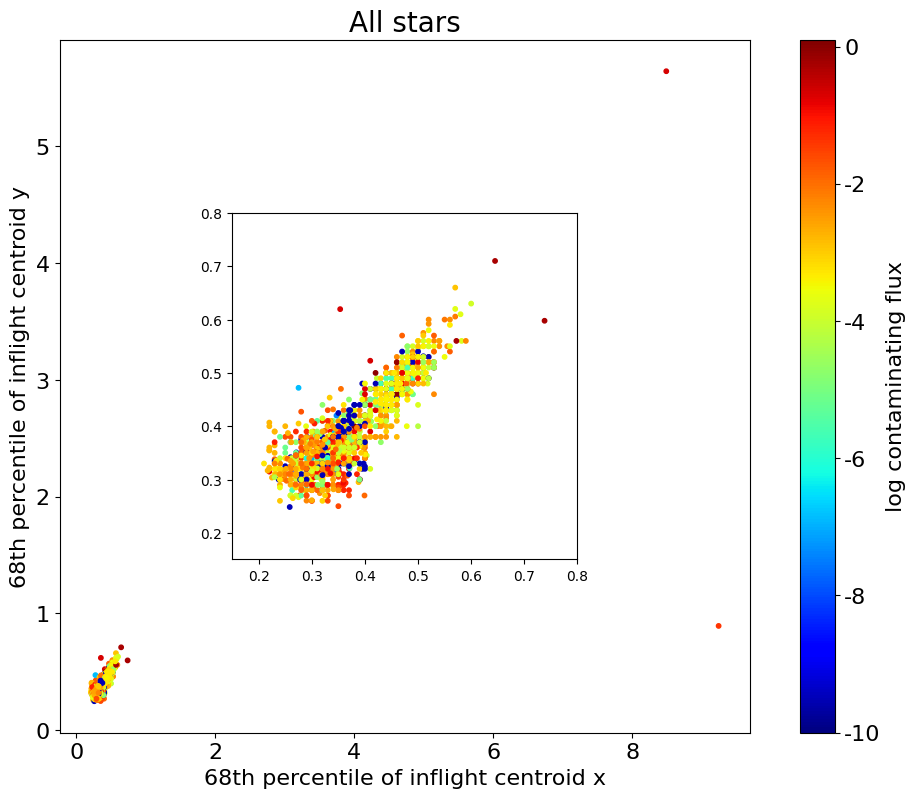}
    \caption{ PITL. The dispersion of the centroid position, as measured by the IFSW, is characterised by the 68th percentile of the distribution, representing a 1-sigma level. This measure of pointing performance consistently remains below 1 arcsecond, demonstrating excellent results. Notably, this performance is achieved almost regardless of the contamination factor, a testament to its robustness. It is worth noting that the presence of four outliers may be attributed to cases involving double stars with similar brightness, causing the algorithm to lock onto and oscillate between the two stars.}
    \label{fig:all_stars_per67_inflight}
\end{figure}

We began by examining the pointing stability. To quantitatively evaluate the tracking, we utilised the 68th percentile, which aligns with the requirement definition. In Fig.~\ref{fig:all_stars_per67_inflight}, the pointing performance of all stars analysed is displayed in relation to contamination. Notably, all stars with contamination levels below $-1$ exhibit pointing deviations well below 1\arcsec\ in both the $X$ and $Y$ directions, ranging from 0.2\arcsec\ to 0.7\arcsec. 

\begin{figure}
    \centering
    \includegraphics[width=\columnwidth]{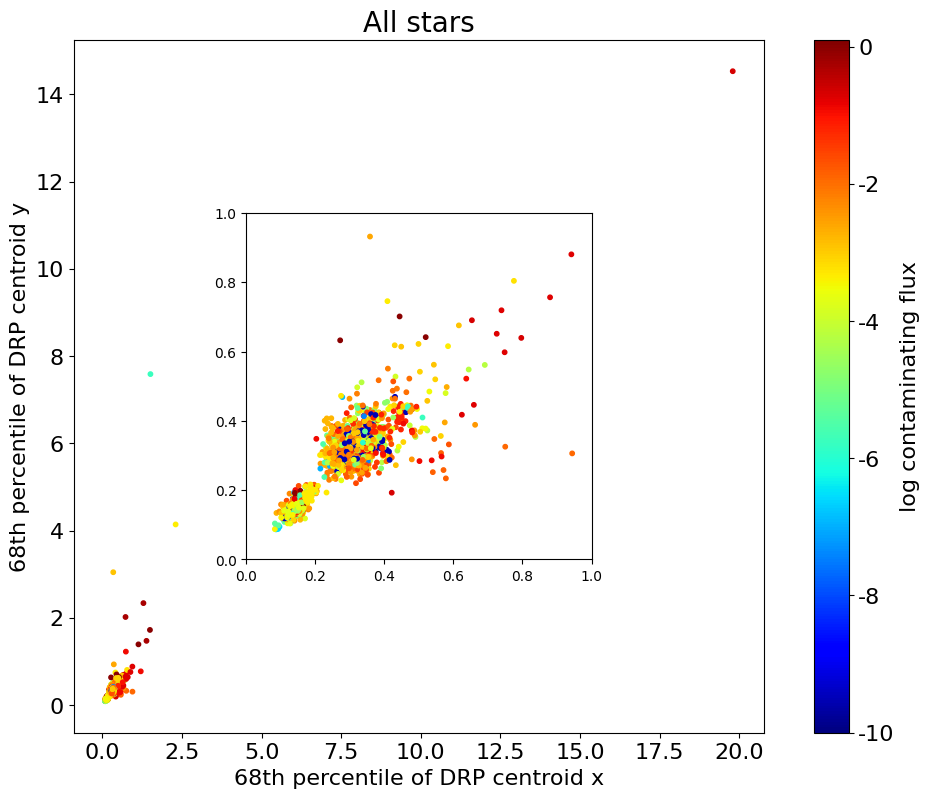}
    \caption{PITL. The dispersion of centroid positions, as measured by the \DRP, is evaluated using the 1-sigma 68th percentile of the distribution. Notably, the pointing performance remains highly favourable, registering less than 1 arcsecond when the contamination factor is below $-2$. This underscores the effective performance of the system in maintaining accurate pointing.}
    \label{fig:all_stars_per67_DRP}
\end{figure}

We next compared the centroids obtained in flight and those computed by the \DRP. The \DRP employs a more advanced algorithm for centroid computation. As illustrated in Fig.~\ref{fig:all_stars_per67_DRP}, this comparison demonstrates a parallel trend of remarkable performance. The pointing precision remains excellent until contamination surpasses approximately $-2$. 

\begin{figure}
    \centering
    \includegraphics[width=\columnwidth]{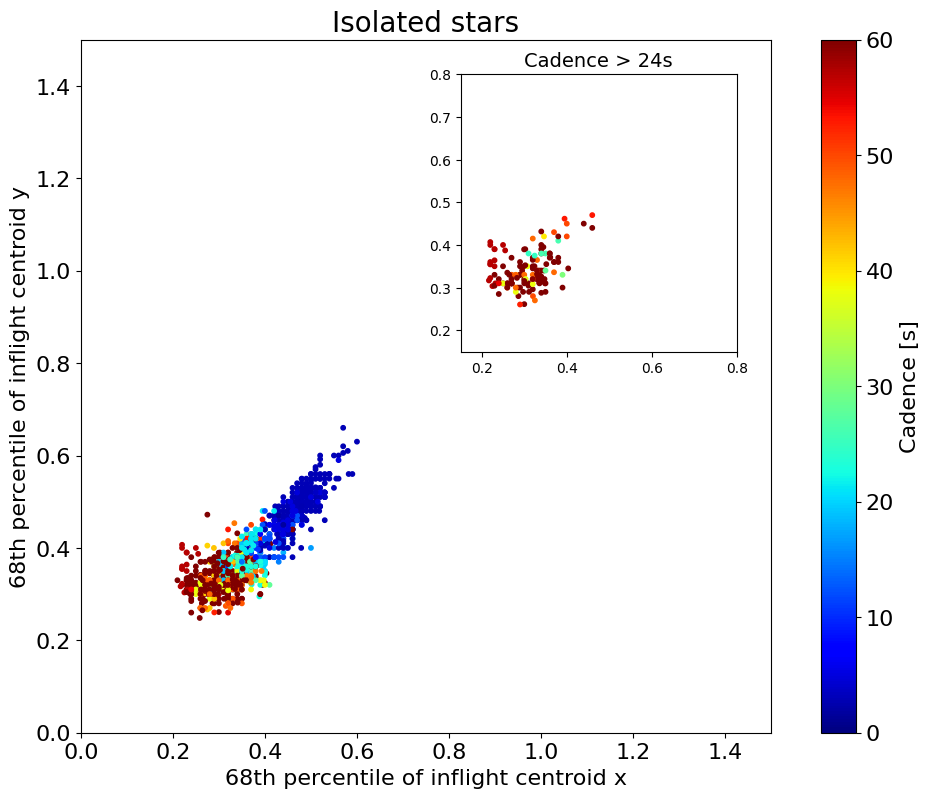}
    \caption{ PITL. Shown is the spread of centroid position measured by the IFSW and quantified using the 1-sigma 68th percentile of the distribution for isolated stars only. The colour code shows the image cadence. The inset shows a subset having only cadences $> 24$ seconds where no imagettes are created, and no image stacking happens. The blue points represent high cadence data, and they sample the high-frequency jitter component that appears to have an amplitude of about 0.5 to 0.6 arcseconds. }
    \label{fig:isolate_per67_inflight}
\end{figure}

\begin{figure}
    \centering
    \includegraphics[width=\columnwidth]{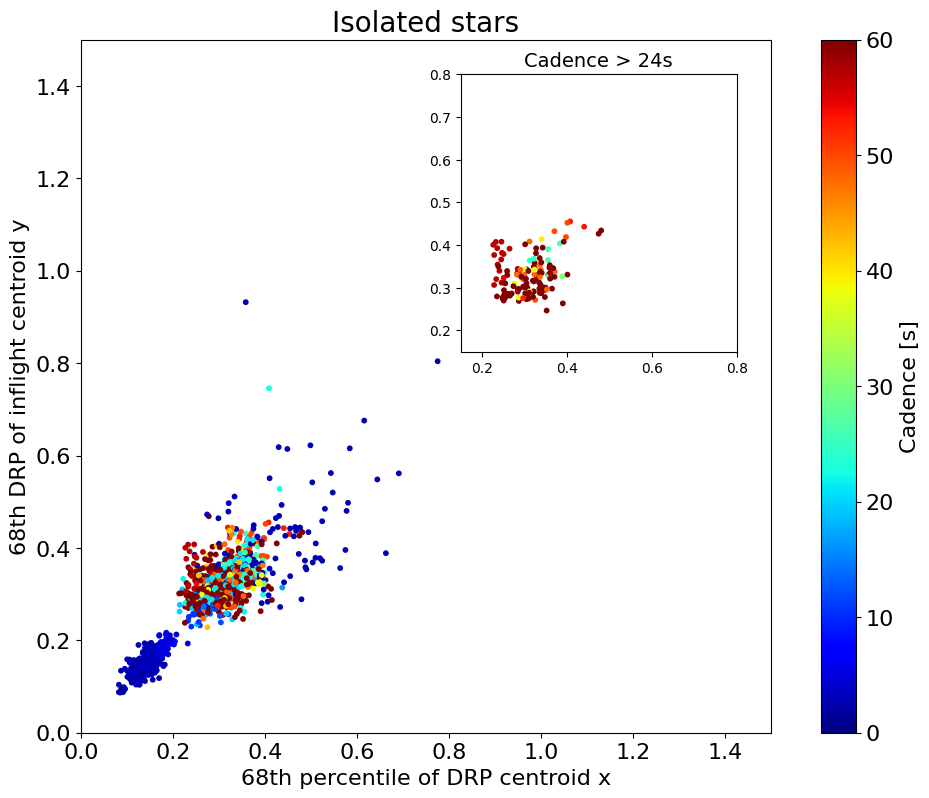}
    \caption{ PITL. Same as Fig.~\ref{fig:isolate_per67_inflight} but for \DRP centroids. High cadence data for the \DRP implies large stacking and that the \DRP centroids are computed on the stacked image. Therefore, the spread is much smaller. The inset shows the long cadence data where no stacking is done, which can be directly compared to IFSW data.}
    \label{fig:isolate_per67_DRP}
\end{figure}

However, it is important to note that their distributions exhibit dissimilar patterns. Both distributions exhibit a central cluster around 0.3--0.4 arcseconds with extended tails towards larger values for the IFSW data and towards smaller values for the \DRP data. To comprehensively analyse this divergence, a focused examination of isolated stars (contamination $<-2$) is performed, with the colour code indicating the raw image cadence. These visualisations are depicted in Figs.~\ref{fig:isolate_per67_inflight} and \ref{fig:isolate_per67_DRP}.

It becomes evident from these plots that the observed distinction originates from cadences faster than 24 seconds. The IFSW resorts to image stacking within this range to mitigate downlink data rates. Consequently, the centroids computed by IFSW are derived from the high-cadence data, while those computed by the \DRP are based on stacked images (i.e.\ slower cadence). For instance, with an exposure time of 3 seconds, 13 images are stacked, leading to a notable enhancement in centroid stability within the co-added frame. This suggests that the jitter comprises a high-frequency component with an amplitude below 0.6 arcseconds, exclusively sampled in brief exposures.

\begin{figure}
    \centering
    \includegraphics[width=\columnwidth]{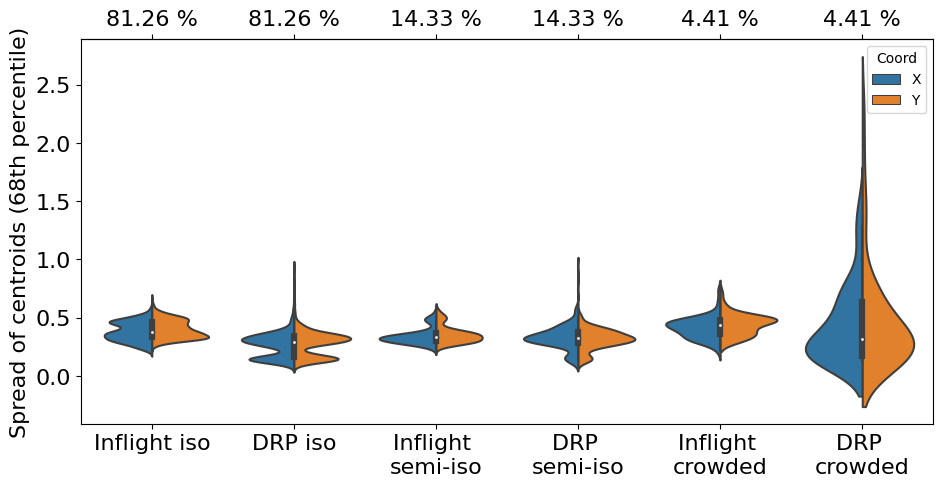}
    \caption{PITL. Violin plots depict the distribution of the centroid's 1-sigma spread in pixels for isolated, semi-isolated, and contaminated stars in both IFSW and \DRP datasets. For isolated stars, a bimodal distribution is evident, characterised by a peak around 0.3 arcseconds shared by both IFSW and \DRP, and a secondary peak with larger/smaller values for IFSW and \DRP, respectively. The shared peak can be attributed to visits with exposure times exceeding 24 seconds, while the discrepancies in the two datasets are due to very high-cadence observations with stacking in the case of \DRP, effectively enhancing precision.}
    \label{fig:violin_iso_semi_iso_xy}
\end{figure}

In order to better visualise the dispersion distribution, Fig.~\ref{fig:violin_iso_semi_iso_xy} offers violin plots illustrating the 68th percentile spreads of isolated, semi-isolated, and contaminated stars. The evident bimodal distribution underscores the impact of stacked versus\ non-stacked images, effectively demonstrating increased jitter at higher frequencies.

\begin{figure}
    \centering
    \includegraphics[width=\columnwidth]{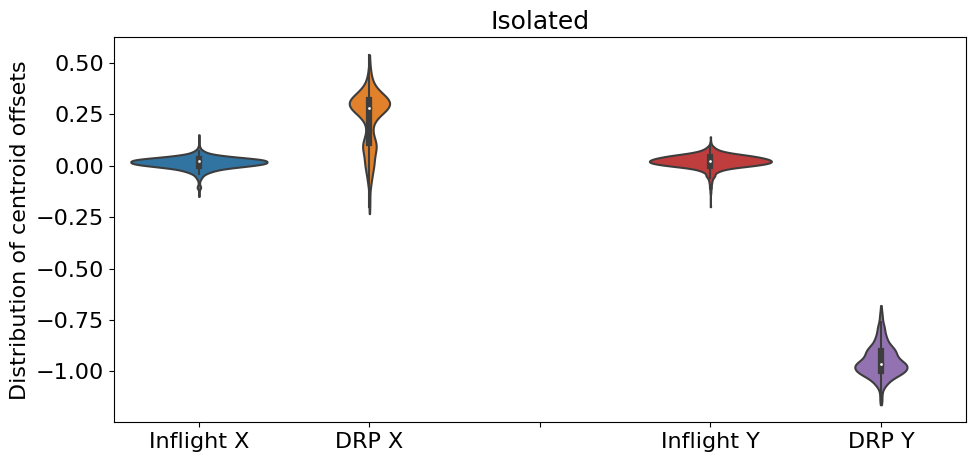}
    \caption{PITL. Shown is the distribution of the median offsets (in pixels) in the $X$ and $Y$ direction calculated by IFSW and \DRP for isolated cases.}
    \label{fig:violin_iso_offset}
\end{figure}

Examining the median positions of IFSW versus\ \DRP measured positions, as depicted in Fig.~\ref{fig:violin_iso_offset}, reveals that the PF consistently positions the star on the target location with sub-pixel accuracy. However, when measuring the centroid with the \DRP, a systematic offset emerges, which is more pronounced in the Y direction due to the asymmetric nature of the PSF along that axis. The magnitude of this offset depends on the target's brightness and spectral type.

\begin{figure}
    \centering
    \includegraphics[width=\columnwidth]{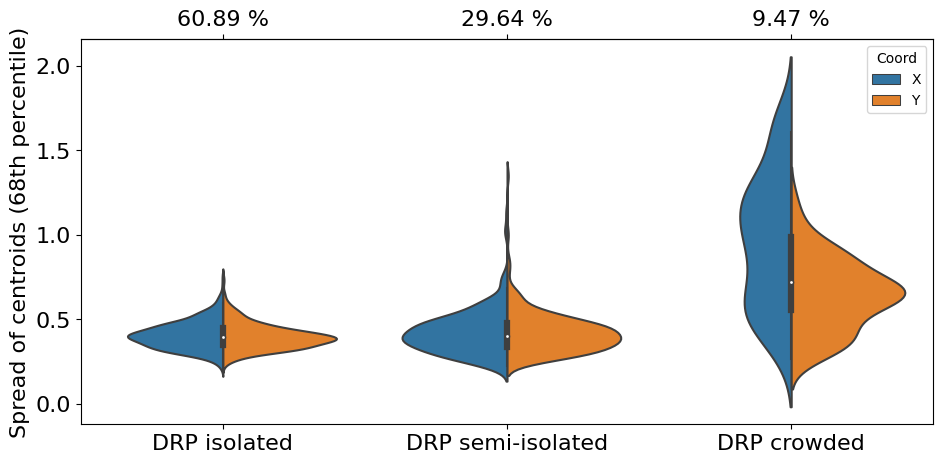}
    \caption{NOPITL. Violin plots illustrate the distribution of the 1-sigma spread of centroids in pixels for different scenarios: isolated, semi-isolated, and contaminated fields. These centroids are calculated using the \DRP's centroid calculation method.}
    \label{fig:violin_iso_semi_iso_xy_nopitl}
\end{figure}

\paragraph{\bf{NOPITL performances}}
When operating in NOPITL mode, the instrument does not transmit PSF offset data to the AOCS, making the tracking system rely solely on star trackers. Consequently, we can evaluate the performance of the \DRP and indirectly assess the stability of the target star's position within the window frame's centre. For isolated stars, the performance is exceptional, with a 1$\sigma$ spread violin plot (Fig.~\ref{fig:violin_iso_semi_iso_xy_nopitl}) showing a precision better than 1\arcsec. In fact, this performance is quite comparable to the PITL mode for low cadence data, with both distributions having modes below 0.4\arcsec.

In the semi-isolated case, both distributions exhibit similar shapes, although they show some outliers reaching up to 1.5\arcsec. Given that there is no feedback to the AOCS, we can assume that the pointing of both cases is identical, and differences are primarily due to imperfect tracking of ground-based positions due to crowding. In crowded fields, the \DRP's ability to track the target star might be slightly compromised (as indicated in the crowded \DRP violin plot in Fig.~\ref{fig:violin_iso_semi_iso_xy_nopitl}), but this does not reflect the SC's pointing precision.

\begin{figure}
    \centering
    \includegraphics[width=\columnwidth]{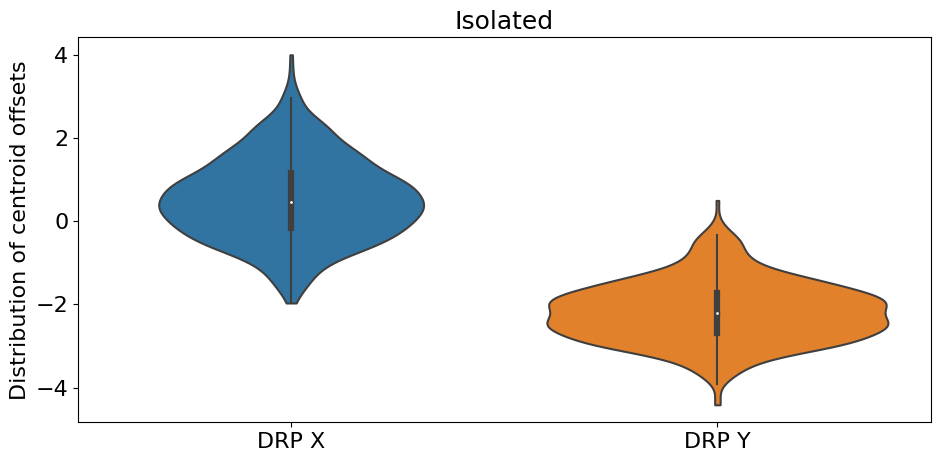}
    \caption{Distribution of the median offsets (in pixels) in $X$ and Y calculated by \DRP for NOPITL data, showing displacement in the location of the PSF. The \DRP centroid algorithm is sensitive to the bright peaks having asymmetrical positions with respect to the geometric centre of the PSF.}
    \label{fig:violin_iso_offset_nopitl}
\end{figure}

Fig.~\ref{fig:violin_iso_offset_nopitl} displays the median offsets that the \DRP has calculated for the observed visits. The plot highlights that the \DRP algorithm's sensitivity to the PSF's shape results in a calculated centre that deviates from the geometric centre. This discrepancy arises due to the asymmetrical nature of the PSF, notably its Y direction, which exhibits two bright peaks in the bottom region and one in the upper part of the PSF. Additionally, a slight misplacement of the bright peak is evident in the X direction. Consequently, the \DRP computes the weighted brightness of the PSF at coordinates of approximately $[+0.5, -2]$. This phenomenon stresses how the algorithm's interpretation of the PSF's shape influences its centre determination, with deviations from the geometric centre in both $X$ and $Y$ directions.

The analysis presented here underscores some interesting findings regarding pointing stability and absolute pointing offset. It was observed that the 1-sigma 68 percentile pointing precision for a single visit using the PITL was comparable to the NOPITL, with PITL exhibiting superior pointing accuracy primarily for high cadence science data. In cases with short exposure times, both methods exhibited similar levels of jitter.
However, when considering the absolute pointing offset, PITL demonstrated a significant advantage over NOPITL in accurately positioning the target star on the same pixel. PITL exhibited a median offset close to zero, with slight variations dependent on the target star's brightness and magnitude. In contrast, NOPITL displayed a considerably wider spread in both the X and Y directions, with a substantial offset in the Y direction, suggesting that the star was not consistently positioned as intended, except for cases when the target was located in the centre of the CCD.
In summary, while jitter levels were largely comparable for long exposure times and exhibited smaller jitter for bright, short cadence targets using PITL, the offset to the commanded target location was consistently close to zero for PITL but could range from 1 to 6 pixels for NOPITL. The observations suggest that the thermo-elastic bias remains relatively constant as long as the same SC attitude is maintained during a visit, thereby having a limited impact on the jitter.

\section{CHEOPS magnitude}
\label{cheops_mag}

A practical definition of the \CHEOPS magnitude can be expressed as follows: A star with a \CHEOPS magnitude of 0 and an effective temperature equivalent to that of Vega should produce an identical quantity of photoelectrons on the CCD as Vega.

To establish the \CHEOPS magnitude ($X$) in relation to the \textit{Gaia} magnitude ($G$), we introduce the following terminology:

\begin{itemize}
    \item $X_0$: \CHEOPS magnitude assigned to Vega, set to 0 by definition to align with the \textit{Gaia} convention;
    \item $G_0$: \textit{Gaia} magnitude of Vega set equal to 0 by convention;
    \item $QE$: Quantum efficiency of the \CHEOPS CCD (wavelength dependent, retrievable from the Mission Archive through the reference file \texttt{REF\_APP\_QE});
    \item $OT$: Optical throughput of the \CHEOPS telescope (wavelength dependent, retrievable from the Mission Archive through the reference file \texttt{REF\_APP\_Throughput});
    \item $ADU$: Analogue-to-digital units read from the CCD; 
    \item $F_0$: Vega photon flux computed by integrating the Vega SED over the wavelength range\footnote{In the spectral distributions, the photon flux at a given wavelength is proportional to the energy multiplied by the wavelength ($E=hc/\lambda$).} of 330 to 1100 nm. This integrated value is then multiplied by the telescope area, considering a circular aperture with a radius of 30 cm. The resulting calculated Vega photon flux is $3.6575\times 10^9$ photons/s.
    \item $F_X$: \CHEOPS flux (number of photoelectrons/s generated in the \CHEOPS CCD, after integration over the global throughput, i.e.\ $OT\times QE$, in the wavelength range of 330 to 1100 nm). 
    \item $F_{0X}$: \CHEOPS flux of Vega (number of photoelectrons/s generated in the \CHEOPS CCD from Vega, after integration over the global throughput, i.e.\ $OT\times QE$).
    \item $F_G$: Flux in the \textit{Gaia} passband (number of photons/s after integration over the \textit{Gaia} passband). 
    \item $F_{0G}$: Flux of Vega in the \textit{Gaia} passband (number of photons/s for Vega after integration over the \textit{Gaia} passband). 
\end{itemize}

The relationship between \CHEOPS and \textit{Gaia} magnitudes can be expressed as follows:

\begin{align*}
    X - X_0 = -2.5 \, \log (F_X / F_{0X}) \\
    G - G_0 = -2.5 \, \log (F_G / F_{0G}).
\end{align*}

\begin{figure}
    \centering
    \includegraphics[width=\columnwidth]{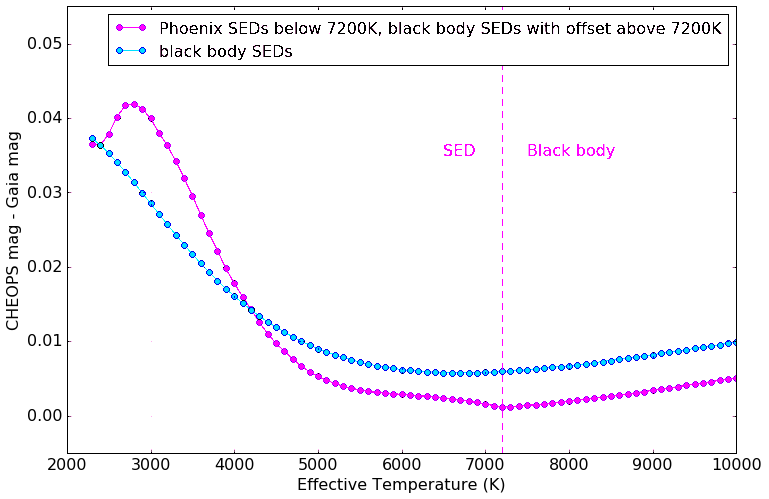}
    \caption{Difference between the \CHEOPS magnitude and \textit{Gaia} magnitude, $X - G$, as a function of the effective temperature of the star, using a SED for stars with \teff$<7200$\.K and the black body spectral distribution for \teff$>7200$\,K (magenta curve). The blue curve shows the same but always uses the black body approximation. }
    \label{fig:XminusG_annex}
\end{figure}

\noindent
Since $X_0 = G_0 = 0$, these simplify to\begin{align*}
    X = -2.5 \, \log (F_X / F_{0X}) \\
    G = -2.5 \, \log (F_G / F_{0G}).
\end{align*}
and hence
\begin{equation*}
    X - G = -2.5 \, \log ( (F_X \, F_{0G}) / (F_{0X} \, F_G) ).
\end{equation*}
$X - G$ is shown as a function of \teff in Fig.~\ref{fig:XminusG_annex} (magenta curve). For \teff$>7200$ K, the value of $X - G$ is decreased by 0.004795 in order to ensure a smooth transition between the use of SEDs (\teff$<7200$ K) and the use of black bodies (\teff$>7200$ K) to model the stellar spectra. We note that $X - G$ is not exactly 0 at the effective temperature of Vega (9600\,K), because $F_X$ and $F_G$ are calculated based on a black body spectrum, whereas $F0_X$ and $F0_G$ are calculated using the measured Vega SED.

\begin{figure}
    \centering
    \includegraphics[width=\columnwidth]{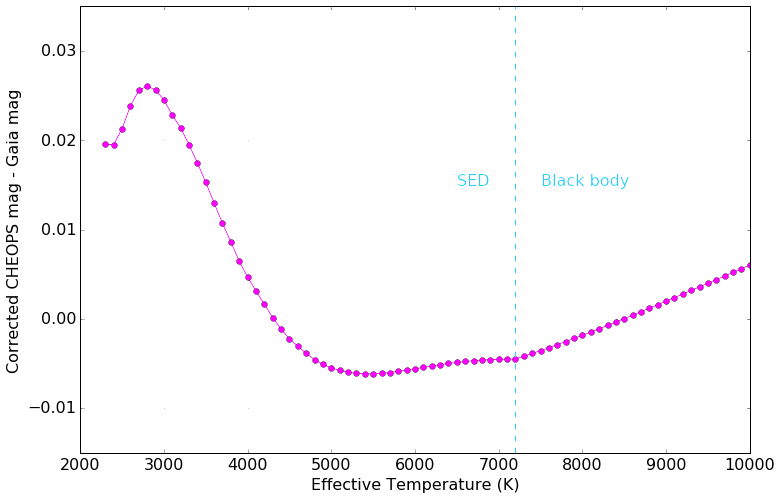}
    \caption{Difference between \CHEOPS magnitude and \textit{Gaia} magnitude, $X - G$, after adjusting empirically for discrepancies between the theoretically expected flux and the actually measured one.}
    \label{fig:XminusG_afterfit_annex}
\end{figure}

Observations indicate that the calculated flux $F_X$ is approximately 7\% higher than the observed flux $F_{\mathrm{obs}}$, with a slight dependence on \teff. To address this discrepancy, a linear regression is conducted to model the ratio $F_{\mathrm{obs}}/F_X$ against \teff. The resulting regression is shown in Fig.~\ref{fig:XminusG_afterfit_annex}, which is generated using both predicted and observed fluxes for 102 non-variable stars that were observed with \CHEOPS during IOC. The outcome of the regression yields the following relationship between the observed and predicted flux:
\noindent
\begin{equation*}
  OP = F_{\mathrm{obs}}/F_X = 0.94604 - 2.0074\times 10^{-6} \times \teff.
\end{equation*}
$F_{0X}$, which defines the overall scale (independent of \teff), is adjusted such that the fit result at the effective temperature of Vega (9600 K) equals 1. Consequently, $F_{0X}$ is multiplied by a scale factor:
\begin{equation*}
    SF =  0.94604 - 2.0074\times 10^{-6} \times 9600 = 0.92676. 
\end{equation*}
Defining $F_{0X}$ in this manner ensures that its physical interpretation corresponds to the number of photoelectrons/s that would be detected by \CHEOPS if Vega were the observed target. This definition also eliminates the necessity for a colour correction for Vega, and it establishes the \CHEOPS magnitude for Vega as zero.

The \CHEOPS magnitude is independent of time, per definition. $F_X$ is calculated considering the global throughput at the beginning of life.
Due to ageing, the received stellar flux decreases with time, as discussed in Sect. \ref{sec:SensitivityLoss}.
In practice, when we consider where the \CHEOPS magnitude is used, we see that this has no impact on the mission performance. The \CHEOPS magnitude is used in the ETC to estimate the incoming flux, which in turn is used to calculate the noise. If we assume that, on average, the flux of a \textit{Gaia} magnitude 9 star (with $\teff=5300$) decreased 18 ppm/day, in 4 years, the equivalent \textit{Gaia} magnitude of the star is 9.03. The estimated noise, according to the ETC, varies in this case from 13 ppm (237.2 ppm) in 6 hours (1 minute) to 13.1 ppm (239.4 ppm), respectively. This variation is below the precision of the ETC. 
The StarMap generation tool uses the \CHEOPS magnitude to simulate the expected performance of the target acquisition algorithm that will occur on board. A 0.03 magnitude difference will not affect operations because the target acquisition has become `obsolete' considering the excellent pointing performances.
The \CHEOPS magnitude is also used by the \DRP to simulate the FoV for the estimation of the contamination and the smearing correction. Such a small variation in the \CHEOPS magnitude also has no impact on the photometry.

\paragraph{\bf File format and usage}
To include the \teff-dependent component of the fit, it is integrated into the colour correction $X - G$ by modifying the formula as follows:
\noindent
\begin{equation*}
X - G = -2.5 \, \log [ ((F_X \, F_{0G})/(F_{0X} \, F_G)) \times (OP/SF) .
 \end{equation*}
The resultant value of $X - G$ as a function of \teff is shown Fig.~\ref{fig:XminusG_afterfit_annex}.

The header of the reference file \texttt{REF\_APP\_FluxConversion} provides the values of $F_{0X}$ and $X0$.
The file provides a \texttt{FITS} table with the following columns:
\begin{itemize}
    \item Column 1: \teff (Kelvin)
    \item Column 2: $X - G$
    \item Column 3: \(\int SED(\teff)\times OT \times QE \,d\lambda\) / \(\int SED(\teff)\times OT\,d\lambda\) (electrons per photon).\end{itemize}
To convert from \textit{Gaia} magnitude to \CHEOPS magnitude: 

\begin{equation*}
X = G + Column2.
\end{equation*}

To convert \CHEOPS magnitude to flux in electrons/s: 
\noindent
\begin{equation*}
    F_X = F_{0X} \times 10^{-0.4\, X}.
\end{equation*}
To convert flux in electrons/s to flux in photons/s: 
\begin{equation*}
F (photons/s) = F_X / Column 3.
\end{equation*}

To convert electrons into $ADU$, the following reference files should be used:   \texttt{REF\_APP\_GainCorrection}  and \texttt{REF\_APP\_CCDLinearisation}. 

\section{Procedure to independently measure \CHEOPS timing precision}
\label{HWVirtiming}

\begin{figure}
    \centering
    \includegraphics[width=\columnwidth]{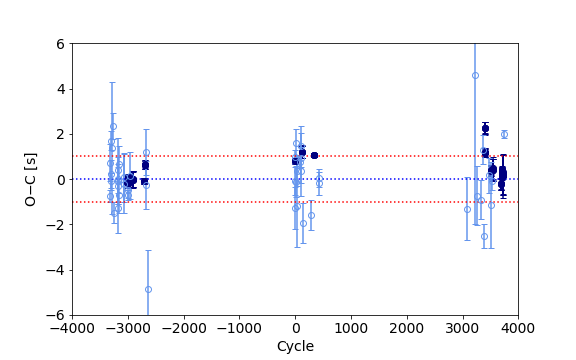}
    \caption{Residuals from the quadratic ephemeris for HW~Vir determined from the ground-based data, shown with open circles (light blue). Times of mid-eclipse measured using \CHEOPS data are shown using filled circles (dark blue).}
    \label{fig:hwvir}
\end{figure}

The ground-based observations of HW Vir are summarised in Table~\ref{table:hwvir-gb}. These observations were conducted using various telescopes, each specified in the table:  SAINT-EX -- 100-cm telescope, San Pedro M\'{a}rtir, Mexico; Szombathely -- 80-cm telescope, Gothard Observatory, Hungary; Konkoly -- 100-cm RCC and 80-cm RC telescopes, Piszk\'{e}s-tet\H{o} Mountain Station, Hungary; Baja Astronomical Observatory -- 80-cm telescope, Baja, Hungary; SPECULOOS -- 100-cm SPECULOOS-South telescope (Io), Paranal, Chile; Asiago -- 122-cm telescope, Asiago, Italy;  SAAO -- SAAO 190-cm telescope, South Africa. The observations were performed with different filters, and light curves were generated from the CCD images using the standard software used at each observatory for this purpose. The times of mid-eclipse ($T_0$) were determined by fitting a symmetric polynomial of the form   
 \[ f = \sum_{i=1}^{5} a_{2i}(t-T_0)^{2i}  \]
to the measured flux from HW Vir (in arbitrary units). Only data collected within 68 seconds of the predicted time of mid-eclipse were used for the fit.  The resulting times of mid-eclipse, along with the residuals from a quadratic ephemeris, are provided in Table~\ref{table:hwvir-gb}. The quadratic ephemeris is given by ${\rm BJD}_{\rm TDB}  =  2\,450\,000 $:
 $$ 9258.815830(3) + 0.116719492(6) \, E + 3.2(8)\times10^{-12} \, E^2$$
relative to ${\rm BJD}_{\rm TDB} = 2\,450\,000 $, where $E$ is the cycle number and numbers in parentheses give the standard error on the final digit of the preceding value. The residuals from this quadratic ephemeris are plotted as a function of cycle number ($E$) in Fig.~\ref{fig:hwvir}, showing that the RMS residual from this ephemeris is 1.4\,s.

The \CHEOPS observations of HW Vir are summarised in Table~\ref{table:hwvir-ch}. The times of mid-eclipse were determined from the light curves obtained using the default aperture size with a radius of 25 pixels. The same method as for the ground-based data was applied to measure the times of mid-eclipse.

During the initial observations conducted as part of the IOC phase, a discrepancy of $-5$ seconds was identified in the mid-eclipse times measured using \CHEOPS compared to the ground-based data. This issue was traced back to an outdated file containing leap-second adjustments to the UTC timescale. The file was updated on March 26, 2020. It was determined that archival data prior to this date required a $+5$ second adjustment to the time stamps provided with the light curve. Consequently, this correction has been applied to the times of mid-eclipse reported in the table. Observations made after March 26, 2020, remain unaffected.

Taking this correction into account, the mean offset of the times of mid-eclipse measured using \CHEOPS, relative to the predicted time of mid-eclipse from the quadratic ephemeris provided above, is $0.46\pm0.14$ seconds, with an RMS residual of 0.60\,s. The agreement between \CHEOPS data and the predicted times demonstrates that the timestamps provided with the light curve data fulfil the requirement for an accuracy of 1 second, as shown in Fig.~\ref{fig:hwvir}.

\begin{table}
\caption[]{Times of mid-eclipse for HW~Vir.}
         \label{table:hwvir-gb}
\begin{tabular}{rrrl}
\hline
\noalign{\smallskip}
\multicolumn{1}{c}{$E$} &\multicolumn{1}{c}{BJD$_{\rm TDB} -2450000$}  & \multicolumn{1}{c}{O$-$C} & Source \\
 & & \multicolumn{1}{c}{[s]} &  \\
\noalign{\smallskip}
\hline
\noalign{\smallskip}
$-3327$ &$ 8870.840247 \pm 0.000003$ &$-0.73$& SAINT-EX  \\
$-3326$ &$ 8870.956983 \pm 0.000010$ &$+0.75$& Szombathely  \\
$-3312$ &$ 8872.591046 \pm 0.000004$ &$-0.07$& Konkoly  \\
$-3303$ &$ 8873.641542 \pm 0.000005$ &$+1.72$& Konkoly  \\
$-3300$ &$ 8873.991684 \pm 0.000007$ &$+0.26$& Szombathely  \\
$-3294$ &$ 8874.692014 \pm 0.000034$ &$+1.41$& Konkoly  \\
$-3274$ &$ 8877.026415 \pm 0.000006$ &$+2.37$& Szombathely  \\
$-3249$ &$ 8879.944358 \pm 0.000005$ &$-1.47$& Szombathely  \\
$-3184$ &$ 8887.531126 \pm 0.000013$ &$-1.28$& SPECULOOS  \\
$-3184$ &$ 8887.531138 \pm 0.000009$ &$-0.29$& Konkoly  \\
$-3175$ &$ 8888.581616 \pm 0.000012$ &$-0.06$& SPECULOOS  \\
$-3175$ &$ 8888.581617 \pm 0.000007$ &$+0.03$& Konkoly  \\
$-3175$ &$ 8888.581621 \pm 0.000016$ &$+0.40$& SPECULOOS  \\
$-3164$ &$ 8889.865522 \pm 0.000009$ &$-0.71$& SAINT-EX  \\
$-3164$ &$ 8889.865538 \pm 0.000009$ &$+0.67$& Szombathely  \\
$-3073$ &$ 8900.487001 \pm 0.000015$ &$-0.21$& SPECULOOS  \\
$-3064$ &$ 8901.537480 \pm 0.000013$ &$+0.07$& SPECULOOS  \\
$-3004$ &$ 8908.540642 \pm 0.000004$ &$-0.48$& SPECULOOS  \\
$-2994$ &$ 8909.707834 \pm 0.000002$ &$-0.71$& SAINT-EX  \\
$-2962$ &$ 8913.442867 \pm 0.000012$ &$+0.13$& SPECULOOS  \\
$-2679$ &$ 8946.474476 \pm 0.000012$ &$-0.29$& SPECULOOS  \\
$-2679$ &$ 8946.474493 \pm 0.000012$ &$+1.15$& Szombathely  \\
$-2645$ &$ 8950.442886 \pm 0.000020$ &$-4.88$& SPECULOOS  \\
$  -12$ &$ 9257.765364 \pm 0.000004$ &$+0.88$& SAINT-EX  \\
$   -5$ &$ 9258.582389 \pm 0.000007$ &$-0.18$& SPECULOOS  \\
$   -4$ &$ 9258.699094 \pm 0.000011$ &$-1.35$& SPECULOOS  \\
$    4$ &$ 9259.632884 \pm 0.000007$ &$+1.51$& SPECULOOS  \\
$   12$ &$ 9260.566620 \pm 0.000008$ &$-0.19$& SPECULOOS  \\
$   21$ &$ 9261.617083 \pm 0.000021$ &$-1.25$& WHT \\
$   38$ &$ 9263.601332 \pm 0.000008$ &$+0.29$& Baja  \\
$   83$ &$ 9268.853714 \pm 0.000003$ &$+0.70$& SAINT-EX  \\
$   98$ &$ 9270.604514 \pm 0.000010$ &$+1.37$& SPECULOOS  \\
$  106$ &$ 9271.538264 \pm 0.000013$ &$+0.85$& SPECULOOS  \\
$  107$ &$ 9271.654977 \pm 0.000013$ &$+0.27$& SPECULOOS  \\
$  132$ &$ 9274.572938 \pm 0.000010$ &$-2.03$& SPECULOOS  \\
$  288$ &$ 9292.781183 \pm 0.000008$ &$-1.64$& SAINT-EX  \\
$  425$ &$ 9308.771773 \pm 0.000005$ &$+0.01$& SAINT-EX  \\
$  417$ &$ 9307.838014 \pm 0.000006$ &$-0.23$& SAINT-EX  \\
$ 3071$ &$ 9617.611554 \pm 0.000016$ &$-0.89$& Asiago  \\
$ 3225$ &$ 9635.586426 \pm 0.000076$ &$+5.04$& Asiago  \\
$ 3250$ &$ 9638.504352 \pm 0.000015$ &$-0.27$& Asiago  \\
$ 3327$ &$ 9647.491752 \pm 0.000010$ &$-0.42$& Asiago  \\
$ 3364$ &$ 9651.810399 \pm 0.000007$ &$+1.80$& Szombathely  \\
$ 3382$ &$ 9653.911306 \pm 0.000006$ &$-2.00$& Szombathely  \\
$ 3465$ &$ 9663.599056 \pm 0.000009$ &$+0.68$& SAINT-EX  \\
$ 3498$ &$ 9667.450785 \pm 0.000022$ &$-0.60$& Asiago  \\
$ 3517$ &$ 9669.668467 \pm 0.000004$ &$+0.43$& SAINT-EX  \\
$ 3737$ &$ 9695.346783 \pm 0.000002$ &$+1.97$& SAAO \\
\noalign{\smallskip}
\hline
\end{tabular}  
\tablefoot{Residuals from the quadratic ephemeris derived from ground-based observations of HW~Vir are given in the column headed $O-C$. }
\end{table}

\begin{table*}
\caption[]{Times of mid-eclipse for HW~Vir measured using \CHEOPS.}
\label{table:hwvir-ch}
\begin{center}
\begin{tabular}{lrrrr}
\hline
\noalign{\smallskip}
\multicolumn{1}{l}{File key} &
\multicolumn{1}{l}{T$_{\rm exp}$ [s]} &
\multicolumn{1}{l}{Cycle} &
\multicolumn{1}{l}{$T_{\rm mid}$ (BJD$_{\rm TDB}$)}& O$-$C [s]  \\
\noalign{\smallskip}
\hline
\noalign{\smallskip}
CH\_PR300006\_TG000101\_V0200 & 10.0 &$ -3018 $&$ 2458906.5564151 \pm 0.0000013 $ &$  -0.11 $ \\
CH\_PR300006\_TG000102\_V0200 & 10.0 &$ -2995 $&$ 2458909.2409654 \pm 0.0000015 $ &$  +0.08 $ \\
CH\_PR300006\_TG000103\_V0200 & 10.0 &$ -2992 $&$ 2458909.5911211 \pm 0.0000026 $ &$  -0.15 $ \\
CH\_PR300051\_TG000301\_V0200 & 30.0 &$ -2911 $&$ 2458919.0454007 \pm 0.0000041 $ &$  -0.02 $ \\
CH\_PR300058\_TG000201\_V0200 & 30.0 &$ -2701 $&$ 2458943.5564912 \pm 0.0000015 $ &$  -0.08 $ \\
CH\_PR300058\_TG000101\_V0200 & 30.0 &$ -2687 $&$ 2458945.1905725 \pm 0.0000021 $ &$  +0.66 $ \\
CH\_PR310080\_TG000301\_V0200 & 30.0 &$   -11 $&$ 2459257.5319233 \pm 0.0000014 $ &$  +0.79 $ \\
CH\_PR310080\_TG000901\_V0200 & 30.0 &$   119 $&$ 2459272.7054623 \pm 0.0000030 $ &$  +1.21 $ \\
CH\_PR310080\_TG001401\_V0200 & 30.0 &$   345 $&$ 2459299.0840663 \pm 0.0000010 $ &$  +1.07 $ \\
CH\_PR320089\_TG001001\_V0200 & 30.0 &$  3397 $&$ 2459655.3119826 \pm 0.0000022 $ &$  +1.18 $ \\
CH\_PR320089\_TG001002\_V0200 & 30.0 &$  3404 $&$ 2459656.1290317 \pm 0.0000030 $ &$  +2.26 $ \\
CH\_PR320089\_TG001003\_V0200 & 30.0 &$  3508 $&$ 2459668.2678384 \pm 0.0000025 $ &$  +0.36 $ \\
CH\_PR320089\_TG001004\_V0200 & 30.0 &$  3551 $&$ 2459673.2867777 \pm 0.0000055 $ &$  +0.40 $ \\
CH\_PR320089\_TG001501\_V0200 & 11.3 &$  3682 $&$ 2459688.5770259 \pm 0.0000028 $ &$  -0.22 $ \\
CH\_PR320089\_TG001401\_V0200 & 30.0 &$  3698 $&$ 2459690.4445438 \pm 0.0000021 $ &$  +0.28 $ \\
CH\_PR320089\_TG001402\_V0200 & 30.0 &$  3708 $&$ 2459691.6117403 \pm 0.0000017 $ &$  +0.40 $ \\
CH\_PR320089\_TG001403\_V0200 & 30.0 &$  3712 $&$ 2459692.0786155 \pm 0.0000022 $ &$  +0.16 $ \\
CH\_PR320089\_TG001404\_V0200 & 30.0 &$  3722 $&$ 2459693.2458109 \pm 0.0000104 $ &$  +0.18 $ \\
CH\_PR320089\_TG001405\_V0200 & 30.0 &$  3727 $&$ 2459693.8294090 \pm 0.0000015 $ &$  +0.23 $ \\
\noalign{\smallskip}
\hline
\end{tabular}  
\end{center}
\tablefoot{Residuals from the quadratic ephemeris derived from ground-based observations of HW~Vir are given in the column headed $O-C$.}
\end{table*}

\section{Calculation of dark, hot, and RTS pixel frames}
\label{sec:CalBadPixels}

During a dark M\&C observation, two sets of full-frame images are acquired: one set consists of ten `60 s frames' with an exposure time of 60 seconds each, and the other set consists of ten `zero frames' with an exposure time of 0.001 seconds each. The purpose of the first set is to capture thermal electrons in hot pixels, while the goal of the second set is to measure the gradient of the dark current during the read-out process. The pointing is intentionally shifted between exposures to prevent field stars from falling onto the same CCD spot.

The initial steps involve calibration and filtering processes. Calibration includes bias, gain, and non-linearity corrections based on standard procedures (as outlined in \citealt{DRPHoyer}). Filtering is applied to discard images with high background flux levels from further analysis. Afterwards, the dark frame is constructed through the following steps:

\begin{enumerate}
    \item A single short frame is generated from the 10 zero-frames by taking their median.
    \item The short frame is subtracted from each of the 60s frames.
    \item The corrected 60s frames are normalised by the exposure time difference (59.999 seconds) in the next step.
    \item The median of these normalised images is computed, resulting in the final dark frame.
    \item The error of the frame is calculated as the standard error of the normalised image cube.
\end{enumerate}

It is important to note that this dark frame includes not only the `true dark current' of the electronics but also the zodiacal light due to the lack of a shutter.
The zodiacal light and bulk thermal dark current must be removed to measure the dark current in the hot pixels. The first three steps of the process are the same as for the dark frame, but in the fourth step, a 2D median filter with a 3$\times$3 pixel kernel is applied to the 60s frames. This step helps eliminate larger background variations caused by structures like stars and aligns the images to a consistent background level. The final step and error calculation remains the same. This resulting frame is used to identify hot pixels and measure their dark current levels. A pixel is categorised as a hot pixel if its flux exceeds 3 electrons per second.

\begin{figure*}
    \centering
    \includegraphics[scale=0.45]{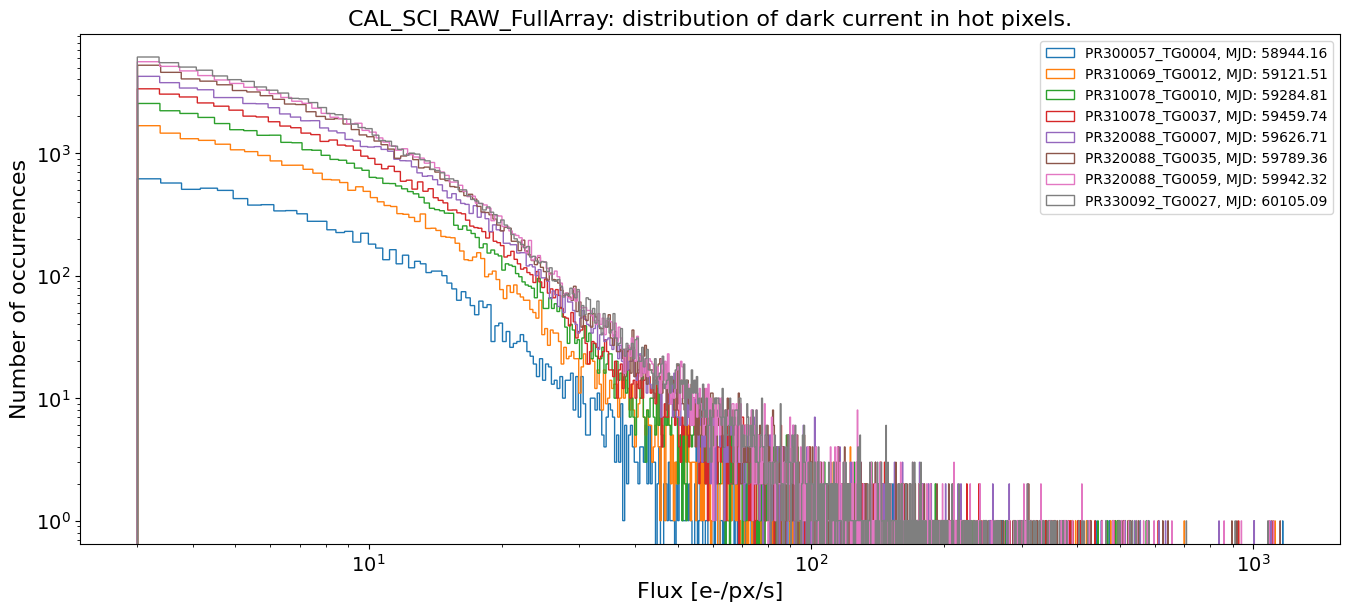}
    \caption{Evolution of the distribution of the dark current in hot pixels. The colours indicate the time when the dark M\&C were taken.}
    \label{fig:hp_dist}
\end{figure*}

The distribution of dark current levels in the hot pixels is depicted in Fig.~\ref{fig:hp_dist}. This distribution spans over more than 1000 days and is presented in 8 snapshots. Notably, the distribution's shape remains consistent over time, indicating that there are no significant distortions or variations. Hot pixels with higher dark current levels are less frequent. The distribution has no prominent value or peak, suggesting unexpectedly higher occurrences of hot pixels at specific dark current values.

\begin{figure}
    \centering
    \includegraphics[width=\columnwidth]{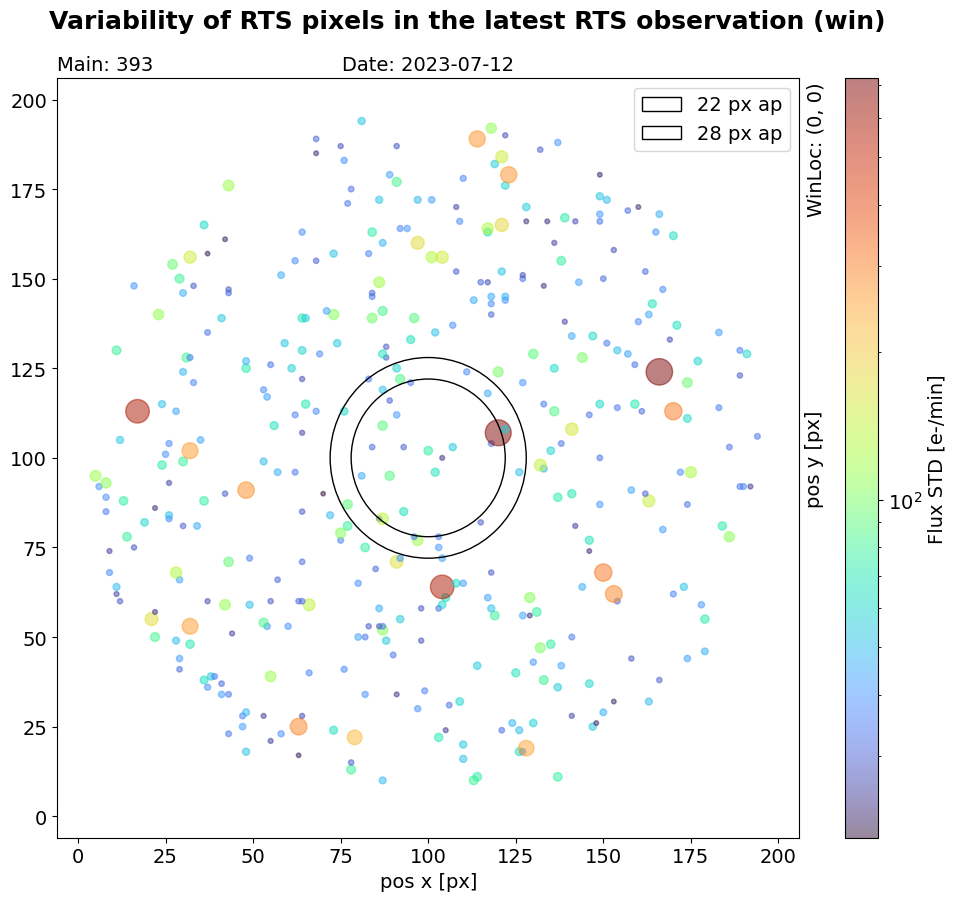}
    \caption{Variability of RTS pixels in the active window frame. The bigger and redder the circle, the more variable the RTS pixel is. The magnitude of the oscillation is calculated for a 1-minute exposure.}
    \label{fig:RTS_rel_osc}
\end{figure}

Observations for the detection of RTS pixels are scheduled four times a year. Each visit involves observing a dark region of the sky for the duration of ten \CHEOPS orbits. However, due to data volume limitations, only the window images corresponding to the active window location are downlinked to the ground.  Within these ten orbits, approximately 1000 images are captured, each with a 60\,s exposure time. It is worth noting that only 50--80\% of these images can be utilised for RTS pixel detection, primarily due to factors such as Earth occultations and images affected by high background contamination, including instances of elevated stray light and SAA crossings. Following the standard calibration process, a subset of images is further processed by calculating the median of each group of five consecutive frames. This step helps eliminate any remaining spatial contamination and cosmic ray effects in the images. Fig.~\ref{fig:RTS_rel_osc} shows the positions and variability of the RTS pixels at the end of the nominal mission in the active window.

\section{Noise due to hot pixels}
\label{sec:NoiseBadPixels}

To identify the most suitable window location, we employed aperture photometry on the hot pixel map across every feasible pixel coordinate on the CCD. This technique involves a 2D convolution of two signals, where the aperture serves as the kernel applied to the hot pixel map. This approach, termed convolutional photometry, employs the \DRP DEFAULT aperture ($r = 25$\,px) to scan the entire CCD (except for a 25-pixel-wide strip at the CCD's edges). For the calculations, we utilised a reference observation from the beginning of the mission (\texttt{CH\_PR300061\_TG000301}) with a low number of hot pixels in the window frame. The target is a relatively quiet star (TYC 5502-1037-1, $G=12$\,mag) with minimal flux variations. This normalisation by the reference star's flux allows us to estimate the noise contribution from hot pixels. The outcome is a map illustrating the anticipated noise at every pixel on the CCD.

Shot noise is the square root of the total flux within the aperture, and a single application of convolutional photometry on the hot pixel map suffices to determine it. Jitter noise, however, is contingent upon pointing stability, which governs how the PSF (and consequently, the aperture) moves during an observation. The changing position of the aperture relative to the hot pixels leads to fluctuations in the number of hot pixels within the aperture and, consequently, in the aperture's flux. This results in noisy light curves. To calculate this noise, multiple aperture photometry measurements must be conducted for each jitter position at every pixel location on the CCD. Subsequently, the standard deviation of the relevant light curves (one for each pixel position on the CCD) is calculated. The positions of the aperture are determined using the centroid algorithm of the \DRP. Finally, the shot and jitter noise maps are combined in quadrature to estimate the expected noise contribution.

\begin{figure}
    \centering
    \includegraphics[width=\columnwidth]{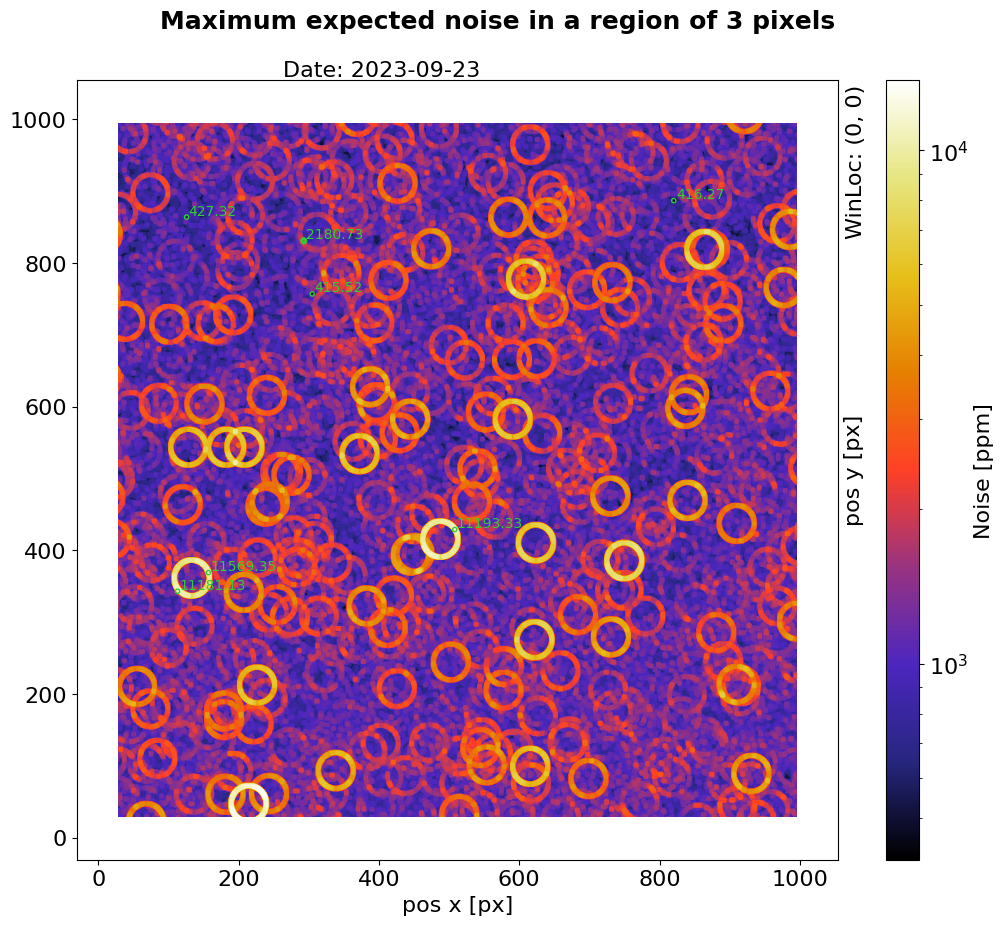}
    \caption{Maximum expected noise in ppm per minute due to hot pixels using the flux of a $G = 12$\,mag star as reference. The estimations comprise two distinct noise components: shot noise and jitter noise. Jitter noise is a consequence of telescope jitter, leading to the PSF's movement on the CCD. Consequently, the edge of the default aperture ($r=25$\,px) intersects hot pixels, inducing fluctuations in flux due to the hot pixels intermittently entering and exiting the aperture. The `noise rings' seen in the image result from the convolution of the exact noise map with a circle of $r = 3$\,px (all potential target locations for the intended one). That is, from all the noise values calculated in the exact map for the pixels within each circle, the maximum one is assigned to each pixel. Green circles indicate the regions of the CCD with the smallest and highest noise attributable to hot pixels.}
    \label{fig:noise_full}
\end{figure}

In an ideal scenario, this `exact' noise map would suffice if the PSF were precisely centred within the window frame and the aperture was perfectly aligned with the PSF's centre. However, practical considerations, such as the PSF's triangular shape and its slight offset from the window centre, the uncertainty between the intended star position in the CCD and the actual one, and the pointing jitter (which will follow a unique path per observation) render this map insufficient for pinpointing the optimal window location. To address this, a circle with a radius of three pixels is defined, encompassing (with margin) all potential PSF real positions. A subsequent noise map is generated to display the mean, minimum, or maximum anticipated noise within this circumscribed region. Fig.~\ref{fig:noise_full} shows the maximum expected noise map of the CCD at the end of the nominal mission for the reference star ($G = 12$\,mag). We note that the location of the active window (coordinates of its bottom left corner: (193, 731)) places the PSF in a low noise region. 

\paragraph{\bf{Noise evolution in the window frame}}
We also monitored the expected noise due to bad pixels at the precise PSF location of the target.  This approach involves creating a simulated observation that includes all the hot pixels within the window frame. We initiated this process using our reference observation as a baseline (target: TYC 5502-1037-1, visit ID: \texttt{CH\_PR300061\_TG000301}) and selected a subset of images with a duration of 300 frames. This choice strikes a balance between reasonable processing time, the high CPU demands, and the lengthy processing times of the \DRP routines we employ for re-processing the data.

To simulate the presence of hot pixels, we generated an image that exclusively contains hot pixels. This was accomplished by extracting the relevant window frame from the comprehensive hot pixel map and aligning it with the current window's location. Subsequently, we introduced Poisson noise to emulate the behaviour of hot pixels. The dark current of these simulated hot pixels was calibrated based on the exposure time of the reference observation. This sequence results in the creation of an image cube with only hot pixels, maintaining dimensions consistent with those of the reference observation.

\begin{figure*}
    \centering
    \includegraphics[scale = 0.45]{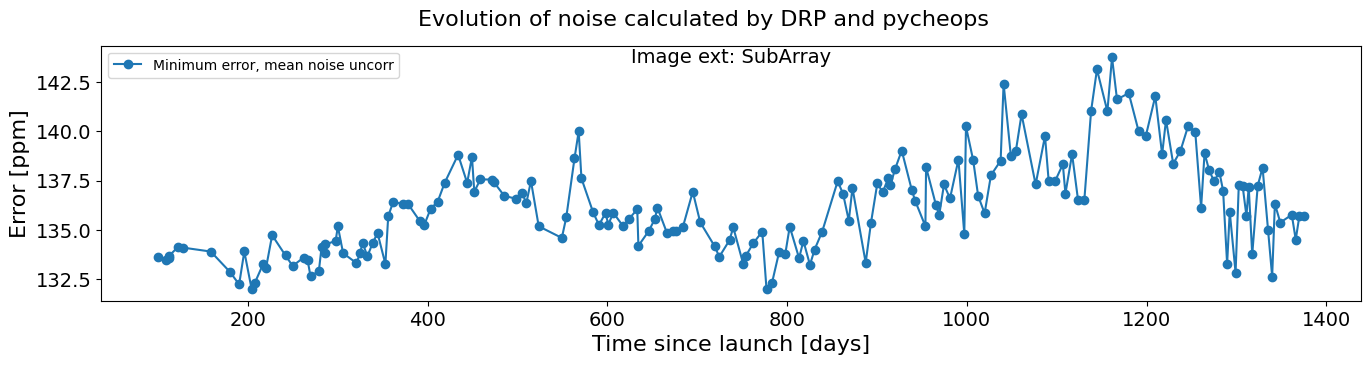}
    \caption{Progression of noise since the beginning of nominal operations, calculated through the \pycheops package according to the minimum error algorithm \citep{Maxtedpycheops}. The figure shows the estimated noise of a 3\,h bin by using a simulated dataset wherein hot pixels were injected into the reference visit and subsequently processed by the \DRP with dark correction module OFF.}
    \label{fig:noise_evo}
\end{figure*}

Following the image generation process, we conducted a basic arithmetic addition between the newly created image cube and the reference observation image cube. This merged image cube was then subjected to the \DRP pipeline for processing and was subsequently analysed using the \pycheops package. We note that the dark correction feature was either disabled or unavailable in earlier versions of the DRP. This was done deliberately to assess the direct effect of hot pixels. We utilised the minimum error algorithm as implemented in \pycheops (\citealt{Maxtedpycheops}) to estimate the noise of the light curve within a 3-hour bin. The outcomes of these analyses are presented as a plot, illustrating the noise evolution over time since the satellite's launch, as shown in Fig.~\ref{fig:noise_evo}. 

\section{Glossary}
\label{Glossary}

Table \ref{table:acronyms} contains the symbols and acronyms used in this paper.

\begin{table*}
\caption[]{Symbols and acronyms.}
\label{table:acronyms}
\begin{center}
\begin{tabular}{llll}
\hline
\noalign{\smallskip}
\multicolumn{1}{l}{Acronym} &
\multicolumn{1}{l}{Definition} &
\multicolumn{1}{l}{Acronym} &
\multicolumn{1}{l}{Definition} \\
\noalign{\smallskip}
\hline  
\noalign{\smallskip}
ACF & Autocorrelation Function & ppm & parts-per-million \\
$ADU$ & Analogue-to-Digital Units & PSF & Point Spread Function \\
AOCS & Attitude and Orbital Control System & PST & Point Source Transmission function \\
CCD & Charged-Coupled Device &  px & pixel\\
CHEOPS &  Characterising ExOPlanet Satellite & QE & Quantum Efficiency \\
CTI & Charge Transfer Inefficiency &  $r$ & Radius \\
\DRP & Data Reduction Pipeline & $R_{star}$ & Radius of the star \\
\DRT & Data Reduction Tool & RMS & Root Mean Square \\
ESA & European Space Agency & RTS & Random Telegraphic Signal  \\
ETC & Exposure Time Calculator & SAA & South Atlantic Anomaly \\
FEE & Front End Electronics & SC & Spacecraft \\
FoV & Field of View & SciReq. & Science Requirement \\
$G$ & \textit{Gaia} $G$ magnitude & SEM & Sensor Electronic Module \\
HK & Housekeeping & SOC & Science Operation Centre \\
IFSW & In-Flight software & \teff & Stellar Effective Temperature \\
IOC & In-Orbit Commissioning  & $t_{\mathrm{exp}}$ & Exposure time \\
LEO & Low-Earth Orbit & UTC & Universal Time Coordinated \\
LoS & Line of Sight & $V$ & Visual magnitude \\ 
LTAN & Local Time of the Ascending Node & VOD & Output transistor Drain Voltage\\
M\&C & Monitoring and Characterisation Programme & VOG & Output Gate Voltage\\
MOC & Mission Operation Centre & VRD & Reset transistor Drain Voltage\\
NOPITL & NO Payload In The Loop & VSS & Substrate Voltage\\  
OBT & Onboard Time \\
OTA & Optical Telescope Assembly \\
PCA & Principal Component Analysis \\
PF & Platform \\
PID & Proportional Integral Derivative  \\
PITL & Payload In The Loop \\
\PIPE & PSF Imagette Photometric Extraction \\

\noalign{\smallskip}
\hline
\end{tabular}  
\end{center}
\end{table*}

\section{List of \CHEOPS visits used to perform the analyses in this paper}
\label{visits}

Table \ref{table:visits} provides a comprehensive list of the \CHEOPS visits employed to generate the results presented in this paper.

\onecolumn
\begin{longtable}{lrrrr}    
\caption[]{\label{table:visits} Visits used for the analyses presented in this paper.}\\
\noalign{\smallskip}
\hline
\noalign{\smallskip}
\multicolumn{1}{l}{Section} &
\multicolumn{1}{l}{File key} &
\multicolumn{1}{l}{Date} &
\multicolumn{1}{l}{Target} &
\multicolumn{1}{l}{Gmag} \\
\noalign{\smallskip}
\hline
\noalign{\smallskip}
Sect.~\ref{sec:MandC} & CH\_PR330096\_TG0007 & 02/03/2023 & HD 88742 & 6.38\\
Sect.~\ref{sec:TempStability} &  CH\_PR100041\_TG000201 & 19/03/2020 & WASP-189 & 6.55 \\
Sect.~\ref{sec:VisBadPixels} & CH\_PR100018\_TG040003 & 20/09/2022 & GJ 908 & 8.15 \\
Sect.~\ref{sec:ETC} and \ref{sec:MCperformance} & CH\_PR300005\_TG000101 & 22/02/2020 & HD 88111 & 8.97 \\
Sect.~\ref{sec:ETC} and \ref{sec:MCperformance} & CH\_PR300005\_TG000501 & 21/03/2020 & TYC 5502-1037-1 & 11.98 \\
Sect.~\ref{sec:ETC} & CH\_PR100011\_TG023501 & 20/04/2020 & GJ 536 & 8.86 \\
Sect.~\ref{sec:ETC} & CH\_PR100011\_TG023701 & 24/04/2020 & GJ 581 & 9.41\\
Sect.~\ref{sec:ETC} & CH\_PR100011\_TG023001 & 03/05/2020 & 61 Vir & 4.48 \\
Sect.~\ref{sec:ETC} & CH\_PR100011\_TG023301 & 12/05/2020 & GJ 3779 & 11.63 \\
Sect.~\ref{sec:ETC} & CH\_PR100011\_TG023901 & 13/05/2020 & GJ 628 &  8.79\\
Sect.~\ref{sec:ETC} & CH\_PR100011\_TG024101 & 08/05/2020 & HD 125612 & 8.16 \\
Sect.~\ref{sec:ETC} & CH\_PR100011\_TG035801 & 24/06/2020 & HD 125612 & 8.16 \\
Sect.~\ref{sec:ETC} & CH\_PR100011\_TG024701 & 22/05/2020 & HD 181433 & 8.09 \\
Sect.~\ref{sec:ETC} & CH\_PR100011\_TG024702 & 01/06/2020 & HD 181433 & 8.09 \\
Sect.~\ref{sec:ETC} & CH\_PR100011\_TG036001 & 26/07/2020 & HD 175986 & 8.14 \\
Sect.~\ref{sec:ETC} & CH\_PR100011\_TG036201 & 17/07/2020 & HD 190360 & 5.33 \\
Sect.~\ref{sec:ETC} & CH\_PR100011\_TG035901 & 13/09/2020 & HD 215152 & 7.8 \\
Sect.~\ref{sec:MCperformance} & CH\_PR310080\_TG000101  & 03/01/2021 & HD 88111 & 8.97 \\
Sect.~\ref{sec:MCperformance} & CH\_PR320089\_TG000101  & 15/02/2022 & HD 88111 & 8.97  \\
Sect.~\ref{sec:MCperformance} & CH\_PR330096\_TG000101  & 15/03/2023 & HD 88111 & 8.97  \\
Sect.~\ref{sec:MCperformance} & CH\_PR310080\_TG000201  & 26/02/2021 & TYC 5502-1037-1 & 11.98 \\
Sect.~\ref{sec:MCperformance} & CH\_PR320089\_TG000301  & 28/02/2022 & TYC 5502-1037-1 & 11.98  \\
Sect.~\ref{sec:MCperformance} & CH\_PR330096\_TG000301  & 31/03/2023 & TYC 5502-1037-1 & 11.98  \\
Sect.~\ref{sec:MCperformance} & CH\_PR100016\_TG010201  & 02/11/2020 & WASP-12 & 11.6  \\
Sect.~\ref{sec:MCperformance} & CH\_PR100016\_TG010202  & 09/11/2020 & WASP-12 & 11.6  \\
Sect.~\ref{sec:MCperformance} & CH\_PR100016\_TG010203  & 10/11/2020 & WASP-12 & 11.6  \\
Sect.~\ref{sec:MCperformance} & CH\_PR100016\_TG010204  & 12/11/2020 & WASP-12 & 11.6  \\
Sect.~\ref{sec:MCperformance} & CH\_PR100016\_TG010205  & 20/11/2020 & WASP-12 & 11.6  \\
Sect.~\ref{sec:MCperformance} & CH\_PR100016\_TG010206  & 21/11/2020 & WASP-12 & 11.6  \\
Sect.~\ref{sec:MCperformance} & CH\_PR100016\_TG010207  & 29/11/2020 & WASP-12 & 11.6  \\
Sect.~\ref{sec:MCperformance} & CH\_PR100016\_TG010208  & 30/11/2020 & WASP-12 & 11.6  \\
Sect.~\ref{sec:MCperformance} & CH\_PR100016\_TG010209  & 04/12/2020 & WASP-12 & 11.6  \\
Sect.~\ref{sec:MCperformance} & CH\_PR100016\_TG010210  & 05/12/2020 & WASP-12 & 11.6  \\
Sect.~\ref{sec:MCperformance} & CH\_PR100016\_TG010211  & 06/12/2020 & WASP-12 & 11.6  \\
Sect.~\ref{sec:MCperformance} & CH\_PR100016\_TG010212  & 09/12/2020 & WASP-12 & 11.6  \\
Sect.~\ref{sec:MCperformance} & CH\_PR100013\_TG001201  & 12/01/2021 & WASP-12 & 11.6  \\
Sect.~\ref{sec:MCperformance} & CH\_PR100013\_TG001701  & 01/11/2021 & WASP-12 & 11.6  \\
Sect.~\ref{sec:MCperformance} & CH\_PR100013\_TG001702  & 10/11/2021 & WASP-12 & 11.6  \\
Sect.~\ref{sec:MCperformance} & CH\_PR100013\_TG001703  & 11/11/2021 & WASP-12 & 11.6  \\
Sect.~\ref{sec:MCperformance} & CH\_PR100016\_TG015001  & 05/12/2021 & WASP-12 & 11.6  \\
Sect.~\ref{sec:MCperformance} & CH\_PR100013\_TG001704  & 05/12/2021 & WASP-12 & 11.6  \\
Sect.~\ref{sec:MCperformance} & CH\_PR100013\_TG001705  & 07/12/2021 & WASP-12 & 11.6  \\
Sect.~\ref{sec:MCperformance} & CH\_PR100016\_TG015002  & 07/12/2021 & WASP-12 & 11.6  \\
Sect.~\ref{sec:MCperformance} & CH\_PR100013\_TG001706  & 15/12/2021 & WASP-12 & 11.6  \\
Sect.~\ref{sec:MCperformance} & CH\_PR100016\_TG015003  & 16/12/2021 & WASP-12 & 11.6  \\
Sect.~\ref{sec:MCperformance} & CH\_PR100013\_TG001707  & 16/12/2021 & WASP-12 & 11.6  \\
Sect.~\ref{sec:MCperformance} & CH\_PR100016\_TG015004  & 17/12/2021 & WASP-12 & 11.6  \\
Sect.~\ref{sec:MCperformance} & CH\_PR100016\_TG015005  & 24/12/2021 & WASP-12 & 11.6  \\
Sect.~\ref{sec:MCperformance} & CH\_PR100013\_TG001708  & 26/12/2021 & WASP-12 & 11.6  \\
Sect.~\ref{sec:MCperformance} & CH\_PR100013\_TG001709  & 29/12/2021 & WASP-12 & 11.6  \\
Sect.~\ref{sec:MCperformance} & CH\_PR100016\_TG015006  & 30/12/2021 & WASP-12 & 11.6  \\
Sect.~\ref{sec:MCperformance} & CH\_PR100013\_TG001710  & 30/12/2021 & WASP-12 & 11.6  \\
Sect.~\ref{sec:MCperformance} & CH\_PR100016\_TG015007  & 31/12/2021 & WASP-12 & 11.6  \\
Sect.~\ref{sec:MCperformance} & CH\_PR100013\_TG001711  & 04/01/2022 & WASP-12 & 11.6  \\
Sect.~\ref{sec:MCperformance} & CH\_PR100013\_TG001712  & 06/01/2022 & WASP-12 & 11.6  \\
Sect.~\ref{sec:MCperformance} & CH\_PR100016\_TG015008  & 07/01/2022 & WASP-12 & 11.6  \\
Sect.~\ref{sec:MCperformance} & CH\_PR100013\_TG001713  & 09/01/2022 & WASP-12 & 11.6  \\
Sect.~\ref{sec:MCperformance} & CH\_PR100013\_TG001714  & 19/01/2022 & WASP-12 & 11.6  \\
Sect.~\ref{sec:MCperformance} & CH\_PR100013\_TG001715  & 26/01/2022 & WASP-12 & 11.6  \\
Sect.~\ref{sec:MCperformance} & CH\_PR100013\_TG001716  & 29/01/2022 & WASP-12 & 11.6  \\
\caption{continued}\\
\hline
\noalign{\smallskip}
\multicolumn{1}{l}{Section} &
\multicolumn{1}{l}{File key} &
\multicolumn{1}{l}{Date} &
\multicolumn{1}{l}{Target} &
\multicolumn{1}{l}{Gmag} \\
\hline
\noalign{\smallskip}
Sect.~\ref{sec:MCperformance} & CH\_PR100013\_TG001717  & 03/02/2022 & WASP-12 & 11.6  \\
Sect.~\ref{sec:MCperformance} & CH\_PR100013\_TG001718  & 04/02/2022 & WASP-12 & 11.6  \\
Sect.~\ref{sec:MCperformance} & CH\_PR100013\_TG001719  & 08/02/2022 & WASP-12 & 11.6  \\
Sect.~\ref{sec:MCperformance} & CH\_PR100013\_TG001720  & 18/02/2022 & WASP-12 & 11.6  \\
Sect.~\ref{sec:MCperformance} & CH\_PR100016\_TG015601  & 23/02/2022 & WASP-12 & 11.6  \\
Sect.~\ref{sec:MCperformance} & CH\_PR100016\_TG015602  & 24/02/2022 & WASP-12 & 11.6  \\
Sect.~\ref{sec:MCperformance} & CH\_PR100016\_TG015603  & 22/11/2022 & WASP-12 & 11.6  \\
Sect.~\ref{sec:MCperformance} & CH\_PR100016\_TG015604  & 23/11/2022 & WASP-12 & 11.6  \\
Sect.~\ref{sec:MCperformance} & CH\_PR330093\_TG000201  & 21/12/2022 & WASP-12 & 11.6  \\
Sect.~\ref{sec:MCperformance} & CH\_PR100016\_TG015605  & 24/12/2022  & WASP-12 & 11.6  \\
Sects.~\ref{sec:RollCorr} and \ref{sec:PCA_ACF_PSF} &  CH\_PR100041\_TG000601 & 23/03/2020 & 55 Cnc & 5.71 \\
Sect.~\ref{sec:RollCorr} & CH\_PR300059\_TG001201 & 06/04/2020 & TYC2502-71-1 & 11.185 \\
Sect.~\ref{sec:BPCorr} & CH\_PR300005\_TG000201 & 27/02/2020 & TYC 5502-1037-1 & 11.98 \\
Sect.~\ref{sec:sat_trails} & CH\_PR100015\_TG009101 & 22/05/2020 & WASP-38 & 9.21 \\
Appendix~\ref{sec:NoiseBadPixels} & CH\_PR300061\_TG000301 & 11/04/2020 & TYC 5502-1037-1 & 11.98 \\
\noalign{\smallskip}
\hline
\end{longtable}

\end{appendix}
\end{document}